\documentclass[twoside,twocolumn,9pt]{article}

\pdfoutput=1

\usepackage{extsizes}
\usepackage[super,sort&compress,comma]{natbib} 
\usepackage[left=1.5cm, right=1.5cm, top=1.785cm, bottom=2.0cm]{geometry}
\usepackage{balance}
\usepackage{mathptmx}
\usepackage{sectsty}
\usepackage{graphicx} 
\usepackage{lastpage}
\usepackage[format=plain,justification=justified,singlelinecheck=false,font={stretch=1.125,small,sf},labelfont=bf,labelsep=space]{caption}
\usepackage{float}
\usepackage{fancyhdr}
\usepackage{fnpos}
\usepackage[english]{babel}
\addto{\captionsenglish}{%
  
}
\usepackage{array}
\usepackage{droidsans}
\usepackage{charter}
\usepackage[T1]{fontenc}
\usepackage[usenames,dvipsnames]{xcolor}
\usepackage{setspace}
\usepackage[compact]{titlesec}
\usepackage{hyperref}
\usepackage{epstopdf}%This line makes .eps figures into .pdf - please comment out if not required.

\definecolor{cream}{RGB}{222,217,201}
\usepackage{isomath}
\usepackage{amsmath}
\usepackage{amsbsy}
\usepackage{amssymb}
\usepackage{amscd}
\usepackage{amsfonts}
\usepackage{graphics}
\usepackage{verbatim}
\usepackage{subfigure}
\usepackage{xspace}
\usepackage{euscript}
\usepackage{alltt}
\usepackage{boxedminipage}
\usepackage{float}
\usepackage{color}
\usepackage[all]{xy}
\usepackage{t1enc}
\usepackage{exscale}
\usepackage{calc}
\usepackage{mdwlist}
\usepackage{units}
\usepackage[normalem]{ulem}
\usepackage[english]{babel}
\usepackage[utf8]{inputenc}
\usepackage{mathtools}
\usepackage{mathrsfs}
\usepackage{bm}
\usepackage[plain]{fancyref}
\usepackage{tabulary}
\usepackage{amsthm}
\usepackage{soul}
\usepackage{todonotes}
\newcommand*{\fancyrefapplabelprefix}{app}
\frefformat{plain}{\fancyrefapplabelprefix}{appendix~#1}
\Frefformat{plain}{\fancyrefapplabelprefix}{Appendix~#1}
\frefformat{main}{\fancyrefapplabelprefix}{appendix~#1}
\Frefformat{main}{\fancyrefapplabelprefix}{Appendix~#1}

\renewcommand{\mathbold}{\bm}
\renewcommand{\mathbf}{\bm}

\usepackage{isomath}
\usepackage{amsmath}
\usepackage{amssymb}
\usepackage{amscd}
\usepackage{amsfonts}

\newcommand{\nhat}{\hat{\mathbold n}}

\newcommand{\beq}{\begin{equation}}
\newcommand{\eeq}{\end{equation}}
\newcommand{\beqs}{\begin{eqnarray}}
\newcommand{\eeqs}{\end{eqnarray}}
\newcommand{\half}{\frac{1}{2}}

\newcommand{\bfchi}{\mathbold {\chi}}

\newcommand{\bfmu}{\mathbold {\mu}}

\newcommand{\erf}{\mathop{\rm erf}\nolimits}

\newcommand{\divergence}{\mathop{\rm div}\nolimits}

\newcommand{\dm}{\ \mathrm{d}}

\newcommand{\bfe}{{\mathbold e}}
\newcommand{\bff}{{\mathbold f}}

\newcommand{\bfm}{{\mathbold m}}
\newcommand{\bfn}{{\mathbold n}}

\newcommand{\bfp}{{\mathbold p}}

\newcommand{\bfr}{{\mathbold r}}

\newcommand{\bfv}{{\mathbold v}}

\newcommand{\bfx}{{\mathbold x}}

\newcommand{\bfE}{{\mathbold E}}

\newcommand{\bfI}{{\mathbold I}}

\newcommand{\bfR}{{\mathbold R}}

\newcommand{\J}{J} % jacobian

\newcommand{\takepartial}[2]{\frac{\partial #1}{\partial #2}}
\newcommand{\takepartialflat}[2]{\partial #1 / \partial #2}

\newcommand{\A}{\mathcal{F}}

\newcommand{\AHelm}{\mathcal{A}}

\newcommand{\T}{T}

\newcommand{\Sent}{\mathcal{S}}

\newcommand{\kB}{k}

\newcommand{\pfunc}{\mathcal{Z}}

\newcommand{\nstates}{\Omega}

\newcommand{\mlen}{b}

\newcommand{\N}{n}

\newcommand{\pop}{m}
\newcommand{\popnum}{m}

\newcommand{\n}{\mathcal{N}}

\newcommand{\density}{\rho}

\newcommand{\force}{\mathfrak{f}}

\newcommand{\rmag}{r}
\newcommand{\rvec}{\bfr}
\newcommand{\rz}{\rmag_3}
\newcommand{\rx}{\rmag_1}

\newcommand{\rdir}{\hat{\rvec}}

\newcommand{\stch}{\gamma}
\newcommand{\stchx}{\stch_1}
\newcommand{\stchz}{\stch_3}

\newcommand{\unitsphere}{\mathbb{S}^2}

\newcommand{\um}{u}

\newcommand{\nvec}{\hat{\bfv}}

\newcommand{\U}{U}

\newcommand{\sussymbol}{\chi}
\newcommand{\sus}[1]{\ifthenelse{#1 < 2}{\sussymbol_{\parallel}}{\sussymbol_{\perp}}}
\newcommand{\sustens}{\boldsymbol{\chi}}
\newcommand{\dsus}{\Delta \sussymbol}

\newcommand{\emag}{E}
\newcommand{\efield}{\bfE}
\newcommand{\edir}{\hat{\efield}}
\newcommand{\ex}{\emag_1}

\newcommand{\ez}{\emag_3}
\newcommand{\ezeromag}{\emag_0}
\newcommand{\ezero}{\efield_0}
\newcommand{\ezerodir}{\hat{\efield}_0}
\newcommand{\ezerox}{\emag_{01}}

\newcommand{\ezeroz}{\emag_{03}}

\newcommand{\dipolemag}{\mu}
\newcommand{\dipole}{\boldsymbol{\dipolemag}}

\newcommand{\chainpolarmag}{p}
\newcommand{\chainpolar}{\bfp}
\newcommand{\chainpolarsm}{\chainpolar_{s\multmag}}
\newcommand{\chainpolarkg}{\chainpolar_{kg}}
\newcommand{\chainpolaras}{\chainpolar_{as}}

\newcommand{\funodimso}{g_{s \unodim}}

\newcommand{\unodim}{\kappa}
\newcommand{\uxnodim}{\unodim_1}
\newcommand{\uznodim}{\unodim_{3}}
\newcommand{\uxznodim}{\unodim_{13}}
\newcommand{\uOnodim}{\unodim_{\perp}}
\newcommand{\uslnodim}{\unodim}
\newcommand{\duzux}{\Delta}
\newcommand{\uzpo}{\zeta}

\newcommand{\etorangle}{\psi}

\newcommand{\x}{\mathbf{x}}
\newcommand{\xk}[1]{\x_{#1}}
\newcommand{\Jij}[2]{J_{#1 #2}}
\newcommand{\froot}{\mathbf{f}}
\newcommand{\frootk}[1]{\froot_{#1}}

\newcommand{\im}{{i\mkern1mu}}

\newcommand{\erfi}{\mathrm{erfi}}

\newcommand{\dirac}{\delta}

\newcommand{\multmag}{\tau}
\newcommand{\mults}{\boldsymbol{\multmag}}
\newcommand{\nmult}{\nu}
\newcommand{\zmult}{\lambda}
\newcommand{\xmult}{\alpha}

\newcommand{\C}{C}

\newcommand*\df{\mathop{}\!\mathrm{d}}

\newcommand{\ds}{\Delta}
\newcommand{\dbar}{\mathrm{d}\hspace*{-0.08em}\bar{}\hspace*{0.1em}}
\newcommand{\dQ}{\dbar Q}
\newcommand{\dW}{\dbar W}

\newcommand{\azi}{\phi}
\newcommand{\polar}{\theta}

\newcommand{\collect}[1]{\begin{Bmatrix}#1\end{Bmatrix}}

\newcommand{\Lang}{\mathcal{L}}
\newcommand{\Langinv}{\Lang^{-1}}
\newcommand{\Langinvs}{\Langinv\left(\stch\right)}
\newcommand{\Langinvso}{\zmult_0}

\newcommand{\csch}{\mbox{ csch }}

\newcommand{\generic}{\Box}

\newcommand{\eone}{\bfe_1}
\newcommand{\etwo}{\bfe_2}
\newcommand{\ethree}{\bfe_3}

\newcommand{\densityso}{\density}
\newcommand{\densitysl}{\density}
\newcommand{\densitysm}{\density_{s\multmag}}
\newcommand{\densitykg}{\density_{KG}}
\newcommand{\weight}{w}
\newcommand{\wsm}{\weight_{s\multmag}}
\newcommand{\wkg}{\weight_{KG}}

\newcommand{\Ukg}{\mathcal{U}_{KG}}

\newcommand{\Skg}{\Sent_{KG}}
\newcommand{\Asm}{\A_{s\multmag}}
\newcommand{\Akg}{\A_{KG}}

\newcommand{\Aas}{\A_{as}}

\newcommand{\I}[1]{I_{#1}}

\newcommand{\ca}{a_1}

\newcommand{\cc}{a_2}
\newcommand{\cd}{a_3}
\newcommand{\xce}{a_4}

\newcommand{\da}{b_1}
\newcommand{\db}{b_2}
\newcommand{\dc}{b_3}

\newcommand{\intoverS}[1]{\int_{0}^{\pi} \df \polar \int_{0}^{2\pi} \df \azi \mbox{ } #1 \sin \polar}
\newcommand{\intoverSns}[1]{\int_{\unitsphere} \df{A} \mbox{ } #1}

\newcommand{\langevin}[1]{\coth#1 - 1/#1}

\newcommand{\zmultzero}{\Langinv\left(\stch\right)}

\newcommand{\erfw}{\erf \left(\sqrt{\uslnodim}\right)}
\newcommand{\Csl}{\left(\N \sqrt{\uslnodim}\right)/\left(2 \pi^{3/2} \erfw\right)}
\newcommand{\zmultsl}{\left(2\sqrt{\pi} \stchz \uslnodim e^{\uslnodim} \erfw\right)/\left(\sqrt{\pi} e^{\uslnodim} \erfw - 2\sqrt{\uslnodim}\right)}
\newcommand{\xmultsl}{\left(4\sqrt{\pi} \stchx \uslnodim e^{\uslnodim} \erfw\right)/\left(\sqrt{\pi} \left(2\uslnodim-1\right)  e^{\uslnodim} \erfw + 2\sqrt{\uslnodim}\right)}

\newcommand{\FigureWidthNumericalGraph}{\linewidth}
\newcommand{\FigureWidthTwoColsNumericalGraph}{0.45 \linewidth}

\newcommand\scalemath[2]{\scalebox{#1}{\mbox{\ensuremath{\displaystyle #2}}}}
\numberwithin{equation}{section}

\begin{document}
\graphicspath{ {./} }

\pagestyle{fancy}
\thispagestyle{plain}
\fancypagestyle{plain}{
%%%HEADER%%%
\renewcommand{\headrulewidth}{0pt}
}
%%%END OF HEADER%%%

%%%PAGE SETUP - Please do not change any commands within this section%%%
\makeFNbottom
\makeatletter
\renewcommand\LARGE{\@setfontsize\LARGE{15pt}{17}}
\renewcommand\Large{\@setfontsize\Large{12pt}{14}}
\renewcommand\large{\@setfontsize\large{10pt}{12}}
\renewcommand\footnotesize{\@setfontsize\footnotesize{7pt}{10}}
\makeatother

\renewcommand{\thefootnote}{\fnsymbol{footnote}}
\renewcommand\footnoterule{\vspace*{1pt}% 
\color{cream}\hrule width 3.5in height 0.4pt \color{black}\vspace*{5pt}} 
\setcounter{secnumdepth}{5}

\makeatletter 
\renewcommand\@biblabel[1]{#1}            
\renewcommand\@makefntext[1]% 
{\noindent\makebox[0pt][r]{\@thefnmark\,}#1}
\makeatother 
\renewcommand{\figurename}{\small{Fig.}~}
\sectionfont{\sffamily\Large}
\subsectionfont{\normalsize}
\subsubsectionfont{\bf}
\setstretch{1.125} %In particular, please do not alter this line.
\setlength{\skip\footins}{0.8cm}
\setlength{\footnotesep}{0.25cm}
\setlength{\jot}{10pt}
\titlespacing*{\section}{0pt}{4pt}{4pt}
\titlespacing*{\subsection}{0pt}{15pt}{1pt}
%%%END OF PAGE SETUP%%%

%%%FOOTER%%%
\fancyfoot{}
\fancyfoot[LO,RE]{\vspace{-7.1pt}\thepage}
\fancyfoot[CO]{\vspace{-7.1pt}\hspace{9.2cm} To appear in Soft Matter (\url{doi.org/10.1039/D0SM00845A})}
\fancyfoot[CE]{\vspace{-7.2pt}\hspace{-10.2cm}To appear in Soft Matter (DOI: 10.1039/D0SM00845A)}
\fancyfoot[RO]{}
\fancyfoot[LE]{}
\fancyhead{}
\renewcommand{\headrulewidth}{0pt} 
\renewcommand{\footrulewidth}{0pt}
\setlength{\arrayrulewidth}{1pt}
\setlength{\columnsep}{6.5mm}
\setlength\bibsep{1pt}
%%%END OF FOOTER%%%

%%%FIGURE SETUP - please do not change any commands within this section%%%
\makeatletter 
\newlength{\figrulesep} 
\setlength{\figrulesep}{0.5\textfloatsep} 

\newcommand{\topfigrule}{\vspace*{-1pt}% 
\noindent{\color{cream}\rule[-\figrulesep]{\columnwidth}{1.5pt}} }

\newcommand{\botfigrule}{\vspace*{-2pt}% 
\noindent{\color{cream}\rule[\figrulesep]{\columnwidth}{1.5pt}} }

\newcommand{\dblfigrule}{\vspace*{-1pt}% 
\noindent{\color{cream}\rule[-\figrulesep]{\textwidth}{1.5pt}} }

\makeatother
%%%END OF FIGURE SETUP%%%

%%%TITLE, AUTHORS AND ABSTRACT%%%
\twocolumn[
  \begin{@twocolumnfalse}
% {\includegraphics[height=30pt]{head_foot/SM}\hfill\raisebox{0pt}[0pt][0pt]{\includegraphics[height=55pt]{head_foot/RSC_LOGO_CMYK}}\\[1ex]
% % \includegraphics[width=18.5cm]{head_foot/header_bar}
% }
\sffamily
\begin{tabular}{m{2.5cm} p{15.5cm} }

  %%% TITLE %%%
& \noindent\LARGE{\textbf{Statistical Mechanical Analysis of the Electromechanical Coupling in an Electrically-Responsive Polymer Chain}} \\
\vspace{0.3cm} & \vspace{0.3cm} \\

 %%% AUTHORS %%%
 & \noindent\large{Matthew Grasinger\textit{$^{a}$}\textit{$^{d}$} and Kaushik Dayal\textit{$^{a}$}\textit{$^{b}$}\textit{$^{c}$}\textit{$^{e}$}} \\
\vspace{0.3cm} & \vspace{0.3cm} \\

 %%% ABSTRACT %%%
\today & \noindent\normalsize{
	Polymeric materials that couple deformation and electrostatics have the potential for use in soft sensors and actuators with potential applications ranging from robotic, biomedical, energy, aerospace and automotive technologies.
	In contrast to the mechanics of polymers that has been studied using statistical mechanics approaches for decades, the coupled response under deformation and electrical field has largely been modeled only phenomenologically at the continuum scale.
	In this work, we examine the physics of the coupled deformation and electrical response of an electrically-responsive polymer chain using statistical mechanics.
	We begin with a simple anisotropic model for the electrostatic dipole response to electric field of a single monomer, and use a separation of energy scales between the electrostatic field energy and the induced dipole field energy to reduce the nonlocal and infinite-dimensional statistical averaging to a simpler local finite-dimensional averaging.
	In this simplified setting, we derive the equations of the most likely monomer orientation density using the maximum term approximation, and a chain free energy is derived using this approximation.
	These equations are investigated numerically and the results provide insight into the physics of electro-mechanically coupled elastomer chains.
	Closed-form approximations are also developed in the limit of small electrical energy with respect to thermal energy; in the limit of small mechanical tension force acting on the chain; and using asymptotic matching for general chain conditions.
} \\
\end{tabular}

 \end{@twocolumnfalse} \vspace{0.6cm}

  ]
%%%END OF TITLE, AUTHORS AND ABSTRACT%%%

%%%FONT SETUP - please do not change any commands within this section
\renewcommand*\rmdefault{bch}\normalfont\upshape
\rmfamily
\section*{}
\vspace{-1cm}

%%%FOOTNOTES%%%

\footnotetext{\textit{$^{a}$~Department of Civil and Environmental Engineering, Carnegie Mellon University}}
\footnotetext{\textit{$^{b}$~Center for Nonlinear Analysis, Carnegie Mellon University}}
\footnotetext{\textit{$^{c}$~Department of Materials Science and Engineering, Carnegie Mellon University}}
\footnotetext{\textit{$^{d}$~Email: grasingerm@gmail.com}}
\footnotetext{\textit{$^{e}$~Email: Kaushik.Dayal@cmu.edu}}

%%%END OF FOOTNOTES%%%

%%%MAIN TEXT%%%%

%%%%%%%%%%%%%%%%%%%%%%%%%%%%%%%%%%%%%%%%%%%%%%%%
%%%%%%%%%%%%%%%%%%%%%%%%%%%%%%%%%%%%%%%%%%%%%%%%
%%%%%%%%%%%%%%%%%%%%%%%%%%%%%%%%%%%%%%%%%%%%%%%%
%%%%%%%%%%%%%%%%%%%%%%%%%%%%%%%%%%%%%%%%%%%%%%%%
\section{Introduction} \label{sec:introduction}

Soft functional polymeric materials that couple between deformation to electric fields, magnetic fields, or illumination, have emerged as leading candidates for sensors and actuators with applications across soft robotics, biomedical devices, biologically inspired robots, advanced prosthetics, and various other technologies \cite{bar-cohen2001electroactive,carpi2011electroactive,kim2007electroactive,huang2012giant,majidi2014soft,bartlett2016stretchable,lopez2014elastic,ware2016localized,erol2019microstructure,castaneda2011homogenization,galipeau2013finite}.
The goal of this paper is to develop a statistical mechanics model of an electrically-responsive polymer chain.
Our motivation is derived from dielectric elastomers, a class of soft elastomers that deform and can provide actuation when subject to electrical loads.
Unlike traditional actuation that can require complex mechanisms and difficult assembly with multiple materials, the actuation mechanism of dielectric elastomers is intrinsic to the material.
Broadly, they behave similarly to that of natural muscles, and stretch when a voltage is applied and contract when it is removed.
In addition, they are naturally soft, lightweight, compliant, easily shaped, and can undergo large recoverable deformations.

However, despite these advantages, electro-responsive polymers are limited by a relatively weak electromechanical coupling, so that large voltages are often required to achieve meaningful actuation~\cite{bar-cohen2001electroactive}.
An understanding of the fundamental physics governing the response of these materials, starting from the level of individual monomers and using statistical mechanics to obtain continuum free energies, can provide new insights and potentially provide new ways to process and tailor the microstructure to obtain enhanced coupling between deformation and electric field.

Prior work in modeling of electro-responsive polymeric materials has consisted primarily of continuum approaches \cite{toupin1956elastic,dorfmann2014nonlinear,henann2013modeling,zurlo2017catastrophic, liu2018emergent,krichen2019liquid,li2015geometrically,darbaniyan2019designing,shmuel2013axisymmetric,fox2008dynamic}.
Broadly, the general form of the energy density is formulated based on considerations of symmetry and building on established models from rubber elasticity.
These approaches provide valuable insights into the behavior in complex geometries and with realistic boundary conditions, i.e., they bridge from the continuum scale to the specimen scale.
However, their starting point at the continuum scale prevents them from being used to understand the role of the network and polymer chain behavior.

On the other hand, there is an extremely rich literature on using statistical mechanics to address the {\em purely mechanical} response of polymer chains and networks and how these give rise to observed rubber elasticity \cite{treloar1975physics,kuhn1942beziehungen,weiner2012statistical,marckmann2006comparison}.
The first works that applied statistical mechanics techniques to the setting of electromechanically coupled polymers appear to be \citet{cohen2016electroelasticity} and \citet{cohen2016electromechanical}.
Broadly, these papers derive the most likely density of monomer orientations; that is, the density of monomer orientations that minimizes the entropy of the chain.
Such a derivation naturally involves optimization with respect to constraints; hence, the authors use the method of Lagrange multipliers.
The equations to find the Lagrange multipliers are formidable, and hence they assume that some quantities are small and use Taylor expansion approximations.
Physically, this results in an approximate density of monomer orientations that is exact when there is no stretch / tension in the chain.

While References \citenum{cohen2016electroelasticity} and \citenum{cohen2016electromechanical} therefore provide an important starting point, we go beyond their approach in two important ways.
First, we examine various regimes, including those in which we do not assume that the stretch / tension in the chain is small.
Second, while those works provide many useful insights, they do not provide an (approximate) expression for the free energy, which is an important contribution of this work.

We note, however, an important body of literature in the statistical mechanics of polyelectrolytes, i.e. studying the statistical mechanics of charged bio-macromolecules typically in solution \cite{shen2017electrostatic,wang2004self,argudo2012dependence}.
While there are several common features between the problem of interest here and the literature on polyelectrolytes, there are also two key differences.
First, the macromolecules considered in the polyelectrolyte literature typically consist of {\em fixed} charge distributions, in the sense that the charge distributions are inseparable from the polymer and are present even in the absence of an applied electric field; in contrast, the monomers in our model are {\em electro-responsive}, i.e., the dipole carried by the monomers in our model depends strongly on the local electric field as well as the orientation of the monomer with respect to the field.
Second, we consider polymer chains that are cross-linked in a network, as opposed to individually in solution.
As a consequence, the ensembles of most relevance are those with a fixed end-to-end vector, because the chain configuration is constrained by the network in which it is embedded, as is typical in the modeling of the mechanical response of polymers \cite{treloar1975physics}.

\begin{figure}[htb!]
	\centering
	\includegraphics[width=\FigureWidthTwoColsNumericalGraph]{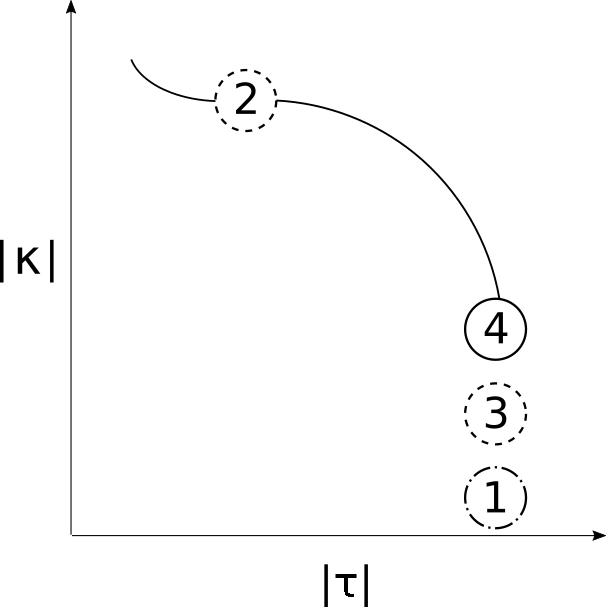}
	\caption{Schematic of the limits for which closed-form approximations have been developed.
		The axis labeled $|\unodim|$ represents the magnitude of the electrical energy with respect to the thermal energy and the axis labeled $|\mults|$ represents the dimensionless tension in the DE chain.
		The approximation labeled (1) corresponds to the monomer density function derived in~\citet{kuhn1942beziehungen} (summarized in \Fref{sec:kuhn-and-grun});
		(2) corresponds to the limit explored in~\citet{cohen2016electroelasticity} and subsequently in \Fref{sec:small-lambda};
		(3) corresponds to the limit investigated in  \Fref{sec:small-omega}; and \Fref{sec:patched} aims to develop an approximation that is valid at (4), (2), and in the space between.}
	\label{fig:current-state}
\end{figure}

In the present work, we employ the maximum-term approach \cite{hill1986statistical,taylor2016non} to derive the most likely density of monomer orientations, and from this an approximation of the chain free energy; that is, we derive a mean-field theory.
Although, as in~\citet{cohen2016electroelasticity}, the Lagrange multipliers are not determined exactly, we investigate their character using using numerical methods; and then proceed to derive closed-form approximations in the limits of (1.) the electrical energy as small compared with thermal energy and (2.) the limit of small chain tension.
Lastly, using the intuition developed from the numerical investigation, we use asymptotic matching to interpolate between the exact solutions in the limit regimes to derive closed-form approximations that remain accurate for many different electrical inputs, chain orientations, and chain stretches.
Fig. \ref{fig:current-state} shows a schematic of the limits in which closed-form approximations have been developed.
The $|\unodim|$-axis represents the magnitude of the electrical energy with respect to the thermal energy.
The $|\mults|$-axis, represents the dimensionless tension in the DE chain.
The approximation labeled (1) corresponds to the classical monomer density function derived in~\citet{kuhn1942beziehungen} (summarized in  \Fref{sec:kuhn-and-grun});
(2) corresponds to the limit explored in~\citet{cohen2016electroelasticity} and subsequently in  \Fref{sec:small-lambda};
(3) corresponds to the limit investigated in  \Fref{sec:small-omega}; and \Fref{sec:patched} aims to develop an approximation that is valid at (4), (2), and in the space between.

\paragraph*{Organization of the Paper}

\begin{itemize}

	\item In \Fref{sec:formulation}, we formulate the potential energy of the polymer chain, accounting for electrical interactions, and making simplifications to reduce from a nonlocal electrostatic problem to a local one;

    \item In \Fref{sec:stat-mech-formulation}, we describe the statistical mechanical averaging to be conducted on the potential energy and the constraints that we apply;
    
	\item In \Fref{sec:numerical}, the governing equations are approximated numerically.
	Observations are made regarding the character of the solutions, with an emphasis on their physical implications.

	\item In \Fref{sec:small-omega}, a closed-form approximation of the governing equations is derived by assuming the chain electrical energy is small compared to thermal energy.
	\item In \Fref{sec:small-lambda}, a closed-form approximation of the governing equations is derived by assuming the chain tension is small.
	\item In \Fref{sec:patched}, we use results developed in previous sections to guide our thinking in developing approximations that are accurate for more general chain conditions.
\end{itemize}

%%% NOMENCLATURE %%%
%\smallskip
%\begin{multicols}{2}
%\printnomenclature
%\end{multicols}

%%%%%%%%%%%%%%%%%%%%%%%%%%%%%%%%%%%%%%%%%%%%%%%%
%%%%%%%%%%%%%%%%%%%%%%%%%%%%%%%%%%%%%%%%%%%%%%%%
%%%%%%%%%%%%%%%%%%%%%%%%%%%%%%%%%%%%%%%%%%%%%%%%
%%%%%%%%%%%%%%%%%%%%%%%%%%%%%%%%%%%%%%%%%%%%%%%%
\section{Formulation of the Potential Energy}
\label{sec:formulation}

We consider a polymer chain, subject to an electric field, and composed of $n$ identical monomers that each carry a dipole.
We use the ensemble with specified temperature $T$, specified average electric field $\bfE_0$ (discussed further in Appendix \ref{sec:constant-E-ensemble}), and specified end-to-end vector $\bfr$.
We assume that the chain is contained in a spatial volume $\Omega$ with boundary $\partial\Omega$ with unit outward normal $\hat{\bfm}$.

The degrees of freedom describing the configuration of the polymer chain are: (1.) the spatial position of the $i$-th monomer, denoted $\bfx_i$;
(2.) the orientation of the $i$-th monomer, denoted $\nhat_i$;
(3.) the point dipole carried by the $i$-th monomer, denoted $\bfmu_i$;
and (4.) the electric field, -$\nabla\phi(\bfx)$, which, for now, is a general function of position $\bfx$.
All of these can be varied independently\footnote[1]{Statistical field theory would be required to treat the averaging over the electric field \cite{kardar2007statistical-fields}, but we make simplifications that allow us to avoid this.}.

We make the following assumptions in our model at this stage.
(1.) Following standard practice, we have implicitly assumed above that only the orientation of the monomers is relevant and that there is no stretching.
That is, stretching of monomers costs energy that is much larger than $kT$, and hence the chain is assumed to be inextensible.
(2.) Again following standard practice, we assume for simplicity that the bending energy -- i.e., energy associated with the change in orientation of the monomers along the chain -- is much less than $kT$ and can hence be neglected.
(3.) While dipole effects are due to electronic and nuclear motion and hence can have interactions between monomers \cite{babaei2017computing}, we assume for simplicity that the dipole induced in a monomer can be modeled without regard to the chain environment; equivalently, we assume that the electrical energy of the monomers can be decomposed additively.

Under these assumptions, the potential energy of a microstate can be written:
\begin{equation}
\label{eqn:energy-1}
    U 
    = 
    \sum_{i=1}^{n} \left( \tilde u(\bfmu_i,\hat\bfn_i) - \bfmu_i \cdot \bfE_i \right) 
    +
    \frac{1}{2} \int_\Omega |\nabla\phi|^2 \dm\Omega
\end{equation}
where we have defined $\bfE_i = - \left. \nabla\phi \right|_{\bfx_i}$ as the local electrical field at the location of the monomer, and where we have chosen to work in Gaussian units (i.e. unit system in which $\epsilon_0$ is unity).
The first term in the summation is the energy $\tilde u$ required to separate charges to form a dipole $\bfmu_i$, and will be discussed further in Section \ref{sec:monomer-E-response}.
The second term in the summation is the energy of interaction between the local electric field at the monomer location and the induced dipole.
The volume integral is the electrical field energy.

%%%%%%%%%%%%%%%%%%%%%%%%%%%%%%%%%%%%%%%%%%%%%%%%
%%%%%%%%%%%%%%%%%%%%%%%%%%%%%%%%%%%%%%%%%%%%%%%%
%%%%%%%%%%%%%%%%%%%%%%%%%%%%%%%%%%%%%%%%%%%%%%%%
%%%%%%%%%%%%%%%%%%%%%%%%%%%%%%%%%%%%%%%%%%%%%%%%
\subsection{Dipole Response of a Single Monomer to an Electric Field}
\label{sec:monomer-E-response}

Following the classical Born-Oppenheimer approximation, we assume that electrons reach their ground state configuration -- under electric field -- rapidly compared to the timescale of the thermal motion of the atoms \cite{yu2016energy}.
That is, thermal effects are assumed to play no role in the response of the electronic structure of the monomer to the electric field.
Since the polarization response of the monomer is essentially due to the reconfiguration of electrons under electric field, this translates to assuming that the polarization response to the electric field is independent of temperature and thermal effects.
While we do not consider quantum effects explicitly, the Born-Oppenheimer approximation justifies neglecting thermal effects in modeling the dipole response of a single monomer to an electric field.
In other words, we are assuming that the first excited state of the electrons has energy much larger than $kT$, and the system is always in the ground state with respect to the electron configuration.

An important implication of this assumption is that the dipole moment of an individual monomer, $\bfmu_i$, is uniquely determined -- through energy minimization -- given $\hat\bfn_i$, $\bfE_i$, and $\bfx_i$.
It is no longer an independent degree of freedom.
To find an expression for $\bfmu_i$ in terms of the other quantities, we differentiate \eqref{eqn:energy-1} with respect to $\bfmu_i$ to obtain the polarization response $\takepartialflat{\tilde u}{\bfmu_i} = \bfE_i$ at the ground state.

In this paper, we model the monomer response through the choice:
\begin{equation} \label{eq:dipole-response}
    \tilde u(\bfmu_i,\hat\bfn_i)
    =
    \half \bfmu_i \cdot \bfchi^{-1}(\hat\bfn_i) \bfmu_i 
    \Rightarrow
    \bfmu_i = \bfchi(\hat\bfn_i) \bfE_i
\end{equation}
where $\bfchi$ is the tensorial polarizability of the monomer.
Following \citet{cohen2016electroelasticity}, we model the tensor $\bfchi$ as transversely isotropic: $\bfchi(\hat\bfn) = \chi_\parallel \hat\bfn\otimes\hat\bfn + \chi_\perp (\bfI - \hat\bfn\otimes\hat\bfn)$.
The non-negative material constants $\chi_\parallel$ and $\chi_\perp$ are measures of the susceptibility along the monomer direction and transverse to the monomer direction respectively.
We refer to monomers with $\chi_\parallel > \chi_\perp$ as uniaxial, and monomers with $\chi_\parallel < \chi_\perp$ as transverse isotropic (TI).
For this choice of $\bfchi$, we have the ground state energy, $u = \tilde u - \bfmu_i \cdot \bfE_i$, as:
\begin{equation}
\label{eq:monomer-energy}
    u(\hat\bfn_i, \bfE_i) 
    =
    -\half\bfmu_i(\hat\bfn_i,\bfE_i)\cdot\bfE_i
    =
    \half \Delta\chi \left( \bfE_i\cdot\hat\bfn \right)^2 - \half \chi_\perp |\bfE_i|^2
\end{equation}
where $\Delta\chi = \chi_\perp - \chi_\parallel$.

If $\chi_\parallel > \chi_\perp$, i.e. the monomer is uniaxial, then the monomer has minimum energy when $\hat\bfn$ is parallel or anti-parallel to $\bfE_i$, and maximum energy when $\hat\bfn$ lies in the plane orthogonal to $\bfE_i$.
If $\chi_\parallel < \chi_\perp$, i.e. the monomer is TI, the situation is reversed: the minimum energy state is when $\hat\bfn$ lies in the plane orthogonal to $\bfE_i$, and the maximum energy state is when $\hat\bfn$ is parallel or anti-parallel to $\bfE_i$.
The interplay between the electrical energy and thermally-driven oscillations will lead to the observed average configurations.

We emphasize that while the relation between the polarization and the local electric field is linear, hence allowing us to easily eliminate either one of these as an independent variable from the model, the resulting energy is nonlinear in $\hat\bfn$.
This is a consequence of frame-indifference and has the same origin as the nonlinearity in nonlinear elasticity.

%%%%%%%%%%%%%%%%%%%%%%%%%%%%%%%%%%%%%%%%%%%%%%%%
%%%%%%%%%%%%%%%%%%%%%%%%%%%%%%%%%%%%%%%%%%%%%%%%
%%%%%%%%%%%%%%%%%%%%%%%%%%%%%%%%%%%%%%%%%%%%%%%%
%%%%%%%%%%%%%%%%%%%%%%%%%%%%%%%%%%%%%%%%%%%%%%%%
\subsection{Multiscale Structure of the Electrical Field Energy, and Consequent Nonlocal-to-Local Decoupling}

Under the assumptions in Section \ref{sec:monomer-E-response},  the potential energy from \eqref{eqn:energy-1} reduces to:
\begin{equation}
\label{eqn:energy-2}
    U 
    = 
    \sum_{i=1}^{n} \left( - \half \bfE_i \cdot \bfchi(\hat\bfn_i) \bfE_i \right) 
    +
    \frac{1}{2} \int_\Omega |\nabla\phi|^2 \dm\Omega
\end{equation}
The boundary conditions on $\phi$ correspond to the specified average electric field ensemble, described in Appendix \ref{sec:constant-E-ensemble}.

The energy posed in \eqref{eqn:energy-2} has a highly nonlocal structure \cite{marshall2014atomistic,james1990frustration,yang2011completely,liu2013energy}.
Physically, this is due to the fact that we need to solve the electrostatics equation to find the field at every monomer location in $\Omega$.
We can see this by examining the ground state with respect to the electric potential.
Taking the variation $\phi \to \phi + \psi$ and requiring this to be $0$ for all variations $\psi$, we find that:
\begin{equation}
\label{eqn:nonlocal-electrostatics}
\begin{split}
    0 = &
    \sum_{i=1}^{n} - \bfchi \left. \bfE \right|_{\bfx_i} \cdot \left. \nabla\psi\right|_{\bfx_i} + \int_\Omega \nabla \phi \cdot \nabla \psi \dm\Omega
    \\
    & \Rightarrow 0 =
    \int_\Omega \sum_{i=1}^{n} \left(- \bfchi \bfE \cdot \nabla\psi \right) \delta_{\bfx_i} \dm\Omega + \int_\Omega \nabla \phi \cdot \nabla \psi \dm\Omega
    \\
    & \Rightarrow 0 =
    \int_\Omega \psi \divergence
    \underbrace{\left[ \sum_{i=1}^{n} \left( \bfchi \bfE \delta_{\bfx_i} \right) \right]}_{\tilde\bfp(\bfx)}
    \dm\Omega - \int_\Omega \psi \divergence \nabla \phi \dm\Omega
    \\
    & \Rightarrow \divergence \nabla \phi = \divergence \tilde\bfp, \quad \text{ subject to boundary conditions.} 
\end{split}
\end{equation}
Here, $\delta_{\bfx_i}$ is the Dirac mass located at $\bfx_i$, and $\tilde\bfp(\bfx)$ is the dipole moment of the chain, treated here as a field through the use of Dirac masses that represent the point dipoles carried by the monomers, to be interpreted in the sense of distributions.

This final equation shows the nature of the nonlocal problem: to evaluate the energy in \eqref{eqn:energy-2}, we need to solve a boundary value problem.
Therefore, statistical mechanical averaging over the energy will need to average over all fields $\phi$ that are consistent with the specified average electric field ensemble.
This is challenging and would require methods of statistical field theory.
To simplify this, we follow \citet{cohen2016electroelasticity} to assume that the average value of the electric field is much larger than the field generated by the monomeric dipoles.
This is motivated by the fact that typical applications of dielectric elastomers use very large externally-applied fields.
Heuristically, we propose to ignore dipole-dipole interactions as being much smaller than the dipole interaction with the external field; note that both of these interactions are naturally present in \eqref{eqn:nonlocal-electrostatics}.
The consequence of this simplification is that we solve an extremely simple PDE to find the electric field, and then conduct statistical averaging while holding this electric field constant. 

While it is straightforward to simply remove these interactions from the field equation, i.e. we can easily solve for the electric field from the electrostatic boundary value problem neglecting dipoles, and then account for the dipoles interacting with the obtained electric field, a critical drawback of such a procedure is the lack of an energetic basis.
As we require a clear energetic basis to perform statistical mechanical averaging, we instead present the argument below that exploits a separation of scales in the contributions to the electrical field energy.
We see that this leads to a simplified local structure for the energy.

We begin by assuming that the energy in \eqref{eqn:energy-2} has a separation of scales such that the energy in forming a dipole by separating charges is of order $kT$, while the stored energy of the field in vacuum is of much higher order.
A separation of energy scales leading to a broken symmetry in coupling has an analogy in the deformation of macroscopic slender structures \cite{audoly2010geometry,kamien2002geometry,witten2007stress,steigmann2008two}.
Stretching costs far more energy than bending in such systems; consequently, ground states can be computed by minimizing the stretching energy to find the centerline curve, and then minimizing the bending energy -- with this centerline curve as a constraint -- to find the configuration of the normal fibers.
Therefore, stretch is insensitive to, but a forcing for, bending.
Translating this to our context, we first minimize with respect to the field energy to find the electric field, and then use this electric field as a constraint in performing the statistical averaging.

The potential energy, with the separation of scales explicitly highlighted, is:
\begin{equation}
\label{eqn:energy-3}
    U 
    = 
    \underbrace{
        \sum_{i=1}^{n} \left( - \half \bfE_i \cdot \bfchi(\hat\bfn_i)  \bfE_i \right)
    }_{\sim kT}
    +
    \underbrace{
        \frac{1}{2} \int_\Omega |\nabla\phi|^2 \dm\Omega
    }_{\gg kT}
\end{equation}
We first minimize the field energy, of order much greater than $kT$, by setting the first variation to zero to obtain $\divergence \nabla \phi = 0$.
Using the specified average electric field ensemble described in Appendix \ref{sec:constant-E-ensemble}, we find that $-\nabla\phi = \bfE_0$.

The final form of the reduced potential energy that we will use for statistical averaging reads simply:
\begin{equation}
\label{eqn:energy-4}
    U 
    = 
   \sum_{i=1}^{n} \left( - \half \bfE_0 \cdot \bfchi(\hat\bfn_i)  \bfE_0 \right)
\end{equation}
where $\bfE_0$ is the specified average electric field.

We highlight two additional important simplifications that are a consequence of the assumption of separation of energy scales.
First, the degrees of freedom required to describe a microstate are vastly reduced from our general starting point.
In particular, the spatial position plays no role, as the electric field is uniform in space; the point dipole is completely specified given $\bfE_0$ and the orientation of the monomer; and the electric field is also completely specified $-\nabla\phi = \bfE_0$.
Therefore, the monomer orientations $\hat\bfn_i$ are the only remaining degrees of freedom over which to conduct statistical averaging.
Also, we do not require statistical field theoretic methods to deal with the averaging over the electric field.
Second, the simplification from having to solve a nonlocal boundary value problem to determine $\nabla\phi$ makes it possible to easily invert the relation between the applied field $\bfE_0$ and the total dipole of the polymer chain, using the expression:
\begin{equation}
    \bfp = \sum_{i=1}^{n} \bfmu_i = \left( \sum_{i=1}^{n} \bfchi(\hat\bfn_i) \right) \bfE_0 
    \Rightarrow
    \bfE_0 = \left( \sum_{i=1}^{n} \bfchi(\hat\bfn_i) \right)^{-1} \bfp
\end{equation}
assuming that the matrix above is invertible.
Without the simplification obtained through the separation of energy scales, this would involve an inverse problem based on the electrostatic PDE.
An important advantage of being able to perform this inversion so easily is that it makes it tractable to perform a Legendre transform to go between the free energy written as a function of $\bfE_0$, which is relatively simpler to evaluate using statistical mechanics, and the free energy written as a function of $\bfp$, which is more convenient for applications since it has a minimum rather than a saddle-point structure.

%%%%%%%%%%%%%%%%%%%%%%%%%%%%%%%%%%%%%%%%%%%%%%%%
%%%%%%%%%%%%%%%%%%%%%%%%%%%%%%%%%%%%%%%%%%%%%%%%
%%%%%%%%%%%%%%%%%%%%%%%%%%%%%%%%%%%%%%%%%%%%%%%%
%%%%%%%%%%%%%%%%%%%%%%%%%%%%%%%%%%%%%%%%%%%%%%%%
\section{Statistical Mechanical Formulation}
\label{sec:stat-mech-formulation}

We now use the potential energy that we formulated in the previous section to conduct statistical averaging.
However, we first provide a brief summary of the classical statistical mechanical theory of non-Gaussian polymer chains in the purely mechanical / entropic setting, to set the framework as well as to contrast with the key new features that are introduced by considering electrostatics.

%%%%%%%%%%%%%%%%%%%%%%%%%%%%%%%%%%%%%%%%%%%%%%%%
%%%%%%%%%%%%%%%%%%%%%%%%%%%%%%%%%%%%%%%%%%%%%%%%
%%%%%%%%%%%%%%%%%%%%%%%%%%%%%%%%%%%%%%%%%%%%%%%%
%%%%%%%%%%%%%%%%%%%%%%%%%%%%%%%%%%%%%%%%%%%%%%%%
\subsection{Summary of Classical Non-Gaussian Mechanical Polymer Chain}
\label{sec:kuhn-and-grun}

In this section, we revisit the classical work of \citet{kuhn1942beziehungen} (see also \citet{treloar1975physics}, Ch. 7) regarding the statistical mechanics of a polymer chain with the aim to later generalize it to the case of a dielectric elastomer chain with combined mechanical and electrical loading.

The monomer length will be denoted by $\mlen$ so that, given the number of monomers in the chain, $\N$, the length of the fully stretched chain is $\N \mlen$.
The end-to-end vector is defined as the vector that connects the beginning of the chain to the end of the chain, and is given by $\displaystyle \bfr = \mlen \sum_{i=1}^{n}  \hat\bfn_i$; we also define $\rmag = |\bfr|$, and the stretch $\stch = \rmag / \N \mlen$.
The polymer chain is idealized as composed of rigid inextensible links that can freely rotate, i.e., all chain configurations have the same potential energy
and therefore are all \emph{equally likely} \cite{hill1986statistical,swendsen2020introduction,kardar2007statistical}.
Therefore, the classical polymer chain, unlike a polymer chain in an electric field, is entirely governed by entropy.

We are ultimately interested in the force-deformation relationship for a single polymer chain.
We obtain the force $\force$ as the derivative of the free energy $\AHelm$ with respect to $r$.
Since the internal energy of the chain is the same for all configurations, only the entropy is relevant.

The entropy of an ensemble in which every microstate has the same internal energy is given by
\begin{equation} \label{eq:entropy}
	\Sent = \kB \log \nstates
\end{equation}
where $\nstates$ is the number of microstates in the ensemble.

The space of all monomer directions makes up the surface of the unit sphere.
We parameterize the surface of the unit sphere in the standard way, so that a unit direction is expressed as $\hat\bfv = \left(\cos \azi \sin \polar, \sin \azi \sin \polar, \cos \polar \right)$ where $\azi$ is the azimuth angle and $\polar$ is the polar angle.
The coordinate system is chosen such that the polar axis is the direction of the end-to-end vector.
Next, we partition the surface into $\n$ patches of area $\sin \polar_i \ds \azi_i \ds \polar_i$ and define the occupation numbers, $\pop_i$, as the number of monomers oriented such that their unit direction lies in the $i$th patch of area.
A polymer configuration is specified by prescribing the direction of each of the $\N$ monomers in the chain.
There are $\frac{\N !}{\Pi_{i=1}^{\n} \pop_i !}$ ways to assign the $\n$ directions to $\N$ monomers.
Consequently,
\begin{equation} \label{eq:nstates-full-sum}
	\nstates = \sum_{{\collect{\pop_i}}'} \frac{\N !}{\Pi_{i=1}^{\n} \pop_i !}
\end{equation}
where the prime in ${\collect{\pop_i}}'$ signifies that the sum is over all distributions that satisfy the constraints
\begin{align} 
\label{eq:cn}
	\N &= \sum_{i=1}^{\n} \pop_i \\
\label{eq:cr} 
    \bfr &= \mlen \sum_{i=1}^{\n} \pop_i \hat\bfv_i
\end{align}
In general, it is difficult to determine each collection of occupation numbers such that \eqref{eq:cn} and \eqref{eq:cr} are satisfied.
So instead we utilize the fact that the quantity of interest is ultimately $\log \nstates$.
The logarithm is a monotonically increasing function of its argument and its derivative goes as the inverse of its argument.
Therefore, if the sum in \eqref{eq:nstates-full-sum} is dominated by one term, then $\log \nstates$ can be approximated by maximizing the logarithm of a generic term\footnote[2]{This approximation is justified  by the central limit theorem \cite{hill1986statistical,weiner2012statistical,krauth2006statistical,swendsen2020introduction,kardar2007statistical}.}.

To perform the maximization, we use the method of Lagrange multipliers.
Thus, we search for stationary points of
\begin{equation*}
	\log \left(\frac{\N !}{\prod_{i=1}^{\n} \pop_i !}\right) + \nmult \left(\N - \sum_{i=1}^{\n} \pop_i\right) + \mults \cdot \left(\frac{\bfr}{\mlen} - \sum_{i=1}^{\n} \pop_i \hat\bfv_i \right)
\end{equation*}
Employing Sterling's approximation, $\log x! \approx x \log x - x$, and setting partials with respect to $\pop_j$ equal to zero, one obtains $\pop_j = \exp\left[\mults \cdot \nvec_j + \nmult + 1\right]$.
We proceed as follows: terms that do not depend on the unit direction are absorbed into an unknown constant $\C$, and we take the limit as $\n \rightarrow \infty$, which results in the transition from a discrete collection of occupation numbers into a continuous density.
The emphasize the distinction, we denote the density by $\density$ (as opposed to $\pop$).
Finally, by symmetry, $\mults$ must be in the direction of the chain stretch:
\begin{equation} \label{eq:density}
	\density(\azi, \polar) = \C \exp \left(\zmult \cos \polar\right)
\end{equation}
where $\zmult$ is the component of $\mults$ in the direction of the polar axis.

The next step is to determine the unknowns, $\C$ and $\zmult$.
To do this, we consider the form of the constraints, \eqref{eq:cn} and \eqref{eq:cr}, in the continuum limit (i.e. $\n \rightarrow \infty$).
The summations over partitions of the unit sphere becomes integrals over the unit sphere
\begin{align} \label{eq:ccn}
	\N = \intoverS{\density\left(\azi, \polar\right)} &= \frac{4 \pi \C}{\zmult} \sinh \zmult \\ \label{eq:ccr}
	\frac{\rmag}{\mlen} = \intoverS{\density\left(\azi, \polar\right) \cos \polar}
						&= \frac{\partial}{\partial \zmult} \left( \frac{4 \pi \C}{\zmult} \sinh \zmult \right)
\end{align}
Dividing \eqref{eq:ccr} by \eqref{eq:ccn} results in the relation
\begin{equation} \label{eq:Linv-gamma}
	\coth \zmult - 1/\zmult = \stch
\end{equation}
The Langevin function appears in many physical problems and is defined as $\Lang \left(x\right):=\langevin{x}$.
Hence, $\C = \frac{\N \zmult}{4 \pi} \csch \zmult$ and $\zmult = \zmultzero$.
Further,
\begin{equation} \label{eq:density-kg}
	\density\left(\azi, \polar\right) = \frac{\N \zmultzero}{4 \pi \sinh\left(\zmultzero\right)}  \exp\left[\zmultzero \cos \polar\right].
\end{equation}
Although a closed form expression of $\Langinv$ does not exist, many accurate approximations have been developed \cite{kroeger2015simple,jedynak2017new}.
Taking \eqref{eq:entropy} to the continuum limit and again, using Stirling's approximation,
\begin{equation*}
	\Sent = \kB \left(\N \log \N - \intoverS{\density \log \density}\right)
\end{equation*}
Thus, remembering \eqref{eq:Linv-gamma},
\begin{equation} \label{eq:entropy-length}
	\Sent = -\kB \N \left[\stch \zmultzero + \log\left(\frac{\zmultzero}{4 \pi \sinh\left(\zmultzero\right)}\right) \right]
\end{equation}
Differentiating \eqref{eq:entropy-length} with respect to $\rmag$ and recognizing 
\begin{equation*}
  \langevin{\left(\zmultzero\right)} = \stch
\end{equation*}
results in
\begin{equation} \label{eq:force-length}
	\force = \frac{\kB \T}{\mlen} \zmultzero
\end{equation}

The chain statistics derived by \citet{kuhn1942beziehungen} and revisited in the current section marked an important development in the modelling of rubber elasticity.
Although the expression for the chain entropy, \eqref{eq:entropy-length}, was obtained by approximating the sum in \eqref{eq:nstates-full-sum} with its maximum-term, the approximation proves to be quite accurate up to the full extension of the chain (i.e. $\stch \rightarrow 1$).
In particular, as $\stch$ approaches unity, both the free energy and the force approach $\infty$; hence, unlike Gaussian chain statistics, \eqref{eq:entropy-length} and \eqref{eq:force-length} have the property that they capture the finite extensibility of the chain.

\subsection{Formulation of the Electromechanical Polymer Chain}
\label{sec:DE-chain}

We are interested in deriving the free energy of an electro-responsive chain as a function of the applied electric field, its end-to-end vector, and its temperature.
In contrast to \Fref{sec:kuhn-and-grun}, the potential energy of a given monomer depends on its orientation and therefore not all microstates in the ensemble are equally likely.
Instead, we must weight each orientation by the Boltzmann factor, $\exp\left(-\beta \U\right)$, where $\beta = 1 /\kB \T$.
As in \Fref{sec:kuhn-and-grun}, we partition the surface of the unit sphere into $\n$ patches of area and define the occupation numbers, $\popnum_i$, as the number of monomers oriented such that their unit direction, $\hat\bfv_i$, lies in the $i$th patch.
The generalization of \eqref{eq:nstates-full-sum} is
\begin{equation} \label{eq:partition-function}
    \pfunc 
    = 
    \sum_{{\collect{\popnum_i}}'} \exp\left[-\beta \U\left(\collect{\popnum_i}\right)\right] \frac{\N !}{\Pi_{i=1}^{\n} \popnum_i !}
\end{equation}
where $\pfunc$ is the partition function.
The potential energy is taken as the sum of the individual monomer energies, i.e. $\U\left(\collect{\popnum_i}\right) = \sum_{i=1}^{\n} \popnum_i \um\left(\nvec_i\right)$.

As in \Fref{sec:kuhn-and-grun}, we notice that enumerating each of the terms in \eqref{eq:partition-function} and evaluating the sum proves to be difficult, and that ultimately we are interested in $\log \pfunc$.
Thus, we approximate the sum by its maximum term.
Since the logarithm is monotonic, we can maximize 
\begin{equation*}
    \log\left[\exp\left(-\beta \sum_{i=1}^{\n} \popnum_i \um\left(\hat\bfv_i\right)\right) \frac{\N !}{\prod_{i=1}^{\n} \popnum_i !}\right]
\end{equation*}
subject to the constraints \eqref{eq:cn} and \eqref{eq:cr}.
Using Stirling's approximation for the $\log \generic!$ terms and the method of Lagrange multipliers to enforce the constraints, the occupation numbers that result in the maximum term are
\begin{equation} 
\label{eq:pop-eap}
\begin{aligned}
    \popnum_j &= \C \exp\left[-\beta \um\left(\hat\bfv_j\right) + \mults \cdot \hat\bfv_j\right] 
    \\
    &= \C \exp\left[-\frac{\beta}{2} \dsus \left(\bfE_0 \cdot \hat\bfv_j\right)^2 + \mults \cdot \hat\bfv_j\right] 
    \\
    &= \C \exp\left[-\unodim \left(\hat\bfE_0 \cdot \hat\bfv_j\right)^2 + \mults \cdot \hat\bfv_j\right]
\end{aligned}
\end{equation}
where all of the terms in the argument of the exponential that did not have a directional dependence were absorbed into the unknown $\C$; and the unknown multiplier $\mults$ is related to the kinematic constraint.
The second step is a result of using \eqref{eq:monomer-energy}; and, in the last step we define the unit direction of the electric field, $\hat\bfE_0 = \bfE_0 / |\bfE_0|$, and define the dimensionless quantity $\unodim = \beta \ezeromag^2 \dsus / 2$, which is a measure of monomer electrical energy with respect to thermal energy.

Taking the limit of $\n \rightarrow \infty$, \eqref{eq:pop-eap} becomes
\begin{equation} \label{eq:density-eap}
    \density\left(\hat\bfv\right) = \C \exp\left[-\unodim \left(\hat\bfE_0 \cdot \hat\bfv\right)^2 + \mults \cdot \hat\bfv\right]
\end{equation}
where  the unknowns, $\C$ and $\mults$, are determined by solving the system of equations that result from taking the discrete constraints, \eqref{eq:cn} and \eqref{eq:cr}, to the continuum limit
\begin{align}
    \label{eq:cn-eap} \N &= \intoverSns{\rho\left(\hat\bfv\right)} \\
    \label{eq:cr-eap} \frac{\rvec}{\mlen} &= \intoverSns{\rho\left(\hat\bfv\right) \hat\bfv} 
\end{align}
and where $\unitsphere$ denotes the surface of the unit sphere.
Once the monomer density function has been approximately determined, one can return to \eqref{eq:partition-function} to find the free energy.

Approximating the sum on the right side of \eqref{eq:partition-function} by its maximum term, taking the logarithm, and using Stirling's approximation:
\begin{equation} \label{eq:A-approx}
    \log \pfunc \approx -\beta \sum_{i=1}^{\n}\popnum_i \um\left(\hat\bfv_i\right) + \N \log \N - \sum_{i=1}^{\n}\popnum_i \log \popnum_i
\end{equation}
Multiplying both sides by $-\kB \T$ and taking the limit of $\n \rightarrow \infty$
\begin{equation} \label{eq:A-approx-eap}
    \A \approx \intoverSns{\left\{\density\left(\hat\bfv\right) \um\left(\hat\bfv\right) + \kB \T \density\left(\hat\bfv\right) \log \left(\density\left(\hat\bfv\right)\right)\right\}} - \N \kB \T \log \N
\end{equation}
we arrive at an expression for an approximation of the free energy.

There are several challenges related to \eqref{eq:cn-eap}, \eqref{eq:cr-eap} and \eqref{eq:A-approx-eap}.
First, the integrals in \eqref{eq:cn-eap}, \eqref{eq:cr-eap} and \eqref{eq:A-approx-eap} are difficult to evaluate in closed-form.
Second, the resulting system of equations will, in general, be nonlinear.
Recall that in \Fref{sec:kuhn-and-grun} the first difficulty was addressed by choosing the coordinate system such that the polar axis was in the direction of the chain end-to-end vector and recognizing that the symmetry of the problem allows one to simplify the $\mults \cdot \nvec$ term in \eqref{eq:density} to $\zmult \cos \polar$.
This symmetry-based argument does not hold in the electro-mechanical setting, as there are now two distinguished directions in the problem, namely $\rvec$ and $\ezero$; consequently, the direction of $\mults$ cannot be determined \textit{a priori}.
For the same reason, explicit solutions are no longer possible due to loss of transverse isotropic symmetry about $\rvec$.
As a result, \eqref{eq:cn-eap} and \eqref{eq:cr-eap} do not appear to be tractable to closed-form solution.
Instead, in \Fref{sec:numerical} we will use numerical methods; and in \Fref{sec:small-omega} and \Fref{sec:small-lambda} we will assume smallness of some parameters, expand in terms of the small parameters, and derive approximate solutions.

%%%%%%%%%%%%%%%%%%%%%%%%%%%%%
%%%%%%%%%%%%%%%%%%%%%%%%%%%%%
%%%%%%%%%%%%%%%%%%%%%%%%%%%%%
%%%%%%%%%%%%%%%%%%%%%%%%%%%%%

\subsection{Thermodynamic Minimum Principles}

A central quantity of interest is the net dipole moment, $\chainpolar$, of the chain, obtained by summation over the dipoles in the chain.
In the continuum limit, it can be obtained via $\density$ with the formula:
\begin{equation} \label{eq:chain-polarization}
	\chainpolar = \intoverSns{\density \dipole}
\end{equation}
We also have that $\takepartialflat{\A}{\bfE_0} = -\chainpolar$, shown in \Fref{app:polarization}.

We highlight the importance of the conjugate relation $\takepartialflat{\A}{\bfE_0} = -\chainpolar$ in terms of our choice of ensemble and minimum principles for the free energy.
By working in a specified average electric field ensemble, the free energy that is obtained corresponds to a thermodynamic system that is linked to a charge reservoir; the analogy is to a constant temperature ensemble that is linked to a reservoir of thermal energy.
That is, charge (respectively, thermal energy) is freely exchanged to maintain the given electric field (respectively, temperature).
Therefore, the free energy in \eqref{eq:A-approx-eap} corresponds to the following thermodynamic potential:
\begin{equation*}
	\A = \U - \T \Sent - \chainpolar \cdot \ezero
\end{equation*}
Now, by Clausius's theorem $\T \df{\Sent} \geq \dQ$.
Similarly, there can only be loss (e.g. friction) from an external force putting work, \hspace{0.1cm}$\dW$, into a thermodynamic system.
Hence, $\ezero \cdot \df{\chainpolar} \geq \dW$.
Using these two inequalities, in confluence with the first law of thermodynamics (i.e. $\df{\U} - \dQ - \dW = 0$), we have that
\begin{equation*}
	\df{\A} = \df{\left(\U - \T \Sent - \chainpolar \cdot \ezero\right)} = \df{\U} - \T \df{\Sent} - \ezero \cdot \df{\chainpolar} \leq 0
\end{equation*}
when $\T$ and $\ezero$ are constant.

The important consequences of the above analysis are:
\\
1. the free energy given by \eqref{eq:A-approx-eap} has a minimum principle when $\T$ and $\ezero$ are constant, and 
\\
2. the free energy $\A$ is the Legendre transform of the Helmholtz free energy $\AHelm$ in the chain net dipole slot, i.e.
\begin{equation*}
    \AHelm = \A + \chainpolar \cdot \efield
\end{equation*}

While it is necessary to properly establish the free energy derived in this work in the context of thermodynamics and minimum energy principles, from here on we will use the shorter ``free energy'' to refer to $\A$.
For a more detailed look at thermodynamic potentials and free energy minimum principles, see~\citet{kardar2007statistical} section 1.7; and, for a review of variational principles in electroelasticity, see~\citet{liu2014energy}.

%%%%%%%%%%%%%%%%%%%%%%%%%%%%%%%%%%%%%%%%%%%%%%%%%%%%
%%%%%%%%%%%%%%%%%%%%%%%%%%%%%%%%%%%%%%%%%%%%%%%%%%%%
%%%%%%%%%%%%%%%%%%%%%%%%%%%%%%%%%%%%%%%%%%%%%%%%%%%%
%%%%%%%%%%%%%%%%%%%%%%%%%%%%%%%%%%%%%%%%%%%%%%%%%%%%

\section{Numerical Solutions} \label{sec:numerical}

%%%%%%%%%%%%%%%%%%%%%%%%%%%%%%%%%%%%%%%%%%%%%%%%%%%%
%%%%%%%%%%%%%%%%%%%%%%%%%%%%%%%%%%%%%%%%%%%%%%%%%%%%
%%%%%%%%%%%%%%%%%%%%%%%%%%%%%%%%%%%%%%%%%%%%%%%%%%%%
%%%%%%%%%%%%%%%%%%%%%%%%%%%%%%%%%%%%%%%%%%%%%%%%%%%%

\subsection{Numerical Methods}

Both evaluating the integrals and solving the resulting nonlinear system of equations given by \eqref{eq:cn-eap} and \eqref{eq:cr-eap} are difficult to do in a closed-form.
Instead we turn to numerical methods to gain insight into the nature of the exact solution, and also to have a measure of accuracy for our closed-form approximations.

For numerical integration, we used the $p$-adaptive algorithm from the cubature package based on Clenshaw-Curtis quadrature rules~\cite{clenshaw1960pcubature}; it is generally well-suited for smooth integrands and integration in low-dimensional space.
%Hence, it is well-suited for \eqref{eq:cn-eap} and \eqref{eq:cr-eap}, as the integrands are infinitely differentiable and the space is two-dimensional.
Newton's method was used to solve the nonlinear system of equations.
The initial guess for Newton's method was the Kuhn and Gr\"{u}n solution from \Fref{sec:kuhn-and-grun}, but in a rotated coordinate system where the polar axis is in the direction $\ezerodir$:
\begin{equation*}
    \xk{0} = \begin{Bmatrix}
    \C_0 \\ \zmult_0 \\ \xmult_0
    \end{Bmatrix} =
    \begin{Bmatrix}
    \frac{\N \zmult}{4 \pi} \csch \left[\zmultzero\right] \\
    \frac{\rz}{\rmag} \zmultzero \\
    \frac{\rx}{\rmag} \zmultzero
    \end{Bmatrix}
\end{equation*}
when $|\unodim| = 0$; recall that $\unodim$ is the nondimensional parameter characterizing the ratio of electrical energy to thermal energy $kT$.
We also have
\begin{equation*}
    \xk{0}' =
    \begin{Bmatrix}
    \Csl \\
    \zmultsl \\
    \xmultsl
    \end{Bmatrix}
\end{equation*}
when $|\unodim| > 1$.
When $|\unodim| \in (0, 1)$ then the initial guess was taken as a linear interpolation between the two guesses\footnote[3]{
    The initial guess $\xk{0}'$ comes from the closed-form approximation derived in \Fref{sec:small-lambda}.
    We leave the details of the derivation until that section.
    },
that is $|\unodim|\xk{0}' + (1 - |\unodim|)\xk{0}$.

A residual tolerance of less than $10^{-10}$ was usually reached within 3-15 iterations.
In instances when Newton's method did not convergence, a series of gradient-free, unconstrained optimization methods were used to approximate a solution.
The SBPLX (based on the Subplex algorithm)~\cite{rowan1990sbplx} and Principle Axis (PRAXIS)~\cite{brent1972praxis} algorithms from the NLopt package~\cite{NLopt} were used, as was a simulated annealing implementation (the implementation was based on~\citet{krauth2006statistical}).
The cost function was taken to be the square root of the sum of the squares of the residuals from \eqref{eq:cn-eap} and \eqref{eq:cr-eap}.

In \Fref{sec:numerics-to-kg}, we present the numerical solution for different electric fields, monomer susceptibilities, and chain stretches and aim to explain some of the physical behavior that is observed.
%%%%%%%%%%%%%%%%%%%%%%%%%%%%%%%%%%%%%%%%%%%%%%%%%%%%
%%%%%%%%%%%%%%%%%%%%%%%%%%%%%%%%%%%%%%%%%%%%%%%%%%%%
%%%%%%%%%%%%%%%%%%%%%%%%%%%%%%%%%%%%%%%%%%%%%%%%%%%%
%%%%%%%%%%%%%%%%%%%%%%%%%%%%%%%%%%%%%%%%%%%%%%%%%%%%

\subsection{Results and Electro-Responsive Chain Physics} \label{sec:numerics-to-kg}

The stiffness of a classical polymer chain is due to thermal fluctuations and the natural tendency of a (constant energy) thermodynamic system to maximize entropy.
For the free energy and stiffness of an electro-responsive polymer chain, we expect both electrostatic energy and thermal fluctuations to play a role.
We aim to determine how each affects the free energy and force-length relationship, and the interplay between their respective contributions.
To this end, we generate numerical solutions for different electric fields, monomer susceptibilities, and chain stretches.
Throughout this section the number of monomers, $\N$, is taken to be $100$, which was seen to be sufficient for convergence to the long-chain limit.
In addition, when $\unodim$ is positive the monomer susceptibilities are $\sus{1} = 0$ and $\sus{2} = 1$; and when $\unodim$ is negative the monomer susceptibilities are $\sus{1} = 1$ and $\sus{2} = 0$.

\begin{comment}
The residuals for \eqref{eq:cn-eap} and \eqref{eq:cr-eap} were calculated for all of the numerical results presented in the current section.
The numerical solutions often converged to a maximum absolute residual of $10^{-10}$ or less; however, in some cases the numerical scheme could not achieve a reasonable level of accuracy.
Those solution results that did not have a maximum absolute residual of most $\N * 0.01$ (or 1.0) are not included in the figures that follow.
From here on, we shall refer to the polymer chain that behaves in accordance with the Kuhn and Gr\"{u}n model (as opposed to say, an electro-responsive chain) as a classical polymer chain. 
In regards to numerics, the numerical solutions are at times plotted along side the Kuhn and Gr\"{u}n solution presented in \Fref{sec:kuhn-and-grun}.
When such calculations require the (approximate) evaluation of the inverse Langevin function, the following approximation is used~\cite{kroeger2015simple}:
\begin{equation*}
\Langinv\left(x\right) \approx \frac{3 x - \frac{x}{5} \left(6 x^2 + x^4 - 2 x^6\right)}{1 - x^2}
\end{equation*}
\end{comment}

\subsubsection{Monomer Orientation Density}

\begin{figure}[htb!]
	\centering
	\includegraphics[width=\FigureWidthNumericalGraph]{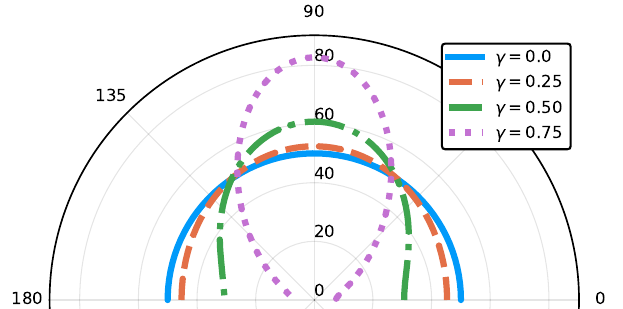}
	\caption{
	    Polar plot of monomer orientation density as a function of polar angle for classical polymer chains -- or equivalently, with $\unodim = 0$ -- being stretched in the $\polar = \pi / 2$ direction.
	    The density is shown for $\stch = 0.0, 0.25, 0.5, 0.75$.
	    The density is uniform for zero stretch and becomes more and more concentrated at $\pi / 2$ as the chain is stretched so that the kinematic constraint may be satisfied.
    }
	\label{fig:density-kg}
\end{figure}

We begin by considering how changing the mechanical and electrical loading of the chain affects the orientation of its monomers.
We first consider the mechanical loading in isolation; that is, how changing the chain end-to-end vector changes the distribution of monomers.
From \eqref{eq:density-kg}, we can see that this relationship is nonlinear.
This nonlinearity is shown graphically by Fig. \ref{fig:density-kg}, which displays the density of monomers as a function of their polar angle.
The density is shown for chains that are being stretched orthogonal to the polar axis and for stretches of $\stch = 0.0, 0.25, 0.5$, and $0.75$.
The density is uniform for zero stretch, as expected, and becomes more and more concentrated at $\pi / 2$ as the chain is stretched so that the kinematic constraint may be satisfied.

\begin{figure}[htb!]
	\centering
	\begin{tabular}{c c}
		\includegraphics[width=\FigureWidthTwoColsNumericalGraph]{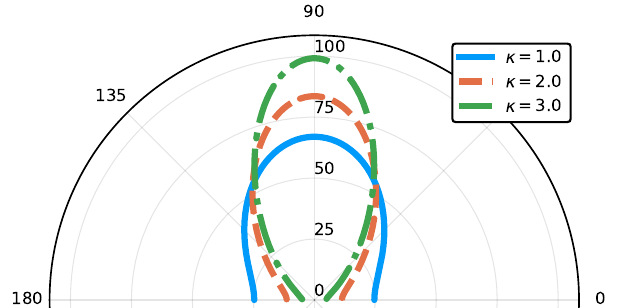} &
		\includegraphics[width=\FigureWidthTwoColsNumericalGraph]{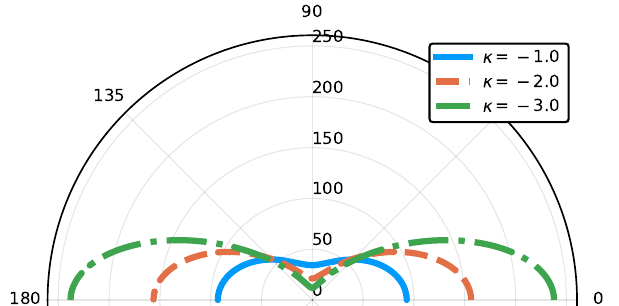}
	\end{tabular}
	\caption{Monomer density as a function of polar angle for TI (left) and uniaxial (right) monomer chains.
		The density is shown for $|\unodim| = 1.0, 2.0, 3.0$.
		The monomer densities bias more towards the direction in which dipoles are aligned with the electric field as $|\unodim|$ increases.
	}
	\label{fig:density-efield}
\end{figure}

We next consider the effect of electric field without any mechanical constraint.
The monomer density function  that minimizes the free energy of an electro-responsive chain is a tradeoff between (1.) aligning monomer dipoles with the electric field so as to minimize the electric potential energy and (2.) a uniform distribution -- which maximizes the entropy.
The nondimensional parameter $\unodim$ is a measure of the electrical energy per monomer relative to the thermal energy per monomer.
As such, it is natural to consider how $\unodim$ influences the monomer density function.
Let the polar axis be aligned with the electric field such that $\polar$ represents the angle between $\nvec$ and $\edir$.
Fig. \ref{fig:density-efield} shows the monomer densities for TI (left) and uniaxial (right) chains with increasing $|\unodim|$.
As $|\unodim|$ increases, in the TI chain case, the monomer density is increasingly biased towards $\polar = \pi / 2$--which is the angle which aligns the TI monomer dipole with the electric field.
Similarly, we see the monomer density is biased toward $0$ and $\pi$ for the uniaxial chains as $|\unodim|$ increases.
Notice that in either case, the monomer density function has a reflection symmetry about $\polar = \pi / 2$.
This is because the electric potential energy is quadratic in $\nvec$.

\begin{figure}[htb!]
	\centering
	\begin{tabular}{c c}
		\includegraphics[width=\FigureWidthTwoColsNumericalGraph]{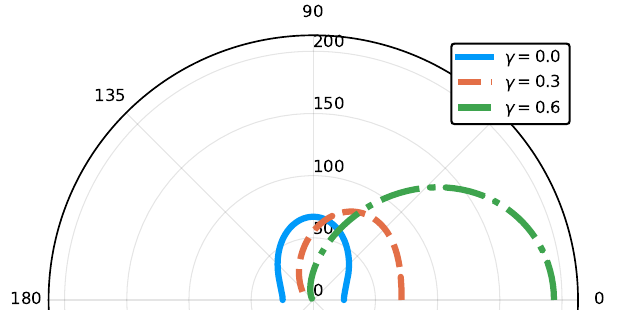} &
		\includegraphics[width=\FigureWidthTwoColsNumericalGraph]{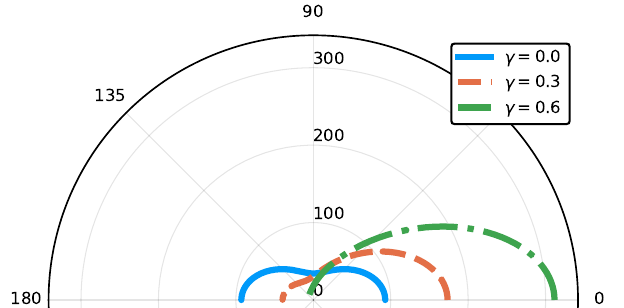} \\
		\includegraphics[width=\FigureWidthTwoColsNumericalGraph]{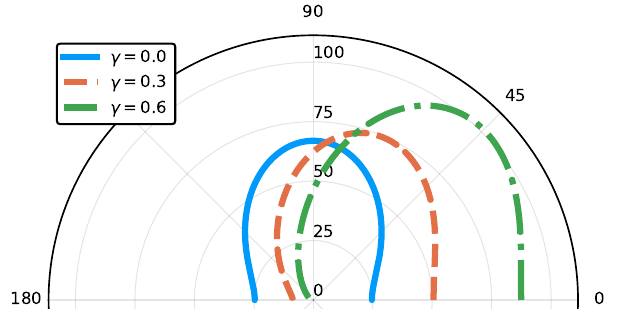} &
		\includegraphics[width=\FigureWidthTwoColsNumericalGraph]{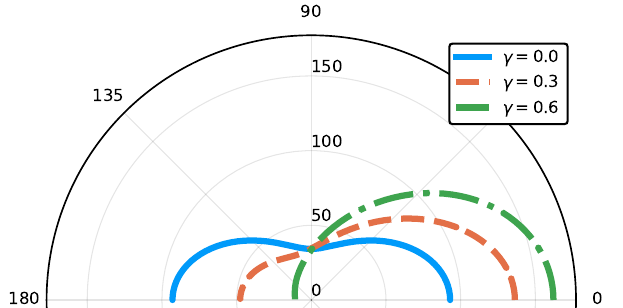} \\
		\includegraphics[width=\FigureWidthTwoColsNumericalGraph]{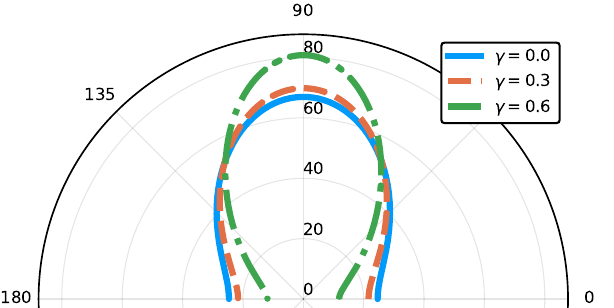} &
		\includegraphics[width=\FigureWidthTwoColsNumericalGraph]{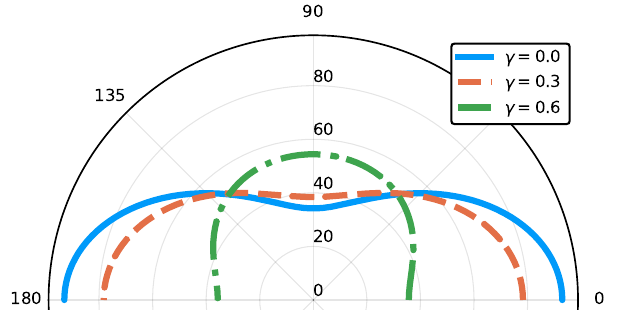}
		
	\end{tabular}
	\caption{Monomer density as a function of polar angle for TI (left) and uniaxial (right)  chains.
		The density is shown for $|\unodim| = 1.0$ and with increasing stretch: $\stch = 0.0, 0.3, 0.6$.
		The stretches are applied along the electric field ($\polar = 0$) in the top row; stretch applied along $\polar = \pi / 4$ in the middle row; and stretch applied along $\polar = \pi / 2$ in the bottom row.
		There is a complex interplay between the electric and geometric influences on the monomer density function.
		The influence transitions from purely electric at $\stch = 0.0$ to purely geometric at $\stch = 1.0$. %, but the exact nature of the transition is nontrivial.
	}
	\label{fig:density-interplay}
\end{figure}

While the geometric influence on the monomer density in Fig. \ref{fig:density-kg} and the electrical influence in Fig. \ref{fig:density-efield} are interesting, we next consider the interplay between these two effects.
Fig. \ref{fig:density-interplay} shows the monomer densities for TI (left) and uniaxial (right) chains at $|\unodim| = 1$ and with increasing stretch along the direction of the electric field (top), at an angle $\pi / 4$ with respect to the field direction (middle), and orthogonal to the direction of the field (bottom).
Clearly there are elements of both a bias toward aligning monomer dipoles and a bias toward the direction of stretch--as the chain is stretched.
However, as the chain approaches its stretched limit, the geometric influence will always eventually begin to dominate the electrical influence.
This is of course because the end-to-end vector constraint must be satisfied.
The transition between the geometrically dominated regime to the electrically dominated regime depends sensitively on the sign and magnitude of $\unodim$, as well as the direction of stretch.

\subsubsection{Free Energy Surface}

\begin{figure}[htb!]
	\centering
	\begin{tabular}{c c}
		\multicolumn{2}{c}{\includegraphics[width=\FigureWidthTwoColsNumericalGraph]{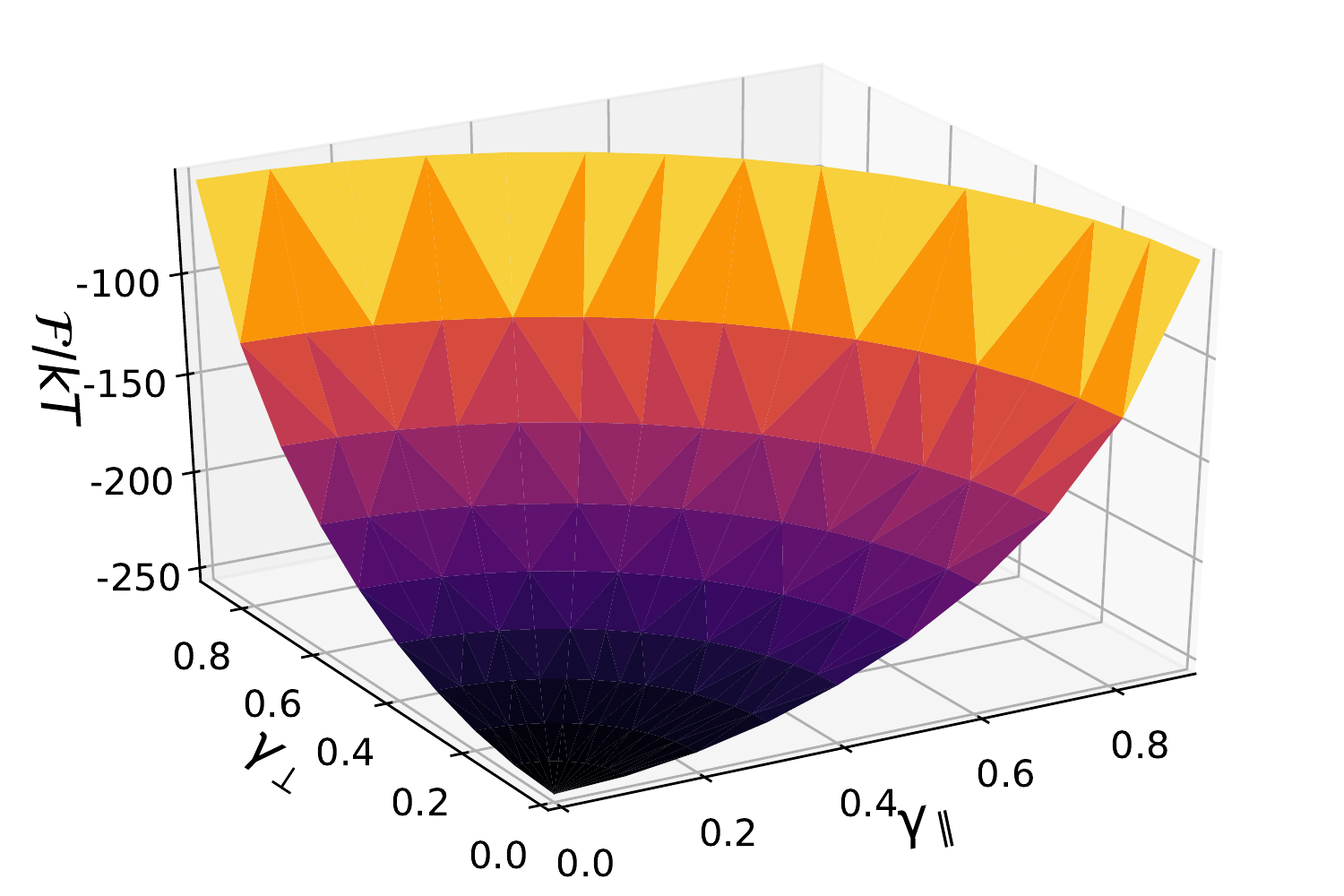}} \\
		\includegraphics[width=\FigureWidthTwoColsNumericalGraph]{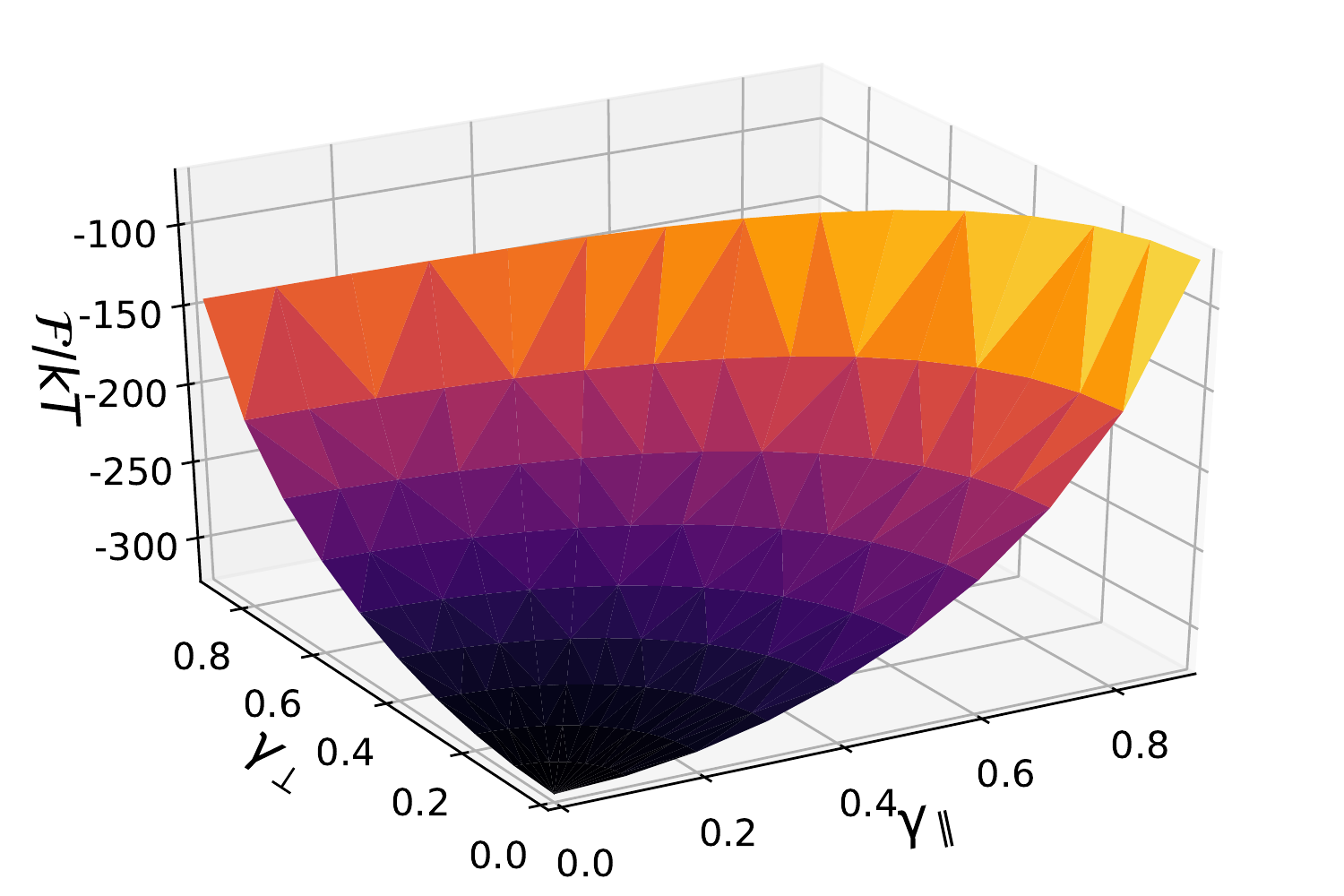} &
		\includegraphics[width=\FigureWidthTwoColsNumericalGraph]{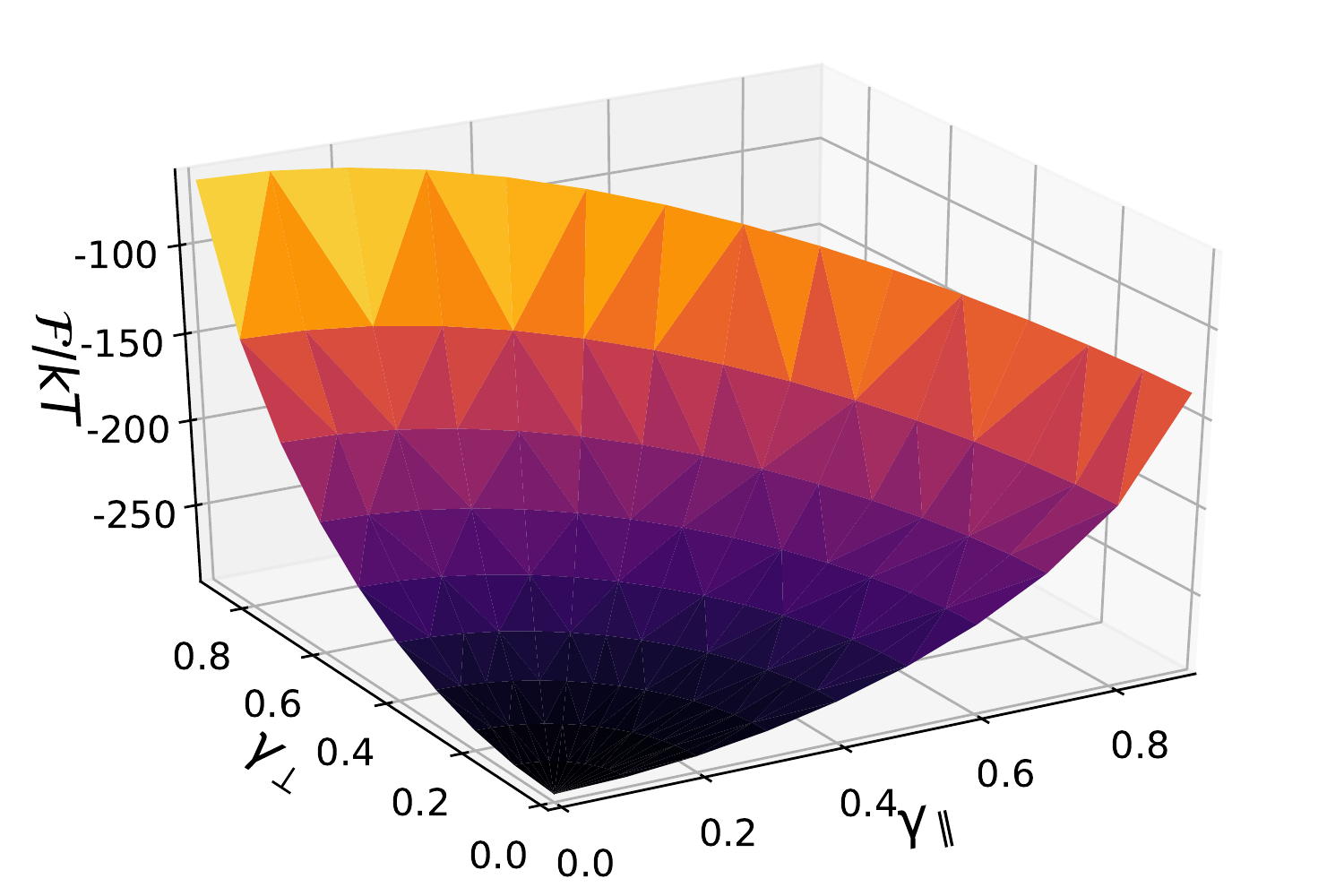}
	\end{tabular}
	\caption{Free energy surfaces in stretch space where the stretch axes are parallel to the direction of stretch, $\stch_\parallel$, and orthogonal to the direction of stretch, $\stch_\perp$.
	Surfaces are shown for a classical polymer chain (top), TI chain ($\unodim = 1.0$) (bottom left) and uniaxial chain ($\unodim = -1.0$) (bottom right).
	The applied electric field breaks the rotational symmetry of that is observed in the classical chain case such that the stiffness of the chain depends on its direction of stretch relative to the field direction.
}
	\label{fig:fe-surfaces}
\end{figure}

% explain the significance of looking at free energy; min principle; derivatives and stiffness
To probe the interplay between the chain end-to-end vector and electrical loading of an electro-responsive chain, we fix the electrical properties of a chain and visualize the free energy as a surface in stretch space.
The free energy surfaces for a classical chain (top), TI chain (bottom-left), and uniaxial chain (bottom-right) are shown in Fig. \ref{fig:fe-surfaces}.
The axes for the independent variables are the stretch parallel to the electric field direction, $\stch_\parallel$, and the stretch perpendicular to the electric field direction, $\stch_\perp$.
A considerable amount of information about the elasticity of, and forces on, a given polymer chain can be derived from these surfaces.
\newcommand{\auxu}{\hat{\mathbf{u}}}
For instance, at a given point in stretch space, $\left(\stch_\parallel, \stch_\perp, \A / \kB \T\right)$, the directional derivative of $\A / \kB \T$ (with respect to the stretch components) represents the tangent stiffness of the chain.
As expected, the classical chain has a rotational symmetry with respect to the chain end-to-end vector--that is, the free energy of a classical chain only depends on the magnitude of the stretch and is invariant with respect to the direction of stretch.
Also, in the absence of mechanical forcing, the equilibrium stretch is zero, i.e. a nonlinear entropic spring with a relaxed length of zero.
Lastly, we can see that the finite extensibility of the chain is captured through the fact that $\A / \kB \T \rightarrow \infty$ as $\stch \rightarrow 1$.

\subsubsection{Mechanical and Electrostatic Torque}
\label{sec:torque}

\begin{figure}[htb!]
	\centering
	\begin{tabular}{c c}
		\includegraphics[width=\FigureWidthTwoColsNumericalGraph]{figs/energy-surface/kappa-010_TI_surface} &
		\includegraphics[width=\FigureWidthTwoColsNumericalGraph]{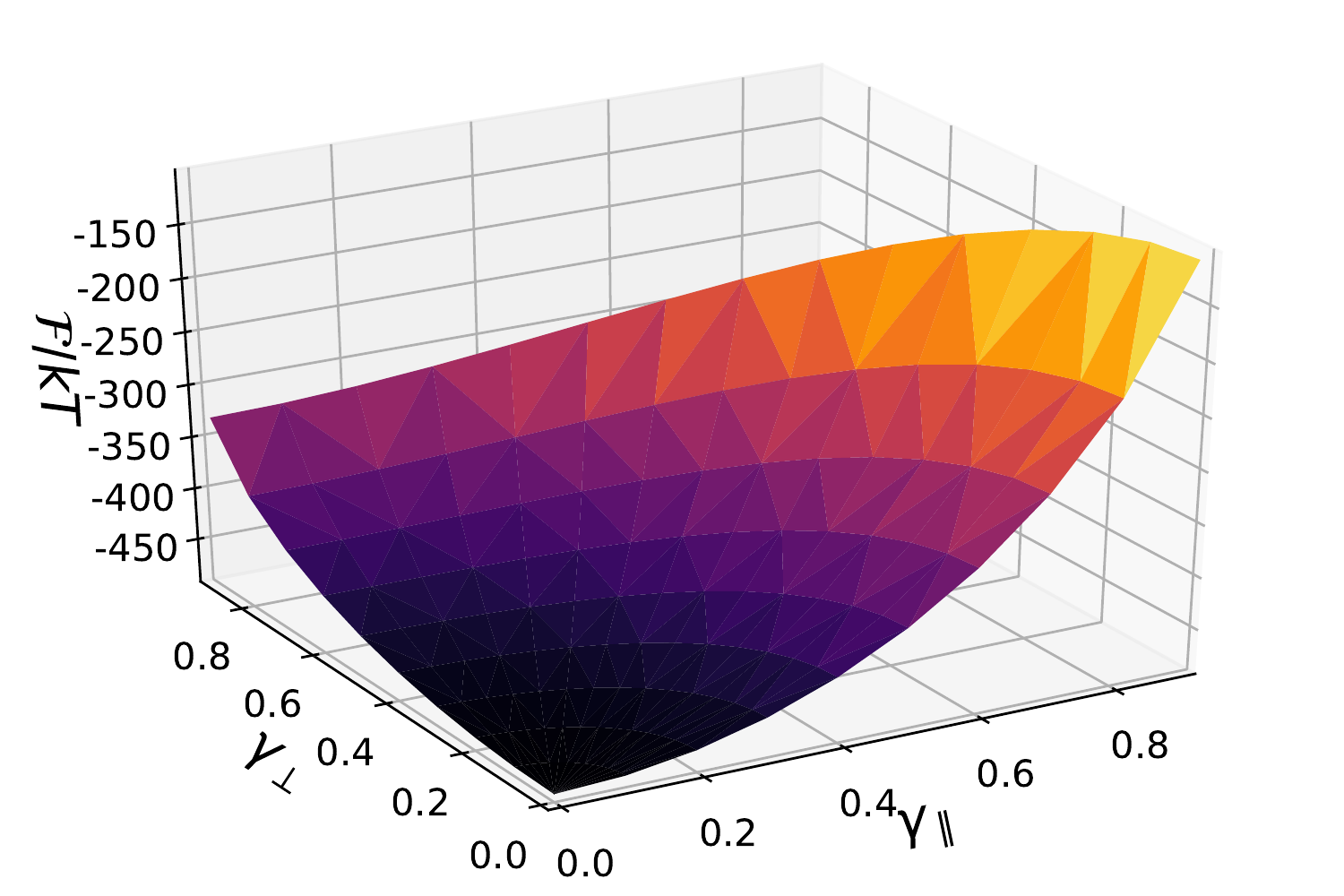}
		\\
		\includegraphics[width=\FigureWidthTwoColsNumericalGraph]{figs/energy-surface/kappa-010_Uni_surface} &
		\includegraphics[width=\FigureWidthTwoColsNumericalGraph]{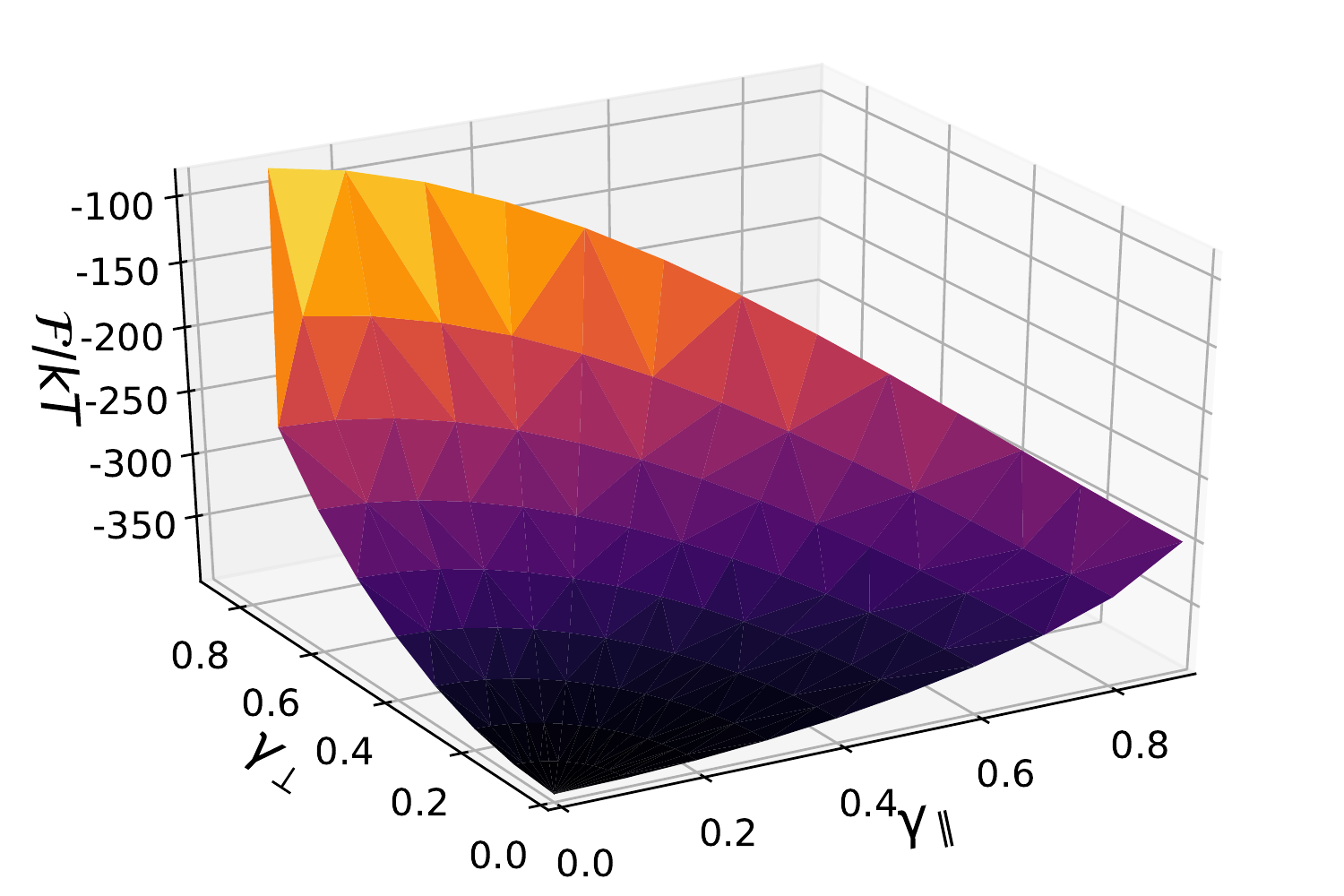}
	\end{tabular}
	\caption{Free energy surfaces in stretch space where the stretch axes are parallel to the direction of stretch, $\stch_\parallel$, and orthogonal to the direction of stretch, $\stch_\perp$.
		Surfaces are shown for a TI chain (top) at $\unodim = 1.0$ (right) and $\unodim = 3.0$ (left); and a uniaxial chain (bottom) at $\unodim = -1.0$ (right) and $\unodim = -3.0$ (left).
		The tilt of the free energy surfaces--which results in a directional dependence on the chain stiffness and an electrostatic torque--increases as $|\unodim|$ increases.}
	\label{fig:fe-kappa-surfaces}
\end{figure}

The rotational symmetry is broken for the electro-responsive polymers when an external electric field is applied.
For instance, the increase in free energy -- from zero stretch to full stretch -- is more gradual when a TI chain is stretched more toward a direction orthogonal to the electric field versus along the direction of the field; and vice versa for a uniaxial chain.
This is because, as the chain is stretched, more and more monomers in the chain are constrained toward the direction of stretch.
Therefore, stretching a TI chain nearly orthogonal to the electric field forces TI monomers near their direction of minimum electrical potential energy.
There is less resistance to such a macroscopic configuration than stretching the same TI chain such that its monomers are forced to align with the electric field.
This is of course because, as mentioned previously, the magnitude of TI dipoles are greater when orthogonal to the field and also, since their dipoles tend to form orthogonal to their axis, such an orientation aligns their dipoles with the field.
This is energetically favorable.
Similar arguments can be made to explain why the uniaxial chain is less stiff when stretched along the field direction and more stiff when stretched orthogonal to the field direction.
Also, as can be seen in Fig. \ref{fig:fe-kappa-surfaces}, the magnitude of this effect increases with respect to $\unodim$.
In fact, the tilt of the free energy surfaces, such that rotational symmetry is broken, has implications beyond an orientational dependent stiffness of the chains.
If, for example, the length of a TI chain is held fixed, but the chain is allowed to rotate, and an external electrical field is applied, then we can see that the chain will spontaneously rotate such that it is orthogonal to the field.
Similarly a uniaxial chain will rotate such that it is oriented along the field direction.

\newcommand{\auxDrho}{\takepartial{\density}{\etorangle}}

In mechanical terms, all of this means that there is an electrostatic torque on the chain.
The torque is equal to the derivative of the free energy with respect to rotation.
We can find a formula for this electrostatic torque.
Let $\etorangle$ denote the angle between $\rdir$ and $\ezerodir$.
Then, to make clear the dependence of various quantities on $\etorangle$, we rewrite \eqref{eq:density-eap} as
\begin{equation*}
    \density = \C\left(\etorangle\right) \exp \left(-\um\left(\nvec; \etorangle\right) / \kB \T + \mults\left(\etorangle\right) \cdot \nvec\right)
\end{equation*}
The chain torque is given by $-\takepartialflat{\A}{\etorangle}$.
We therefore take the partial derivative of \eqref{eq:A-approx-eap} with respect to $\etorangle$:
\begin{equation} \label{eq:dA1}
\begin{split}
	\takepartial{\A}{\etorangle} &= \intoverSns{\left(\auxDrho \um + \density \takepartial{\um}{\etorangle} + \kB \T \auxDrho \log \density + \kB \T \auxDrho\right)} \\
	&= \intoverSns{\left(\density \takepartial{\um}{\etorangle} + \kB \T \auxDrho\left(\log \C + \mults \cdot \nvec + 1\right)\right)}
\end{split}
\end{equation}
%where, because of the continuity of the integrand, we can exchange the order of integration and differentiation.
\newcommand{\auxarbconst}{C'}
To simplify, we take derivatives of both sides of \eqref{eq:cn-eap} to find that $\displaystyle \intoverSns{\auxDrho} = 0$.
Using this in \eqref{eq:dA1}, we get:
\begin{equation} \label{eq:dA2}
\begin{split}
	\takepartial{\A}{\etorangle} &= \intoverSns{\left(\density \takepartial{\um}{\etorangle} + \kB \T \auxDrho \mults \cdot \nvec\right)} \\
	&= \intoverSns{\left(\density \takepartial{\um}{\etorangle}\right)} + \kB \T  \left(\intoverSns{\left(\auxDrho \nvec\right)}\right) \cdot \mults \\
	&= \intoverSns{\left(\density \takepartial{\um}{\etorangle}\right)} + \frac{\kB \T}{\mlen} \mults \cdot \takepartial{\rvec}{\etorangle}
\end{split}
\end{equation}
The physical interpretation of the above equality is as follows: the first term on the right side is the electrostatic torque on the chain, and the second term is the mechanical torque due to the force required to enforce kinematic constraint.
When the end-to-end vector is constrained, the two terms must balance.
To focus on the electrostatic contribution to the torque, we fix $\rvec$ and use  \eqref{eq:monomer-energy} to obtain:
\begin{equation*}
	\takepartial{\A}{\etorangle} = \frac{\ezeromag^2 \dsus}{2}\intoverSns{\density \takepartial{}{\etorangle}\left(\ezerodir \cdot \nvec\right)^2}
\end{equation*}
Using $\omega$ to denote the angle between $\ezerodir$ and $\nvec$, we can write: 
%write $\omega = \etorangle + \varphi$.
\begin{equation} \label{eq:dA3}
\begin{split}
	\takepartial{\A}{\etorangle} &= -\ezeromag^2 \dsus \intoverSns{\density \cos \omega \sin \omega} \\
	&= -\ezeromag^2 \dsus \intoverSns{\density \left(\ezerodir \cdot \nvec\right) \left|\nvec \times \ezerodir\right|}
\end{split}
\end{equation}
Lastly, operating the cross-product with $\ezero$ on \eqref{eq:chain-polarization}, and then interchanging the order of integration and the cross product, we can write \eqref{eq:dA3} as:
\begin{equation}
	-\takepartial{\A}{\etorangle} = \left|\chainpolar \times \ezero\right|
\end{equation}
Thus, the electrostatic chain torque is equivalent to the torque on a point dipole, with dipole vector $\chainpolar$, in an applied electric field.
From this, we should expect then that the degree of ``tilt'' of the free energy surfaces in stretch space scales with $\sqrt{\unodim}$, which agrees with Fig. \ref{fig:fe-kappa-surfaces}.

\begin{figure}[htb!]
	\centering
	\includegraphics[width=\FigureWidthNumericalGraph]{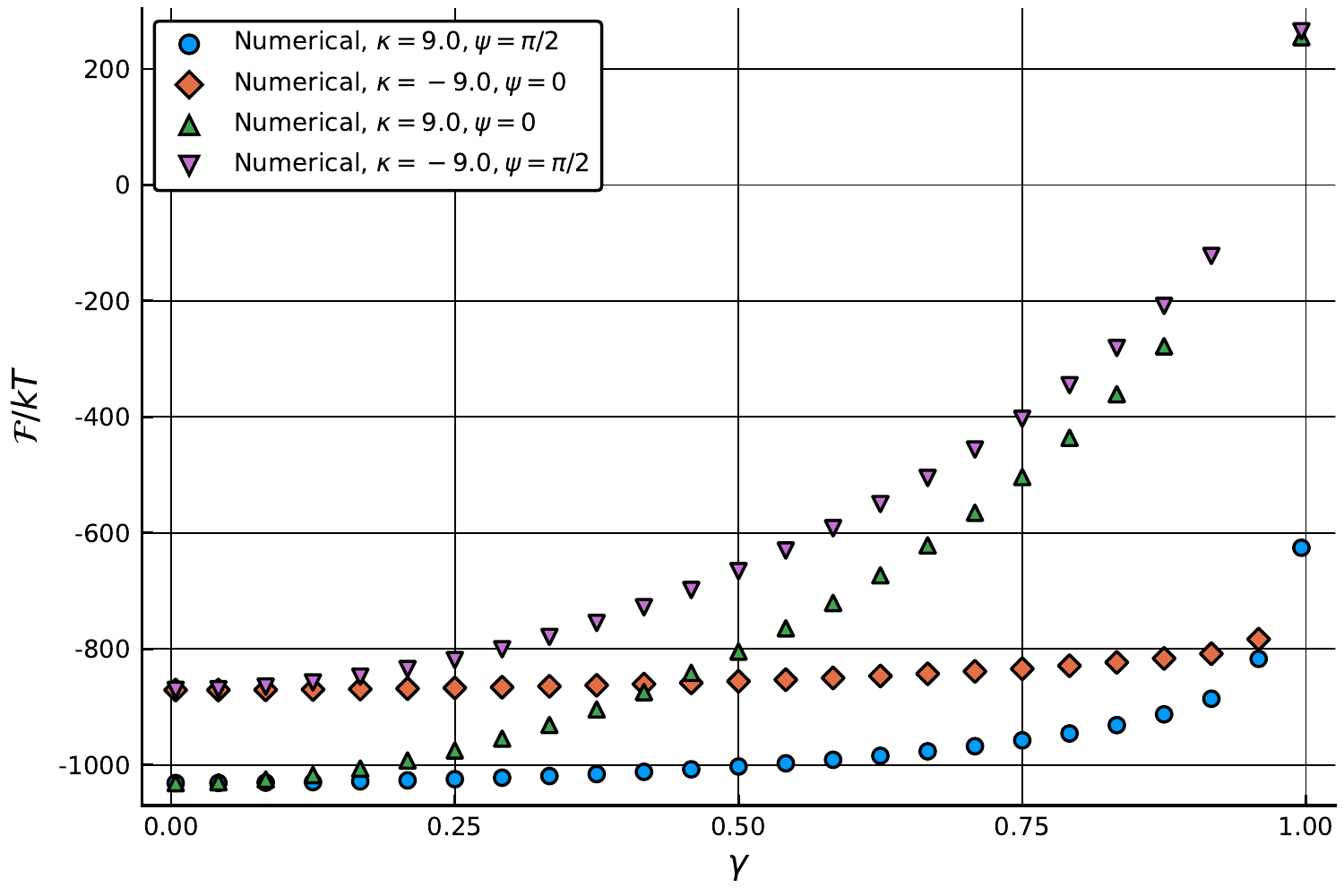}
	\caption{Comparison of TI and uniaxial chains being stretched in the direction of their respective energy maximum orientations, $\etorangle = \frac{\pi}{2}$ and $0$, respectively; and TI and uniaxial chains being stretched in the direction of their respective energy minimum, $\etorangle = 0$ and $\frac{\pi}{2}$, respectively. Notice the subtle differences in behavior between the TI and uniaxial chains as a result of the fact that the minimum energy orientation of a uniaxial monomer is when $\nvec = \pm \ezerodir$, where as the minimum energy orientation of a TI monomer occurs when $\nvec \cdot \ezerodir = 0$.
		The uniaxial minimum is only two discrete directions but the TI minimum orientation describes a plane in which $\nvec$ can rotate and the TI monomer still be at an energy minimum.
		The differences in the electrostatic monomer responses lead to a difference in the overall chain behaviors.}
	\label{fig:numerical-E0-3}
\end{figure}

Before closing the discussion of how $\unodim$, chain stretch, and chain orientation with respect to $\ezerodir$ affect the free energy of an electro-responsive chain, we point out a subtle but important detail.
Upon inspection of the free energy surfaces (e.g. Fig. \ref{fig:fe-kappa-surfaces}), one may be tempted to conclude that a symmetry exists such that changing the sign of $\unodim$ and rotating $\etorangle$ by $\frac{\pi}{2}$ results in the same chain statistics.
However, this is not the case.
The $\A / \kB \T$ - $\stch$ relation for TI and uniaxial chains oriented at $\etorangle = 0$ and $\etorangle = \frac{\pi}{2}$ are shown in Fig. \ref{fig:numerical-E0-3}.
Despite the fact that the depth of the electrical potential well is the same for TI and uniaxial monomers (when $|\dsus|$ is the same for each), the zero stretch free energy of the TI chains are lower.
This can be explained by the fact that the minimum energy orientation of a uniaxial monomer is a minimum when $\nvec = \pm \ezerodir$ where as the minimum energy orientation of a TI monomer occurs when $\nvec \cdot \ezerodir = 0$.
The uniaxial case is only two discrete directions but the TI case describes a plane in which $\nvec$ can rotate and the TI monomer still be at an energy minimum.
Thus, there is a larger space of directions in which TI monomers can be oriented which are also energetically favored (at or near a potential well), meaning the entropy is able to be larger and the entropic contribution to the free energy is able to be more negative (compared with uniaxial monomers).
Also, in particular, notice that as the (TI, $\etorangle = 0$) and (uniaxial, $\etorangle = \frac{\pi}{2}$) chains approach their fully stretched limits, the curves begin to meet.
This is because (1.) the kinematic constraint is forcing the monomers of each chain into or near their maximum energy state, which in this case is the same amount of energy, and (2.) regardless of the direction of stretch or type of monomer, the entropic term approaches infinite as $\stch \rightarrow 1$.

\subsubsection{Net Chain Dipole} \label{sec:numerical-dipole}

\begin{figure}[htb!]
	\centering
	\begin{tabulary}{\linewidth}{c c}
		\includegraphics[width=\FigureWidthTwoColsNumericalGraph]{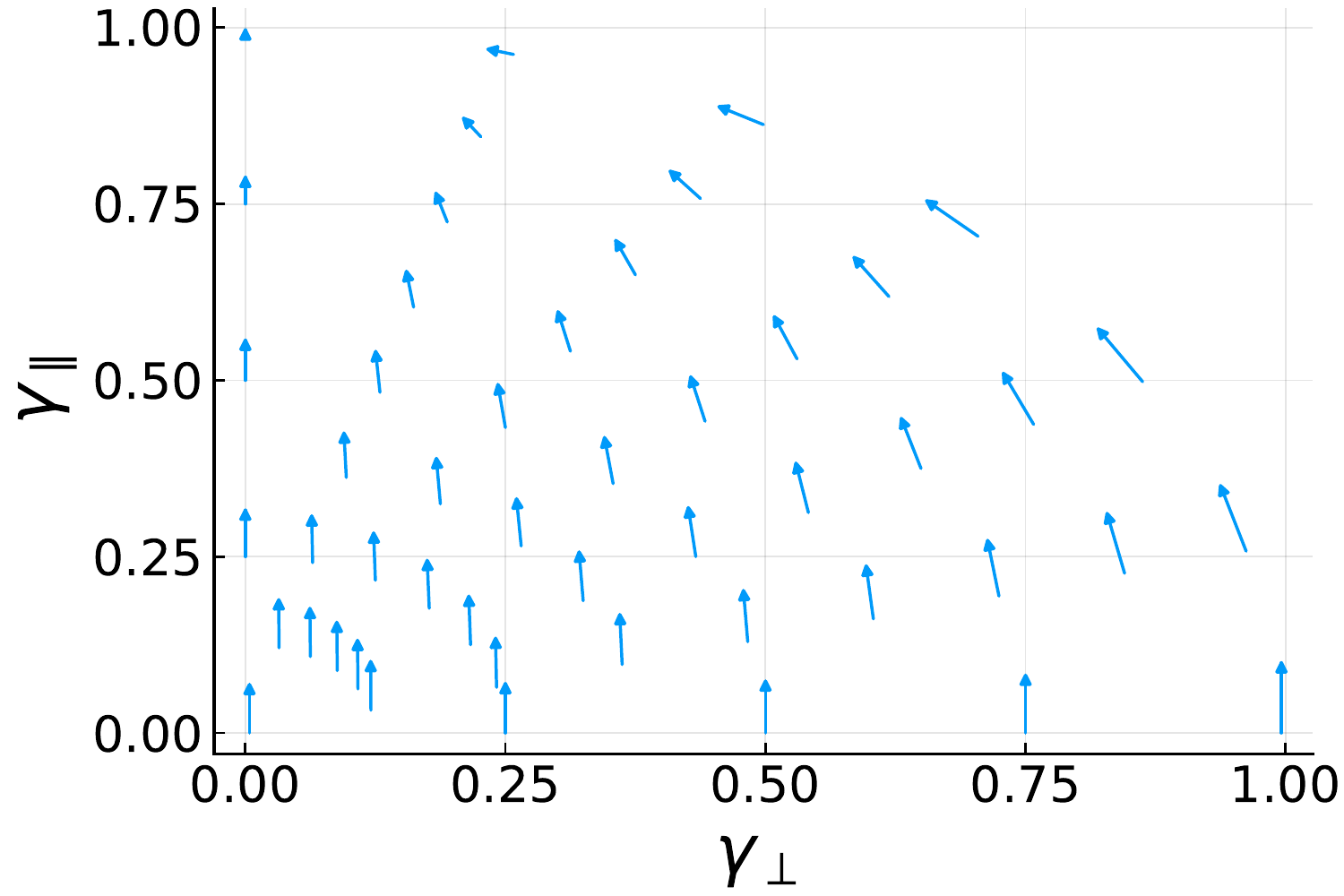}
		&
		\includegraphics[width=\FigureWidthTwoColsNumericalGraph]{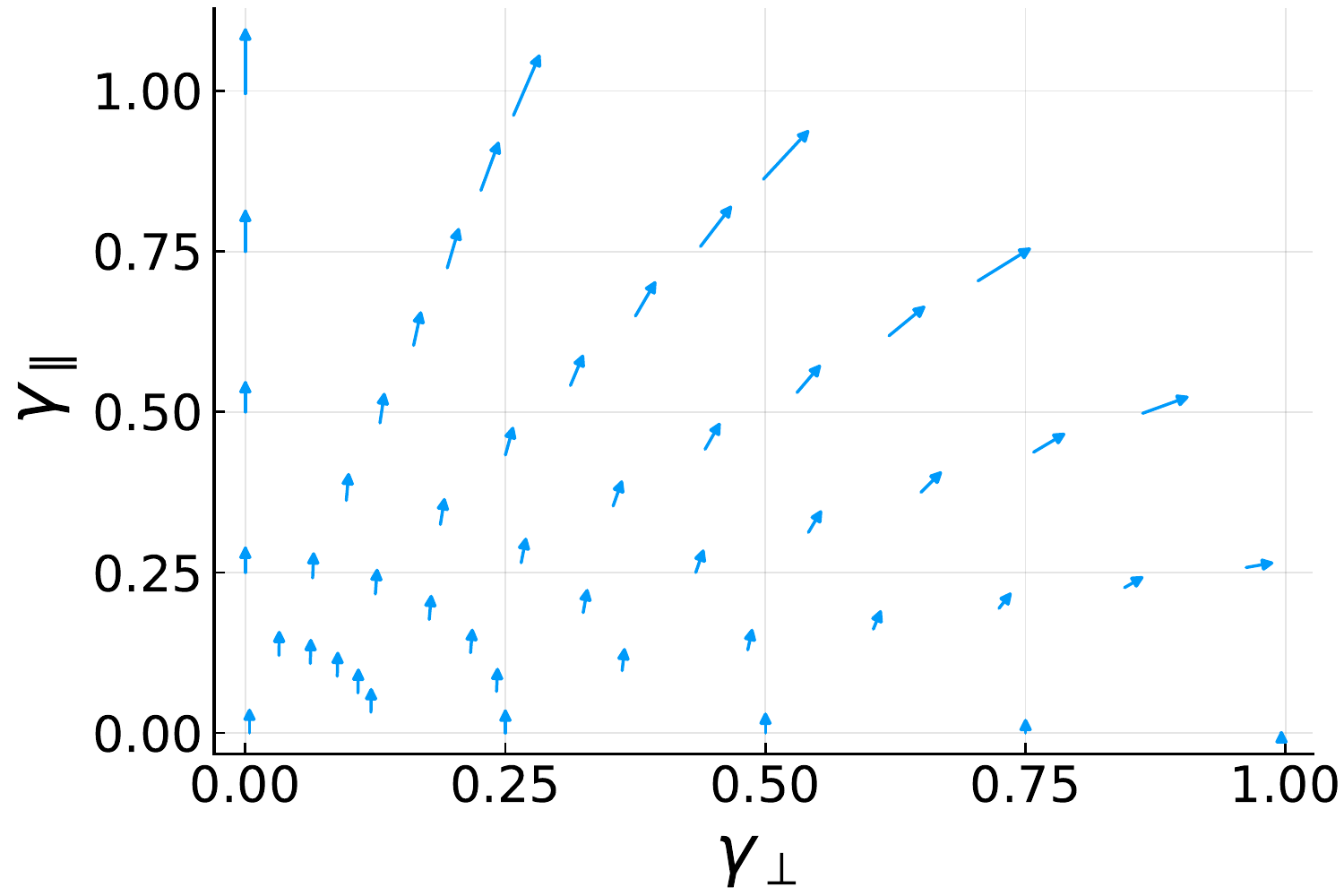}
		\\
		
		\includegraphics[width=\FigureWidthTwoColsNumericalGraph]{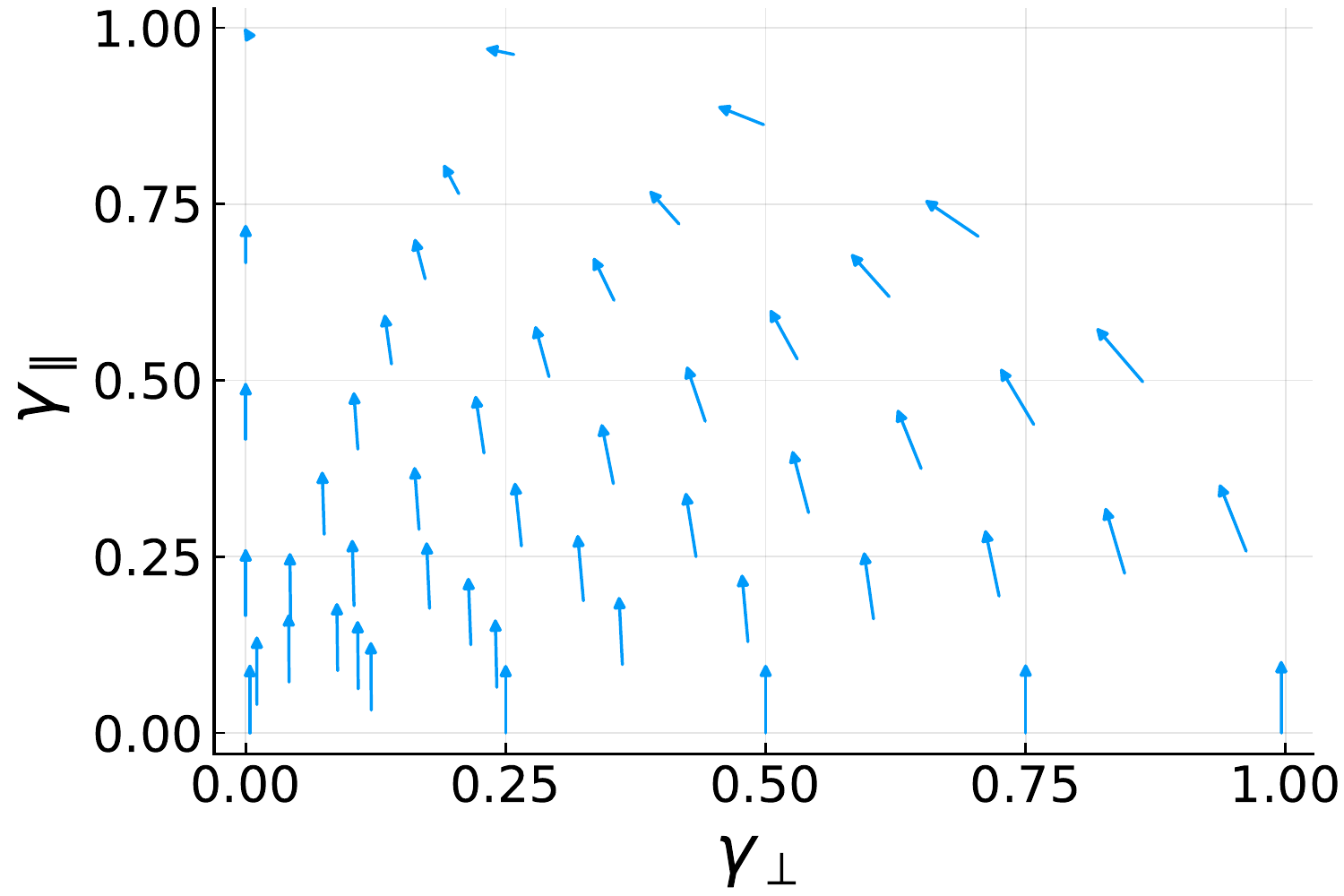}
		&
		\includegraphics[width=\FigureWidthTwoColsNumericalGraph]{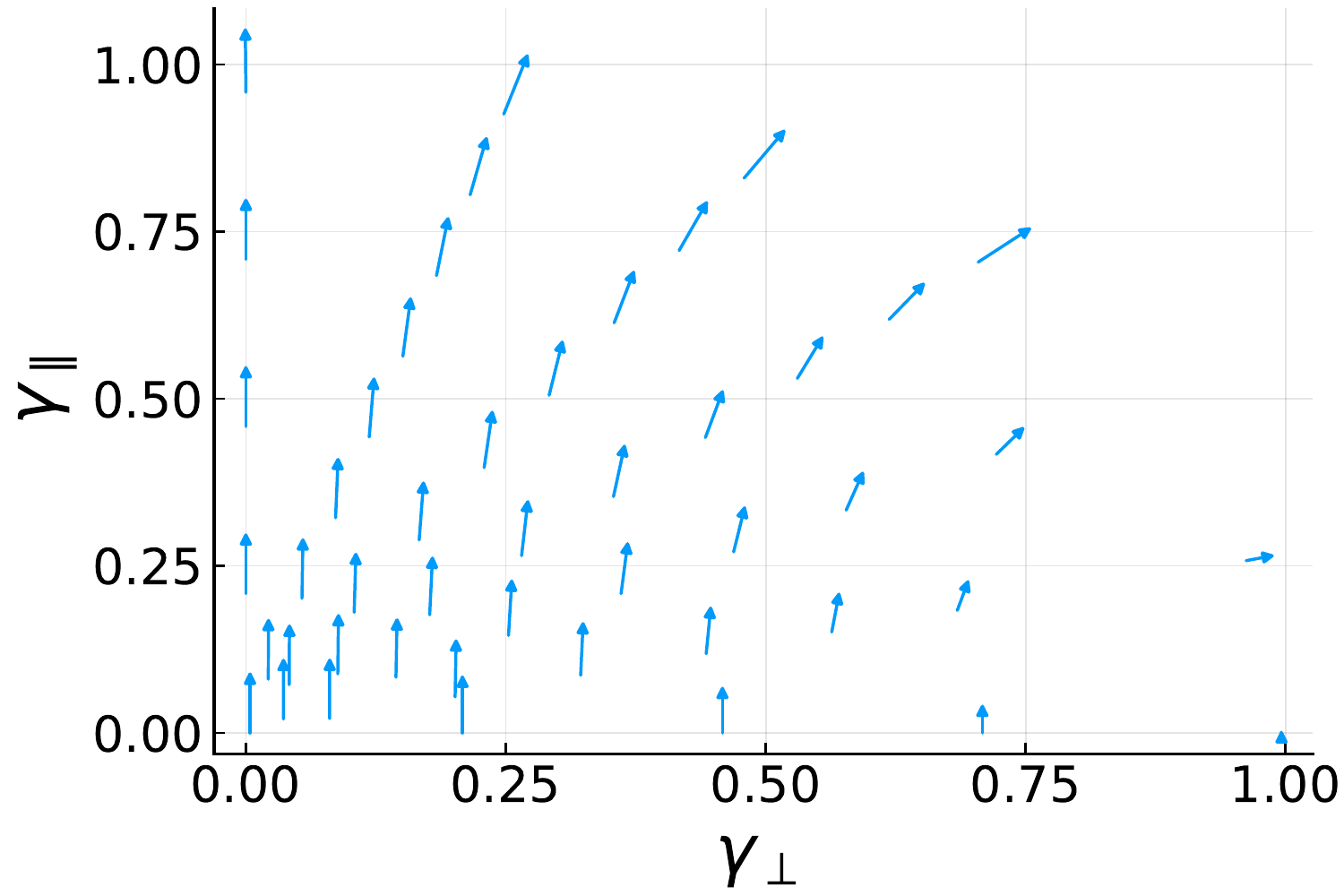}
		\\
	\end{tabulary}
	\caption{Net chain dipole for $\unodim = 0.25$ (top left), $\unodim = -0.25$ (top right), $\unodim = 9.0$ (bottom left), and $\unodim = -9.0$ (bottom right) at different stretches and orientations.
		The net chain dipole vectors are scaled such that each vector is given by $\chainpolar / \left( 10 \N \sqrt{|\unodim| \kB \T} \right)$.}
	\label{fig:mus}
\end{figure}

Thus far our focus has been on the influence of the chain electrostatics on the chain mechanics.
However, we also see that there is similarly a mechanical influence on the net chain dipole.
Fig. \ref{fig:mus} shows the net chain dipole for TI chains (e.g. $\unodim = 0.25$ (top left), $\unodim = 9.0$ (top right)) and uniaxial chains (e.g. $\unodim = -0.25$ (bottom left), $\unodim = -9.0$ (bottom right)) at different stretches and orientations.
The horizontal and vertical coordinates of (the base of) each net dipole vector represents the chain stretch in the direction orthogonal and parallel to the electric field, respectively; and the net chain dipoles are scaled such that each vector is given by $\chainpolar / \left( 10 \N \sqrt{|\unodim| \kB \T} \right)$, where the factor of $10$ is included purely for the convenience of not having vectors overlap each other.
At small stretches (0.0 to 0.25) the net dipole is in the direction of the electric field, as expected.
However, as the stretch increases toward its limit, monomers are forced to be oriented in the direction of chain stretch which influences the direction of magnitude of the net chain dipole.
For TI chains (left), because $\sus{1} < \sus{2}$, the magnitude of the net dipole decreases with increasing $\stch_\parallel$ and increases with $\stch_\perp$; and vice versa for the uniaxial chains (right).

Further, note, comparing the TI chains to uniaxial chains, the small stretch net chain dipole is (approximately) two times greater for TI chains than for uniaxial chains at the same $|\unodim|$.
This difference can again be explained by again considering the fact that a TI monomer has a plane of directions in which it attains its maximum $|\dipole|$, whereas a uniaxial monomer only has two discrete directions in which it attains its maximum $|\dipole|$.
Hence, in the balance between the internal energy and entropy terms in the chain free energy, TI monomers are able attain a larger net dipole because a chain with a larger density of monomers oriented nearly orthogonal to the electric field will have a larger entropy.
And since the difference in $|\chainpolar|$ between TI and uniaxial (of the same $|\unodim|$) is due to entropy, one can see that it vanishes as the chain is nears its fully stretched limit.

Lastly, we note the effect of $|\unodim|$ on the net chain dipole.
In Fig. \ref{fig:mus}, comparing $\unodim = 0.25$ and $\unodim = 9.0$ (or alternatively, $\unodim = -0.25$ and $\unodim = -9.0$) one can see that as $|\unodim|$ increases the small stretch magnitude of $|\chainpolar| / \left( 10 \N \sqrt{|\unodim| \kB \T} \right)$ increases.
This is because equilibrium corresponds to a minimization of free energy.
When $\stch \rightarrow 0$, the influence of chain orientation (with respect to the electric field) vanishes and instead the density of monomer orientations is determined by a balance between the electrostatic energy and entropy terms in the free energy.
As $|\unodim|$ increases, the influence of the electrostatic energy term increases relative to the entropic term such that $|\chainpolar| / \left( 10 \N \sqrt{|\unodim| \kB \T} \right)$ increases.

\subsection{Summary}

In general, with regard to the statistics of an electro-responsive polymer chain, there are three competing factors: (1.) electrical energy--which would induce monomer dipoles and have monomers rotate to align their respective dipoles with the electric field (2.) thermal energy--which prefers monomers to be oriented in a uniform random manner and (3.) the kinematic constraint of the end-to-end vector.
The quantity $|\unodim|$ is a measure of the influence of (1.) versus (2.).
Whereas the quantity $|\mults|$, and hence $\stch$, are measures of the influence of (3.).
The stiffness of a polymer chain is related to the slope of the free energy with respect to stretch.
For an electro-responsive chain, its stiffness depends on $\unodim$, the current stretch, and its orientation with respect to the electric field.
More specifically, TI chains exhibit a larger stiffness when stretched in or near the direction of the electric field, or opposite the direction of the electric field, than when stretched in or near a direction orthogonal to the electric field.
This is because the electrostatic energy minimum of a TI monomer is in the plane orthogonal to the electric field.
For a chain of uniaxial monomers, the larger stiffness occurs when stretched in or near a direction orthogonal to $\ezerodir$ as opposed to in the direction $\ezerodir$.
The effect of orientation on chain stiffness increases with respect to $|\unodim|$ and vanishes when $\unodim \rightarrow 0$.
A related, and perhaps even more significant, effect of the dependence of the chain direction on its free energy is that \emph{electro-responsive chains experience an electrostatic torque in an applied field}.

There are two regimes to the net chain dipole.
When the chain stretch is small ($\stch < 0.25$), the net chain dipole is in the direction of the electric field and its magnitude increases with $\N$ and $\ezeromag$.
In addition, all other parameters equal, the small stretch net dipole for $\sus{1} = 0, \sus{2} = 1$ (TI monomers) is twice that of $\sus{1} = 1, \sus{2} = 0$ (uniaxial monomers).
As the stretch increases, the net chain dipole approaches $\sustens\left(\rdir\right) \ezero$ and depends on the monomer dipole susceptibilities, $\sus{1}$ and $\sus{2}$.

% highlight important aspects: stiffness, torque
% define closed-dielectric free energy; derive its min principle; explain physical sig. (usable work for blah)

%%%%%%%%%%%%%%%%%%%%%%%%%%%%%%%%%%%%%%%%%%%%%%%%%%%%%%%%%%
%%%%%%%%%%%%%%%%%%%%%%%%%%%%%%%%%%%%%%%%%%%%%%%%%%%%%%%%%%
%%%%%%%%%%%%%%%%%%%%%%%%%%%%%%%%%%%%%%%%%%%%%%%%%%%%%%%%%%
%%%%%%%%%%%%%%%%%%%%%%%%%%%%%%%%%%%%%%%%%%%%%%%%%%%%%%%%%%

\section{Closed-form Solution in the Small Electrical Energy Limit} \label{sec:small-omega}

In this section, we examine the limit with $|\unodim|$ and $|\xmult|$ small.
Physically, the small $|\unodim|$ limit corresponds to the magnitude of the electrical energy of the system being much less than the thermal energy.
Further, $|\xmult|$ is the component of the chain force $\mults$ orthogonal to the direction of stretch; since in the classical limit with no electrical interactions, the stretch and $\mults$ are aligned, the misalignment will be small when the electrical contributions are small compared to $kT$.

Assuming $|\unodim|$ and $|\xmult|$ are small enables us to perform a Taylor expansion of the monomer orientation density function such that $\unodim$ and $\xmult$ are no longer in the argument of the exponential, and consequently obtain an approximation of the monomer density function that allows for a straight forward evaluation of the integrals given by \eqref{eq:cn-eap} and \eqref{eq:cr-eap}.
However, the system of equations that resulted from this approximate form are nonlinear.
To obtain an approximate closed-form solution, we use the $\unodim = 0$ solution (\Fref{sec:kuhn-and-grun}) as an initial guess and perform a single step of the Newton iteration scheme by hand.
We will see below that this closed-form approximation is accurate, in comparison to numerical solutions, for $|\uznodim|, |\uxnodim|, |\uxznodim| \leq 0.25$.
When $|\uznodim|, |\uxnodim|, |\uxznodim| \geq 1.0$, it is not accurate and further predicts unphysical features such as a nonconvex dependence of the free energy on stretch.

We begin by defining our coordinate system such that:
(1.) the polar axis and $\bfe_3$ are taken to be in the direction of $\rvec$, and
(2.) $\ezero$ lies in the plane spanned by $\eone$ and $\ethree$.

Because $\rvec$ and $\ezero$ both lie in the $\eone$, $\ethree$-plane, $\mults$ is also in the $\eone$, $\ethree$-plane; this can be assumed by symmetry, using that $\mults$ has no bias towards either of $+\etwo$ and $-\etwo$, and hence lies in the $\eone-\ethree$ plane.  
However, symmetries could break for various reasons, so we examine the combination of \eqref{eq:density-eap} and \eqref{eq:cr-eap}:
\begin{equation}
    \frac{\rvec}{\mlen} 
    = 
    \C \intoverSns{\exp\left[-\unodim \left(\hat\bfE_0 \cdot \hat\bfv\right)^2 + \mults \cdot \hat\bfv\right] \hat\bfv} 
\end{equation}
Writing out $\hat\bfv,\bfE_0,\mults$ in components, we see from symmetry that the component of $\hat\bfv$ in the $\etwo$ direction cancels out in the integration.
Using this, we write $\mults$ = $\left(\xmult, 0, \zmult\right)$.
Using \eqref{eq:monomer-energy} and \eqref{eq:density}, and expressing the result in terms of $\azi$ and $\polar$, we find
\begin{equation}
	\density\left(\azi, \polar\right) = \C \exp\left[-\funodimso\left(\azi, \polar\right) + \xmult \cos \azi \sin \polar + \zmult \cos \polar\right]
\end{equation}
where $\dsus = \sus{2} - \sus{1}$, define the dimensionless parameters
\begin{align*}
	\uznodim &= \beta \ez^2 \left(\dsus\right) / 2 \\
	\uxnodim &= \beta \ex^2 \left(\dsus\right) / 2 \\
	\uxznodim &= \beta \ex\ez \left(\dsus\right) / 2 = \sqrt{\uznodim \uxnodim}
\end{align*}
that are a measure of the electrical energy per monomer with respect to thermal energy per monomer (related to the $\ethree$ direction, $\eone$ direction, and interaction of the $\ez$ and $\ex$ components of the electric field, respectively), $\funodimso\left(\azi, \polar\right) = \uznodim \cos^2 \polar + 2 \uxznodim \cos \azi \cos \polar \sin \polar + \uxnodim \cos^2 \azi \sin^2 \polar$, and where terms in the argument of the exponential that were independent of $\azi$ and $\polar$ were absorbed into the unknown $\C$.

We proceed by assuming that $\ezeromag^2 \left(\dsus\right) \ll \kB \T$.
As a result of this assumption, and the choice of coordinate system, we have that $\xmult$ should also be small, regardless of the amount of chain stretch because $\xmult$ is a component of the force orthogonal to the direction of stretch.
Rewrite the exponential as $\exp\left[-\funodimso + \xmult \cos \azi \sin \polar\right] \times \exp\left[\zmult \cos \polar\right]$, then Taylor expand the first exponential in the product up to first order to obtain the approximate density function
\begin{equation} \label{eq:density-small-om}
	\densityso\left(\azi, \polar\right) \approx \C \left[1 - \funodimso\left(\azi, \polar\right) + \xmult \cos \azi \sin \polar\right] \exp\left(\zmult \cos \polar\right).
\end{equation}
Substituting the approximate density function into \eqref{eq:cn-eap} and \eqref{eq:cr-eap} and integrating results in the system of equations
\begin{align} \label{eq:eap-so-cn}
	\N &= \frac{4 \pi \C}{\zmult^3} \left[ \duzux\zmult\cosh\zmult - \left(\duzux-\uzpo\zmult^2\right) \sinh \zmult \right] \\ \label{eq:eap-so-crz}
	\frac{\rmag}{\mlen} &= -\frac{4 \pi \C}{\zmult^4} \left[ \left(3\duzux - \uzpo\zmult^2\right)\zmult\cosh\zmult + \left(-3\duzux - \left(\duzux-\uzpo\right)\zmult^2\right) \sinh \zmult \right] \\ \label{eq:eap-so-crx}
	0 &= -\frac{4 \pi \C}{\zmult^4} \left[ -\left(\xmult\zmult + 6\uxznodim\right)\zmult\cosh\zmult + \left(\xmult\zmult+2\left(3+\zmult^2\right)\uxznodim\right) \sinh \zmult \right]
\end{align}
where we define $\duzux = 2\uznodim - \uxnodim$ and $\uzpo = 1 - \uznodim$ for brevity.
Also, let $\I{1}$, $\I{2}$, and $\I{3}$ be defined as the right hand side of \eqref{eq:eap-so-cn}, \eqref{eq:eap-so-crz}, and \eqref{eq:eap-so-crx}, respectively.
The system \eqref{eq:eap-so-cn}--\eqref{eq:eap-so-crx} is clearly nonlinear.
In principle, one could approximate a solution to \eqref{eq:eap-so-cn}--\eqref{eq:eap-so-crx} by perturbation methods; see for example, \citet{bender2013advanced} Ch. 7 or \citet{hinch1991perturbation} Ch. 1.
However, power series expansions sometimes suffer from a slow rate of convergence and a limited radius of convergence.
In addition, for the case of the system \eqref{eq:eap-so-cn}--\eqref{eq:eap-so-crx}, there are multiple possibilities for the small parameter.
The dimensionless parameters $\uxnodim$, $\uznodim$, and $\uxznodim$ are all assumed small; however, they are inter-related in a complex way and not independent.
%For instance, one cannot take $\uxnodim = \uznodim =\uxznodim = \smallparam$ because $\uxnodim$ and $\uznodim$ can be varied independently of each other nor are they totally independent.

Therefore, we instead look to iterative methods that, while are typically used in numerical analysis, are well-suited for the current problem.
As in \citet{debruijn1981asymptotic} (Section 2.6), we use Newton's method to obtain an approximate solution to \eqref{eq:eap-so-cn}--\eqref{eq:eap-so-crx}.
Recall that in \Fref{sec:kuhn-and-grun} we effectively derived a solution for $\C$ and $\zmult$ in the absence of an electric field.
Since we are interested in the limit $\ezeromag^2 \dsus \ll \kB \T$, we can use the result of \Fref{sec:kuhn-and-grun} as an initial guess and calculate a correction using a Newton iteration.
Two advantages of Newton's method are, first, that its rate of convergence is quadratic, and, second, that the resulting approximation is rational instead of a power series.
Although rational approximations can be obtained by other means, such as Pad\'{e} approximations (see \citet{bender2013advanced} Ch. 8), they generally converge faster than power series and better capture behavior near singularities, for example, a fully stretched chain.

Let $\x = \left(\C, \zmult, \xmult\right)$ be the vector of unknowns, $\froot = \left(\N - \I{1}, \frac{\rmag}{\mlen} - \I{2}, \I{3}\right)$ be the vector of residuals, and $\Jij{i}{j} = \partial \frootk{i}/\partial \xk{j}$ be the Jacobian matrix.
The initial guess will be denoted by $\x_0 = \left(\C_0, \zmult_0, \xmult_0\right)$ so that $\froot_0 = \froot\left(\x_0\right)$ and $\J_0 = \J\left(\x_0\right)$ are the vector of residuals and Jacobian matrix evaluated at the initial guess.
Then
\begin{equation} \label{eq:newton-raphson}
	\x \approx \x_0 - \J_0^{-1} \froot_0
\end{equation}
As mentioned previously, because we are interested in an approximation that is accurate in the limit $\ezeromag^2 \sussymbol \ll \kB \T$, we take the exact solution for $\ezeromag^2 \sussymbol = 0$ as the initial guess, which is precisely the classical Kuhn and Grun solution. 
Thus, $\zmult_0 = \Langinvs$ and $\C_0$ is set as the right-hand side of \eqref{eq:ccn}.
To determine $\xmult_0$, we substitute $\C_0$ and $\zmult_0$ into \eqref{eq:eap-so-crx} and, remembering $\left(\langevin{\Langinvs}\right) = \stch$, solve for $\xmult$ to obtain
\begin{equation}
	\xmult_0 = -2 \uxznodim \left(\frac{3}{\Langinvs} - \frac{1}{\stch}\right)
\end{equation}
Evaluating $\J$ and $\froot$ at $\x_0$ and simplifying results in
%\begin{equation}
%	\J_0 = \Langinvso^{-3}\scalemath{0.75}{\begin{bmatrix}
%	
%		4 \pi \Langinvso \left(\uzpo\Langinvso - \duzux\stch\right)\sinh\Langinvso & 
%		\N \Langinvso \left(\uzpo \stch \Langinvso^2 + \duzux\left(3\stch-\Langinvso\right)\right) &
%		0
%		\\
%		
%		4 \pi \left(\uzpo \stch \Langinvso^2 + \duzux\left(3\stch-\Langinvso\right)\sinh\Langinvso\right) &
%		\N \left(\uzpo \Langinvso^3 - \left(\uzpo + \duzux\right)\Langinvso^2\stch + 4\duzux\left(\Langinvso - 3\stch\right) \right) &
%		0
%		\\
%		
%		0 &
%		\frac{2 \N \uxznodim}{\stch} \left(\stch^2\left(\Langinvso^2+3\right)-\Langinvso^2+2\Langinvso\stch\right) &
%		\N \Langinvso^2 \stch
%		
%	\end{bmatrix}}
%\end{equation}
\begin{equation}
	\J_0 = \Langinvso^{-3}\scalemath{1.0}{\begin{bmatrix}
	
		4 \pi \Langinvso \xce \sinh\Langinvso & 
		\N \Langinvso \left(\cc - \ca\right) &
		0
		\\
		
		4 \pi \left(\cc - \ca\right) \sinh\Langinvso &
		\N \left(\xce \Langinvso^2 + 4\ca - 2\cc\right)
		0
		\\
		
		0 &
		2 \N \cd \uxznodim / \stch
		 &
		\N \Langinvso^2 \stch
		
	\end{bmatrix}}
\end{equation}
and
%\begin{equation}
%	\froot_0 = \begin{Bmatrix}
%		\N \left(\uznodim - \frac{\duzux \stch}{\Langinvso}\right) \\
%		\frac{\N}{\Langinvso^2}\left(\Langinvso^2 \stch \uznodim - \duzux \Langinvso + 3\duzux \stch\right) \\
%		0
%		\end{Bmatrix}
%\end{equation}
\begin{equation}
	\froot_0 = \N \begin{Bmatrix}
		\xce / \Langinvso - 1 \\
		\stch \left( \uzpo - 1 \right) - \ca / \Langinvso^2 \\
		0
		\end{Bmatrix}
\end{equation}
respectively, where
$    \ca = \left(3\stch\duzux-\duzux\Langinvso\right), 
    %\cb &= \Langinvso \uzpo - \stch \duzux \\
    \cc = \stch \Langinvso^2 \uzpo,
    \cd = \stch^2 \left(\Langinvso^2 + 3\right) + 2\stch\Langinvso - \Langinvso^2, 
    \xce = \Langinvso \uzpo + \stch\duzux
$.
%Their exact physical significance is not obvious; however, one should pause to notice that: $\ca$, $\cc$, and $\xce$ consist of terms involving products of dimensionless energy (e.g. $\duzux$, $\uzpo$, etc.) and stretch terms (e.g. $\stch$, $\Langinvso$), and $\cd$ is a parameter that depends entirely on stretch.
Substituting $\x_0$, $\J_0$, and $\froot_0$ into \eqref{eq:newton-raphson} results in the approximation
\begin{align}
	\C &\approx \frac{\N \dc \csch \Langinvso}{4\pi \da} \\
	\zmult &\approx \Langinvso\left(1 - \db / \da\right) \\
	\xmult &\approx -\frac{2\left(\ca\da\stch + \cd \db \duzux\right)\uxznodim}{\da \stch^2 \duzux \Langinvso}
\end{align}
where
\begin{align*}
\da &= \ca^2+\cc^2+2\cc \xce-2\ca\left(\cc+2\xce\right)-\xce^2\Langinvso^2 = \frac{\Langinvso^6\csch\Langinvso}{4\pi\N^2\stch}|\J_0|\\
\db &= \cc \left(\xce - \Langinvso\right) + \Langinvso \left(\ca + \xce \stch \Langinvso \uznodim\right) \\
\dc &= \xce \Langinvso^3 - \cc^2 - \cc \Langinvso \left(\stch \Langinvso \uznodim + 2\right) + \ca\left[\cc \Langinvso\left(4+\stch\Langinvso\uznodim\right)\right]
\end{align*}
%TODO: Consider: should some of this be moved to an appendix?

Having obtained an approximate solution for the unknowns $\C$ and $\mults$, we turn our attention to the free energy.
Substituting the approximate density function \eqref{eq:density-small-om} into \eqref{eq:A-approx}, and using a Taylor expansion
\begin{equation*}
    \log\left(1 + \funodimso + \xmult \sin \polar \cos \azi\right) \approx \funodimso + \xmult \sin \polar \cos \azi
\end{equation*}
and integrating results in
\begin{equation} \label{eq:A-approx-so}
\begin{split}
	\A 
	=
	\frac{k T}{\zmult^4}\Bigg(
	    & -4\pi \C \sinh \zmult \Big[\xmult^2\zmult + 2\xmult\left(\zmult^2 + 3 \right)\uxznodim + \zmult\log\C\left(\duzux - \zmult^2 \uzpo \right) -
	\\
      & \duzux\zmult\left(\uOnodim + 3\right) + \zmult^3 \left(\uOnodim + \uxnodim - \uznodim\left(\uOnodim + 3\right) + 1\right) \Big] + 
	\\ 
		& 4\pi \C\zmult\cosh\zmult \Big[6\xmult\uxznodim + \zmult\left(\xmult^2 + 2\uznodim\log \C - \uznodim\left(\zmult^2 + 2\uOnodim + 6 \right) + \zmult^2 \right) +
	\\
    & \zmult\uxnodim\left(\uOnodim - \log \C + 3\right) \Big] - \zmult^4 \N\log \N \Bigg)
\end{split}
\end{equation}
where $\uOnodim = \sus{2} \ezeromag^2 / 2 \kB \T$ is the non-dimensional analog of the constant energy term in \eqref{eq:monomer-energy}.

\begin{figure}[htb!]
	\centering
	\begin{tabulary}{\linewidth}{c c}
		\includegraphics[width=\FigureWidthTwoColsNumericalGraph]{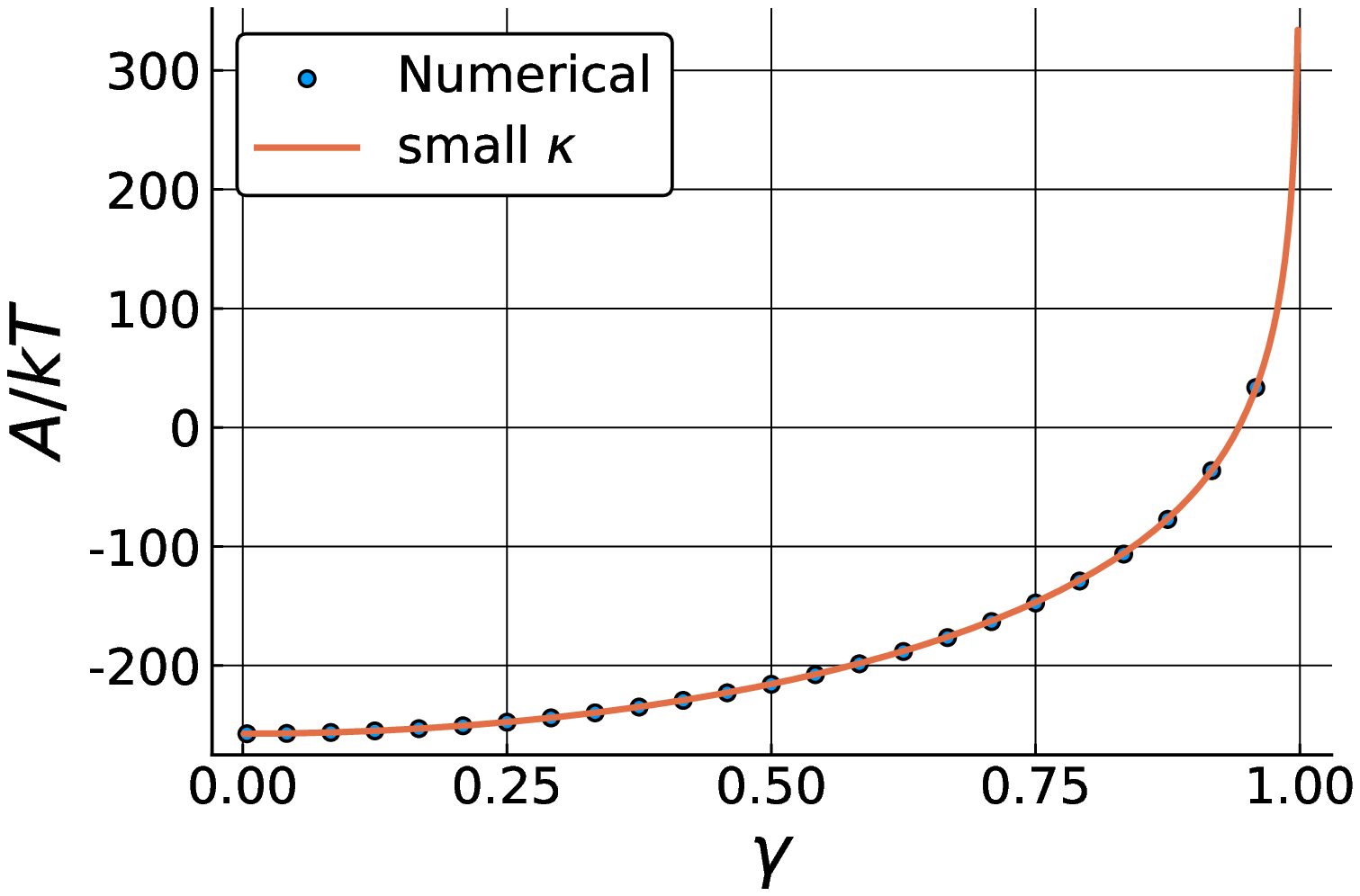}
		&
		\includegraphics[width=\FigureWidthTwoColsNumericalGraph]{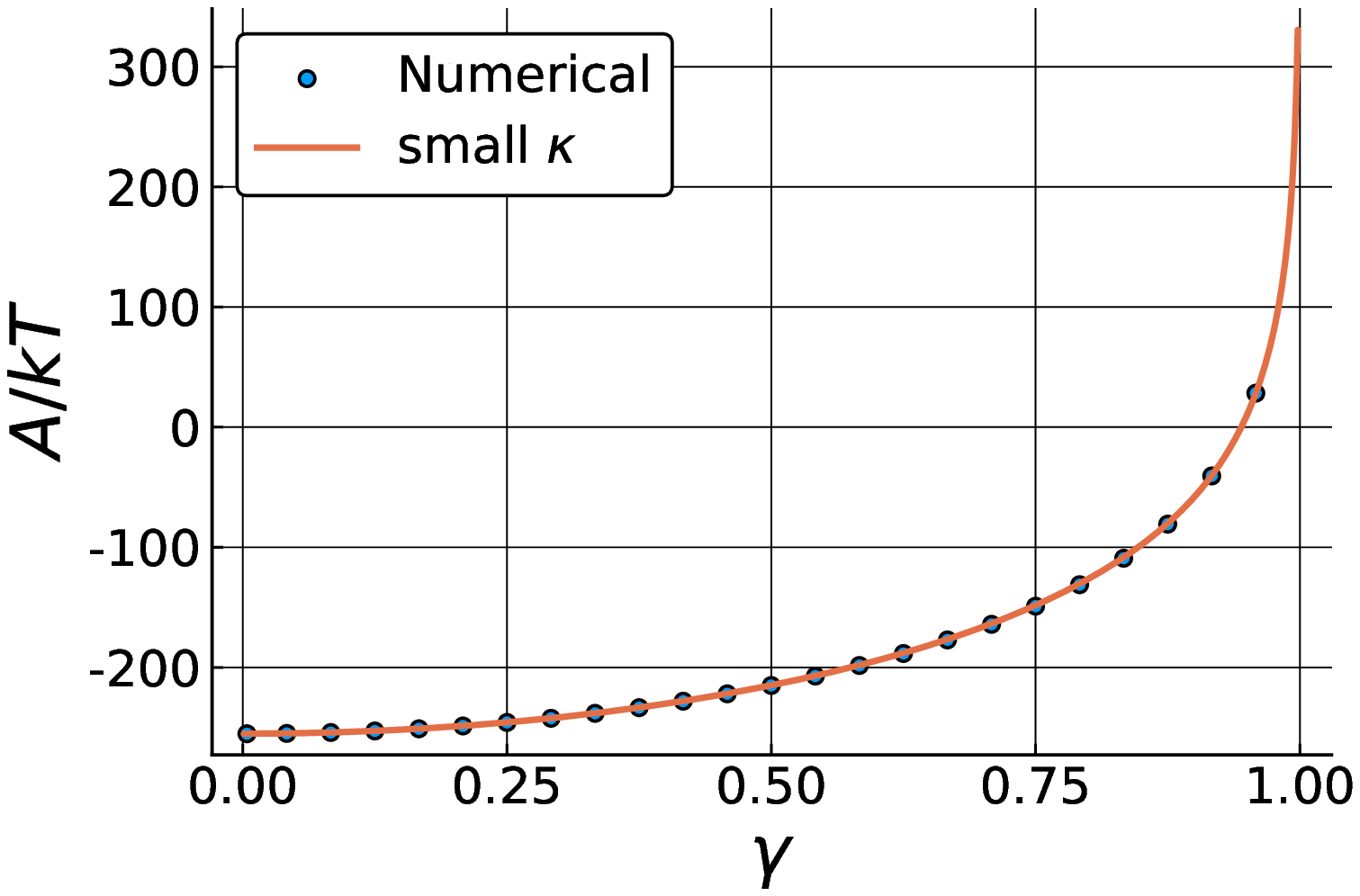}
		\\
		\includegraphics[width=\FigureWidthTwoColsNumericalGraph]{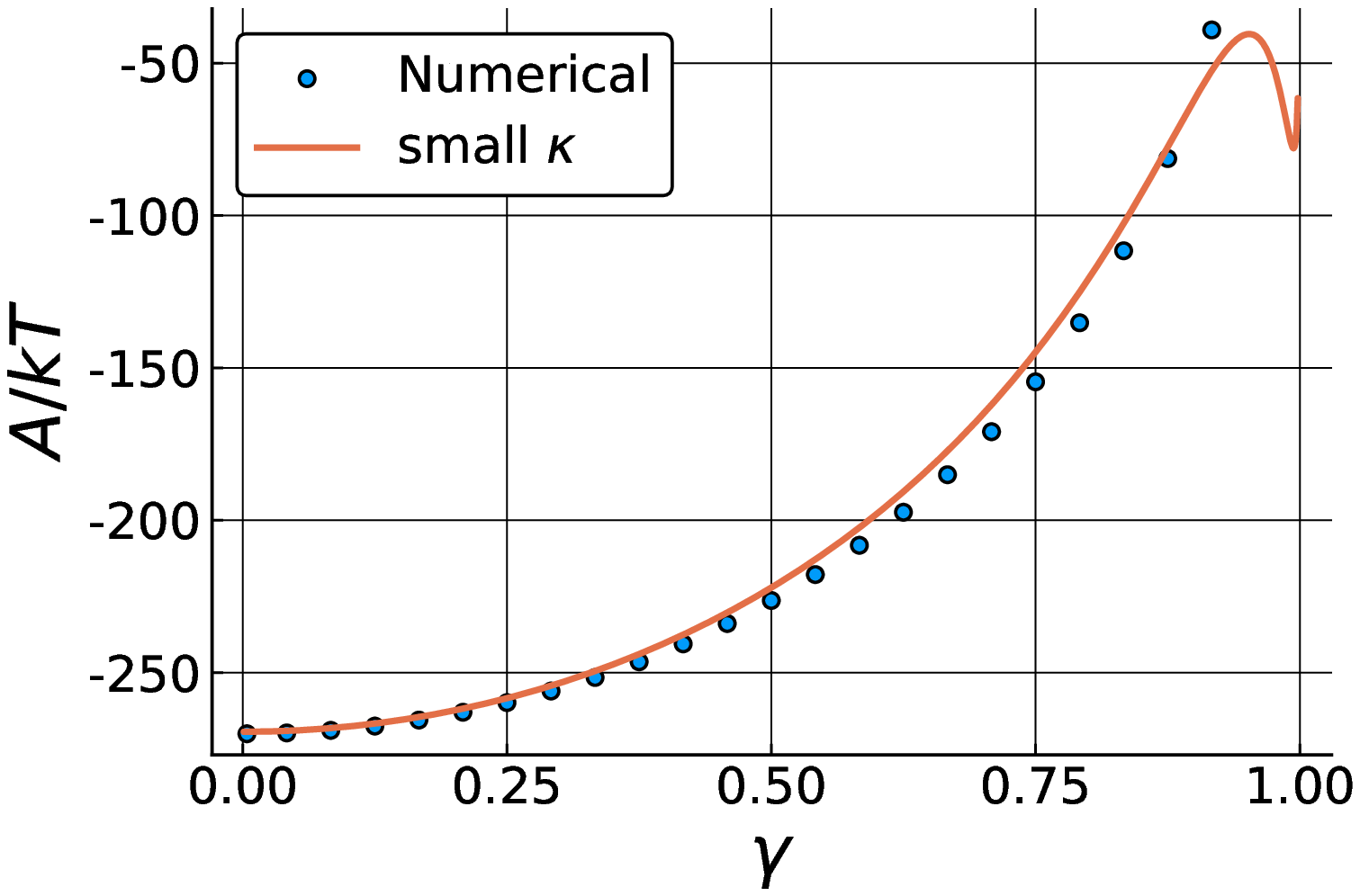}
		&
		\includegraphics[width=\FigureWidthTwoColsNumericalGraph]{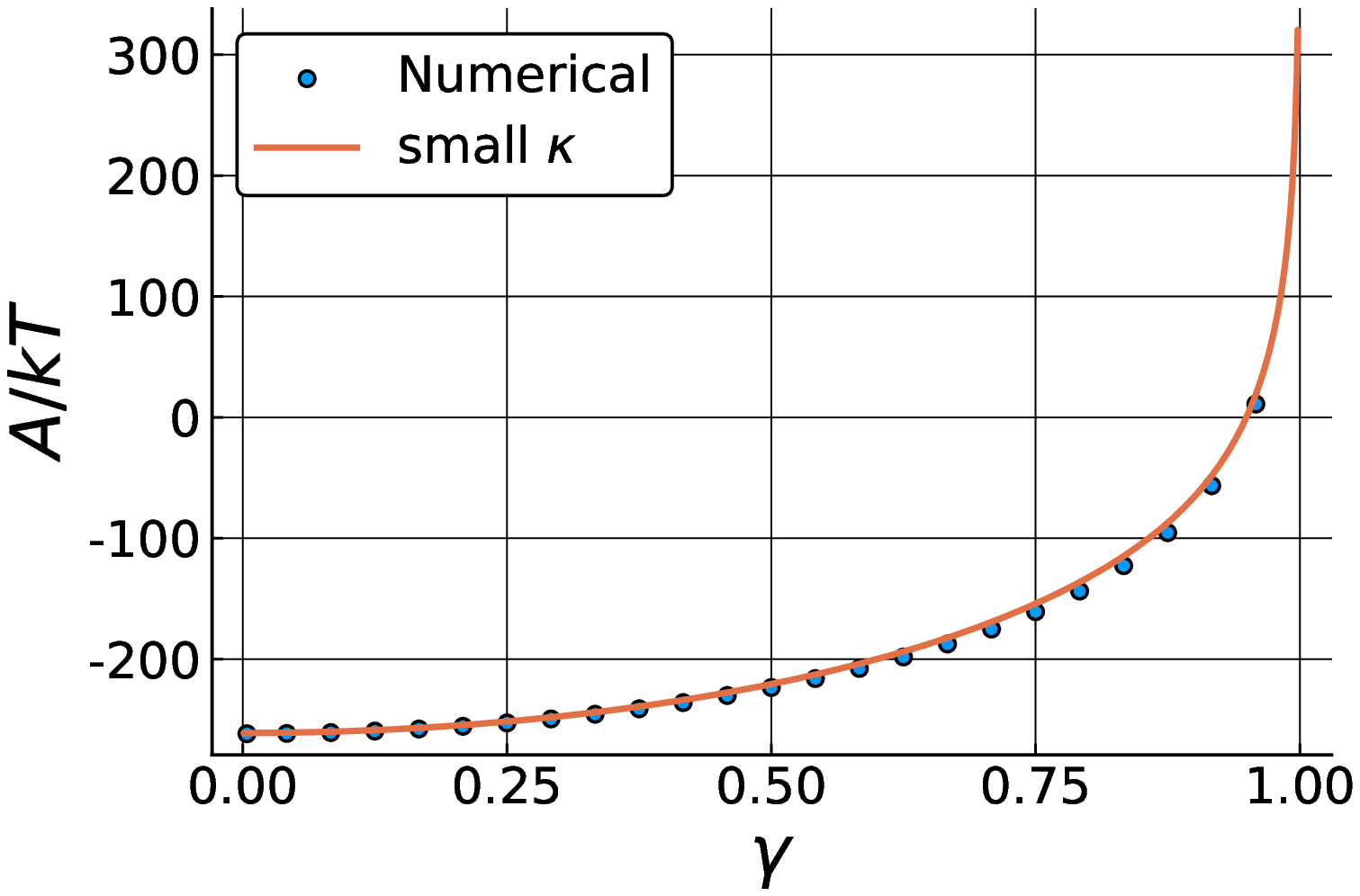}
		\\
		\includegraphics[width=\FigureWidthTwoColsNumericalGraph]{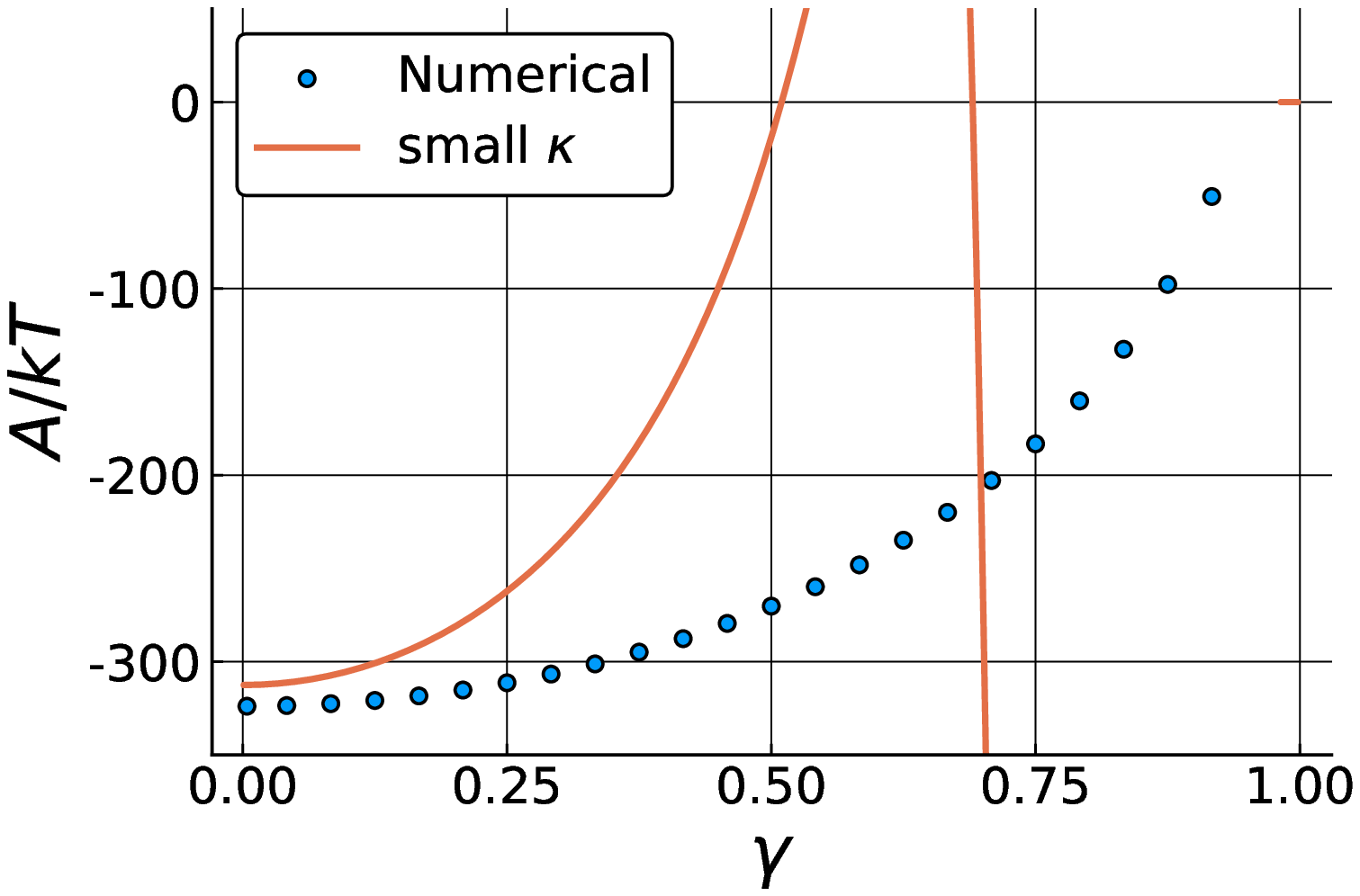}
		&
		\includegraphics[width=\FigureWidthTwoColsNumericalGraph]{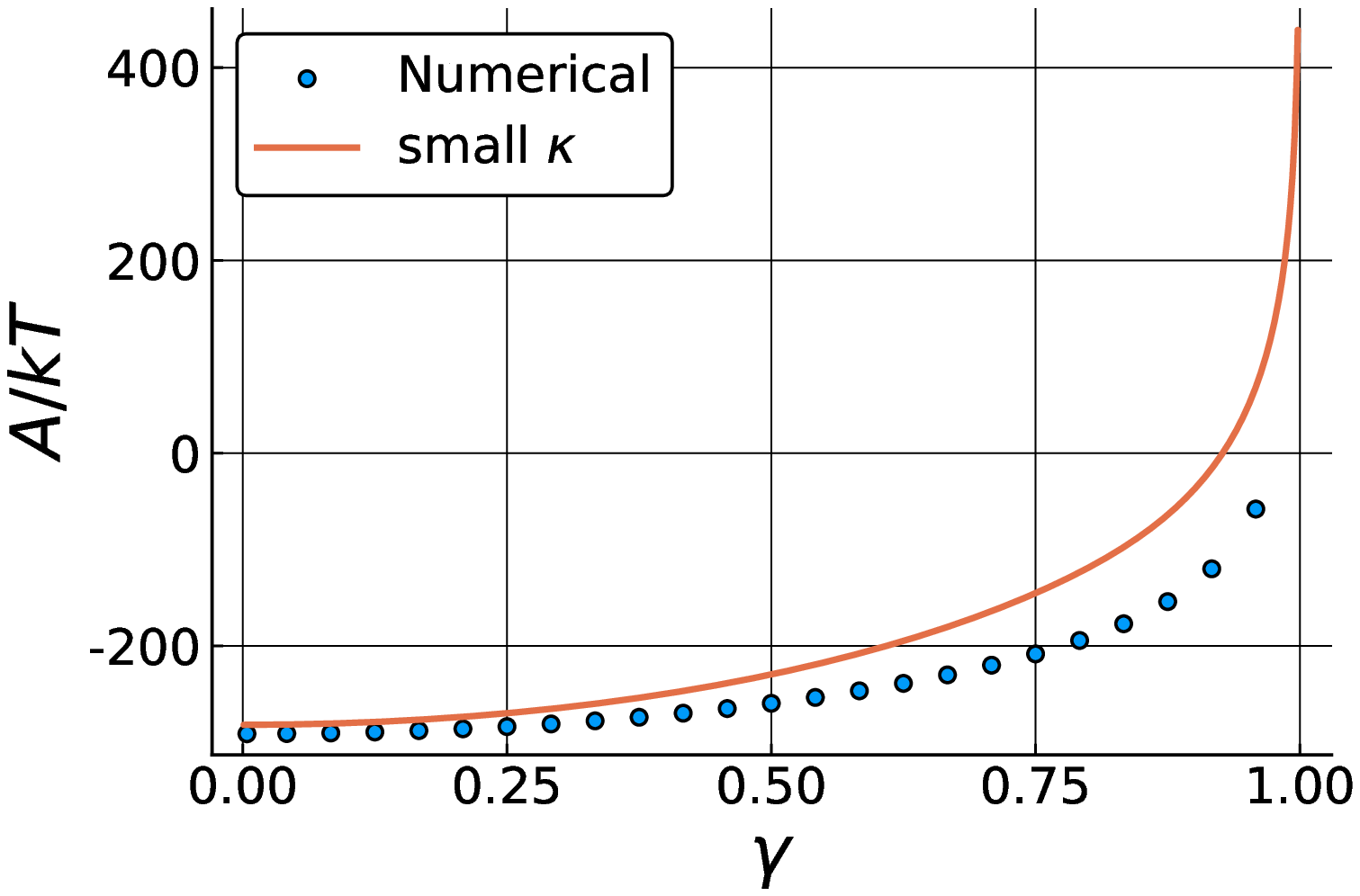}
		\\
		\includegraphics[width=\FigureWidthTwoColsNumericalGraph]{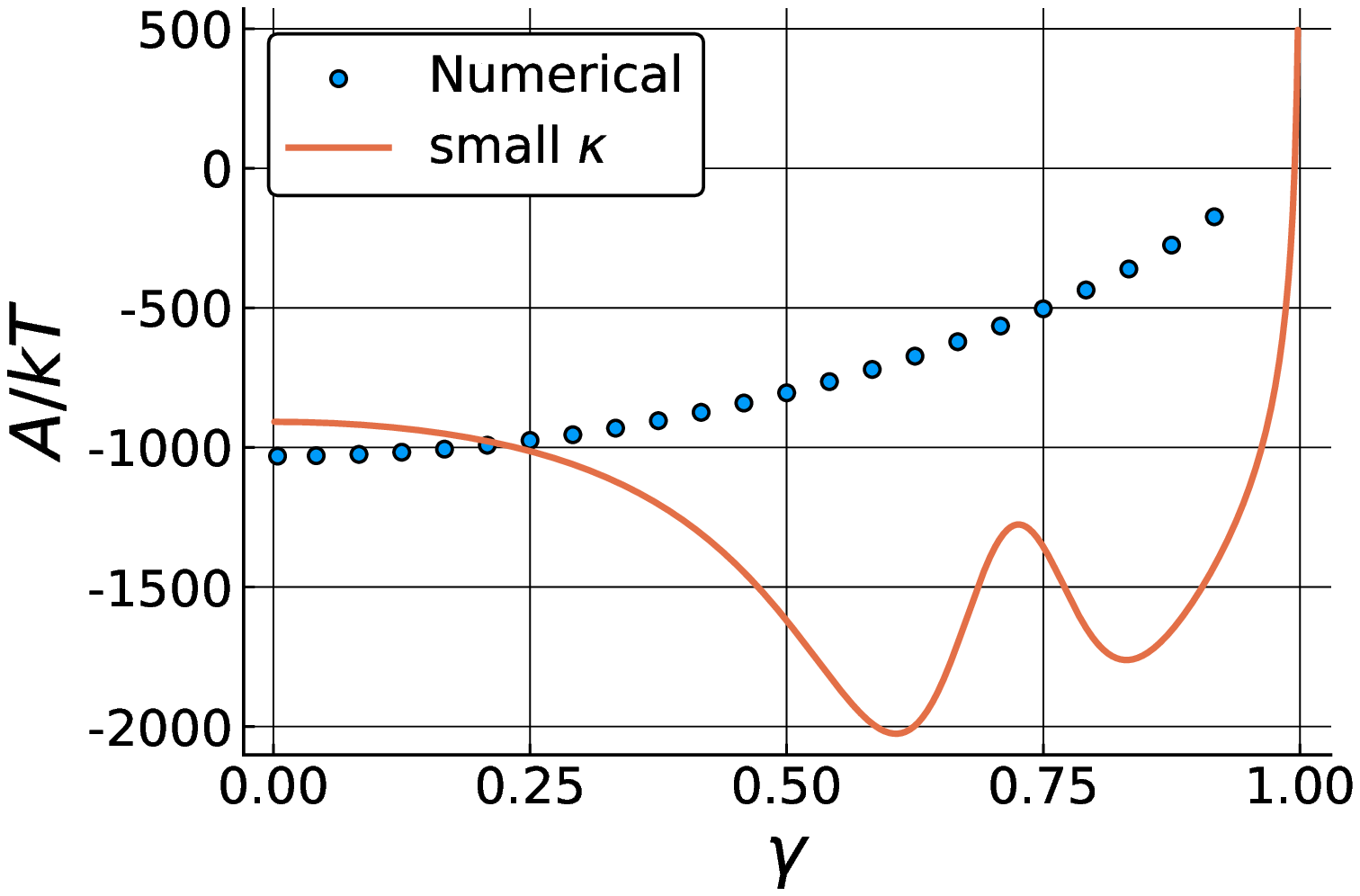}
		&
		\includegraphics[width=\FigureWidthTwoColsNumericalGraph]{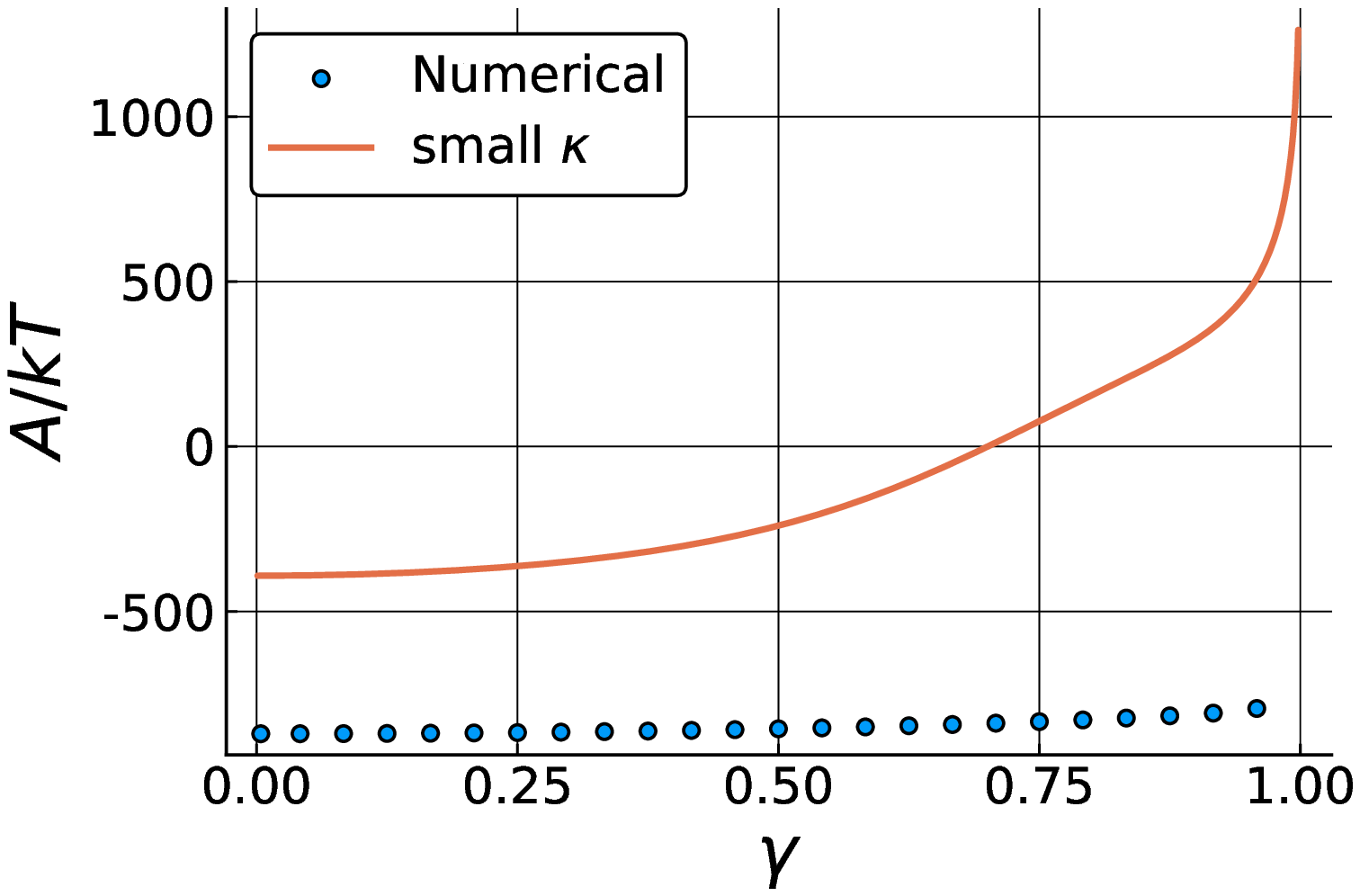}
	\end{tabulary}
	\caption{Comparison of the predicted $\A / \kB \T$ with $\stch$ relationship using the small $\unodim$ approximation and the numerical solutions.
		TI chains appear on the right and uniaxial chains on the left;
		$|\uznodim| = 0.0625, 0.25, 1.0, 9.0$ (top row, middle-top, middle-bottom, bottom); $\uxnodim = \uxznodim = 0.0$}
	\label{fig:so-A-Ez}
\end{figure}

Next, we compare the closed-form approximate solutions derived here to the numerical solutions obtained in  \Fref{sec:numerical}.
Fig. \ref{fig:so-A-Ez} compares $\A / \kB \T$ as a function of stretch.
The left plots in Fig. \ref{fig:so-A-Ez} show TI DE chains, and uniaxial chains are shown on the right.
The parameter $|\uznodim|$ increases from the top of the figure to bottom ($|\uznodim| = 0.0625, 0.25, 1.0, 9.0$, respectively; $\uxnodim = \uxznodim = 0.0$).
Recall that $|\unodim|, |\uznodim|, |\uxnodim|$, etc. are measures of electrical energy with respect to thermal energy; that is, increasing $|\uznodim|$ corresponds to an increase in the electric field and/or a decrease in temperature.
One can see that the approximation is accurate for smaller values of $|\uznodim|$ (e.g. $0.0625, 0.25$), but that it can be very inaccurate for $|\uznodim| = 1.0$ (and larger).
In addition, even for moderate $|\uznodim|$ (i.e. $|\uznodim| \ge 0.25$), the small $\unodim$ approximation for TI monomers predicts a nonconvex $\A / \kB \T$ vs $\stch$ curves, that can imply instability, phase transitions \cite{grekas2019cells} and so on that are simply not observed in the numerical solution.
As discussed in \Fref{sec:numerical}, one would not expect such phase transitions to be physical since (1.) stretch can only cause an increase in entropy, and (2.) a decrease in the electrical energy term cannot be greater than the increase in the entropy term, since otherwise the monomers would have taken such a configuration before stretching.

\begin{figure}[htb!]
	\centering
	\begin{tabulary}{\linewidth}{c c}
		\includegraphics[width=\FigureWidthTwoColsNumericalGraph]{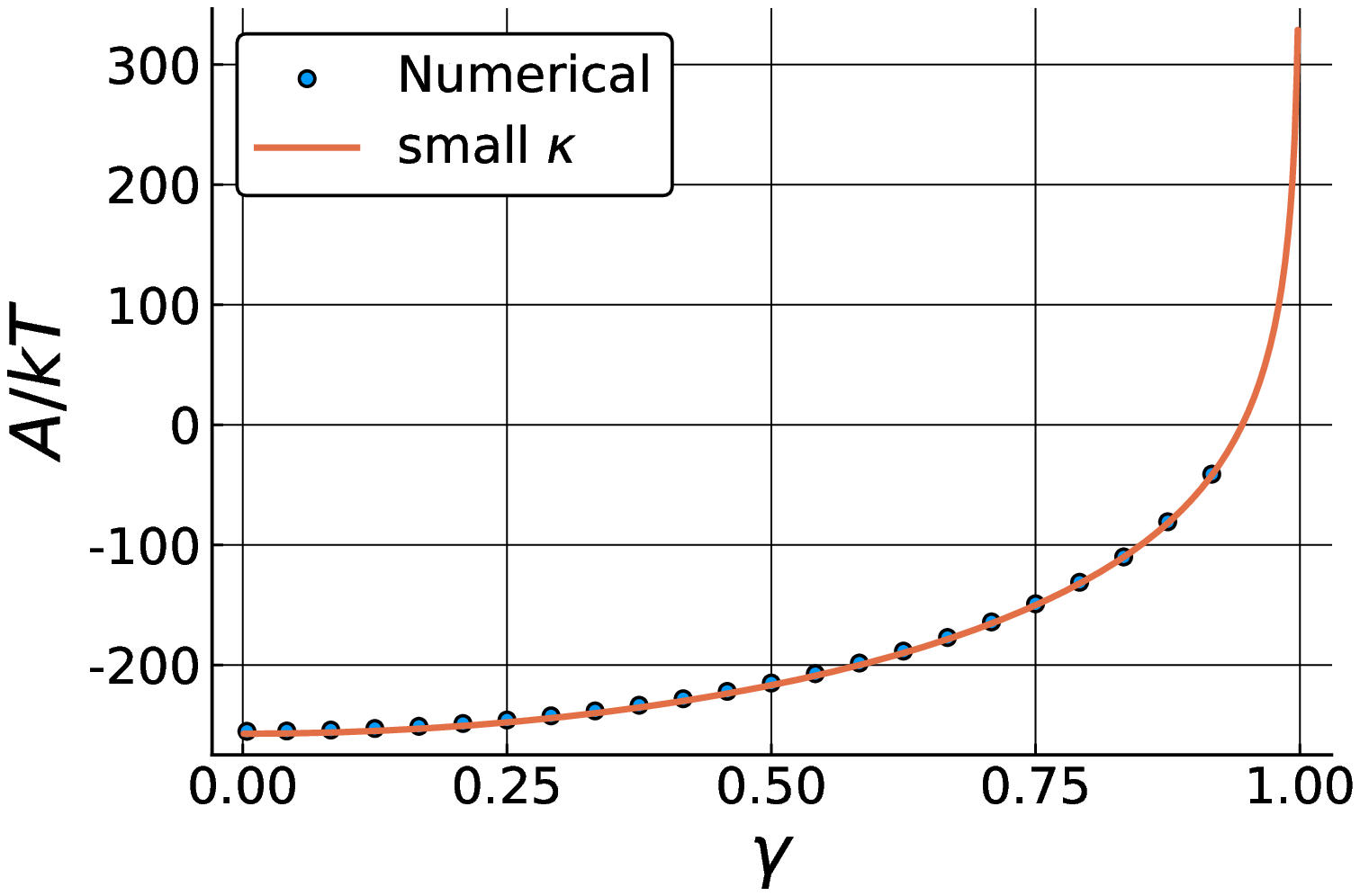}
		&
		\includegraphics[width=\FigureWidthTwoColsNumericalGraph]{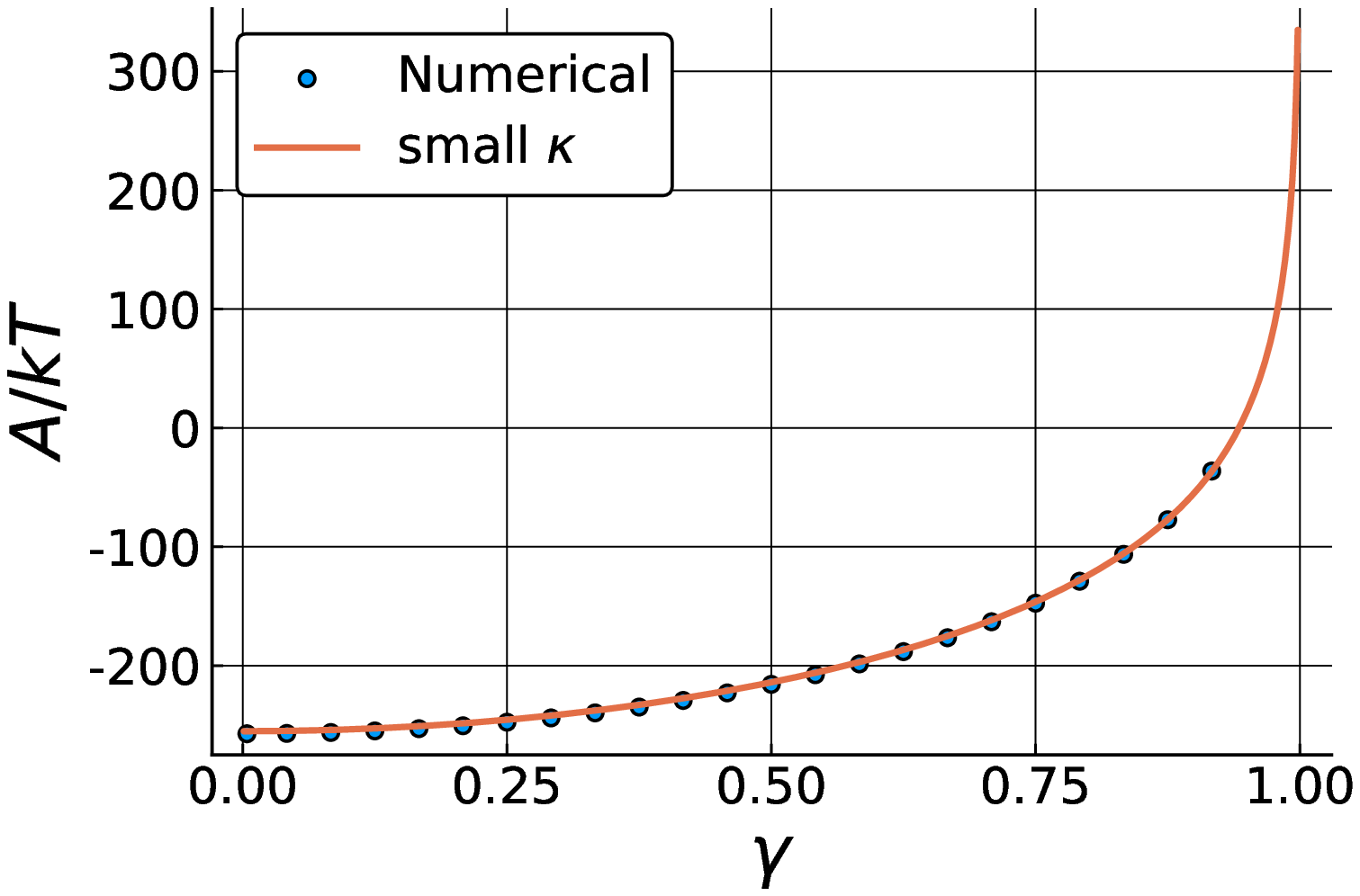}
		\\
		\includegraphics[width=\FigureWidthTwoColsNumericalGraph]{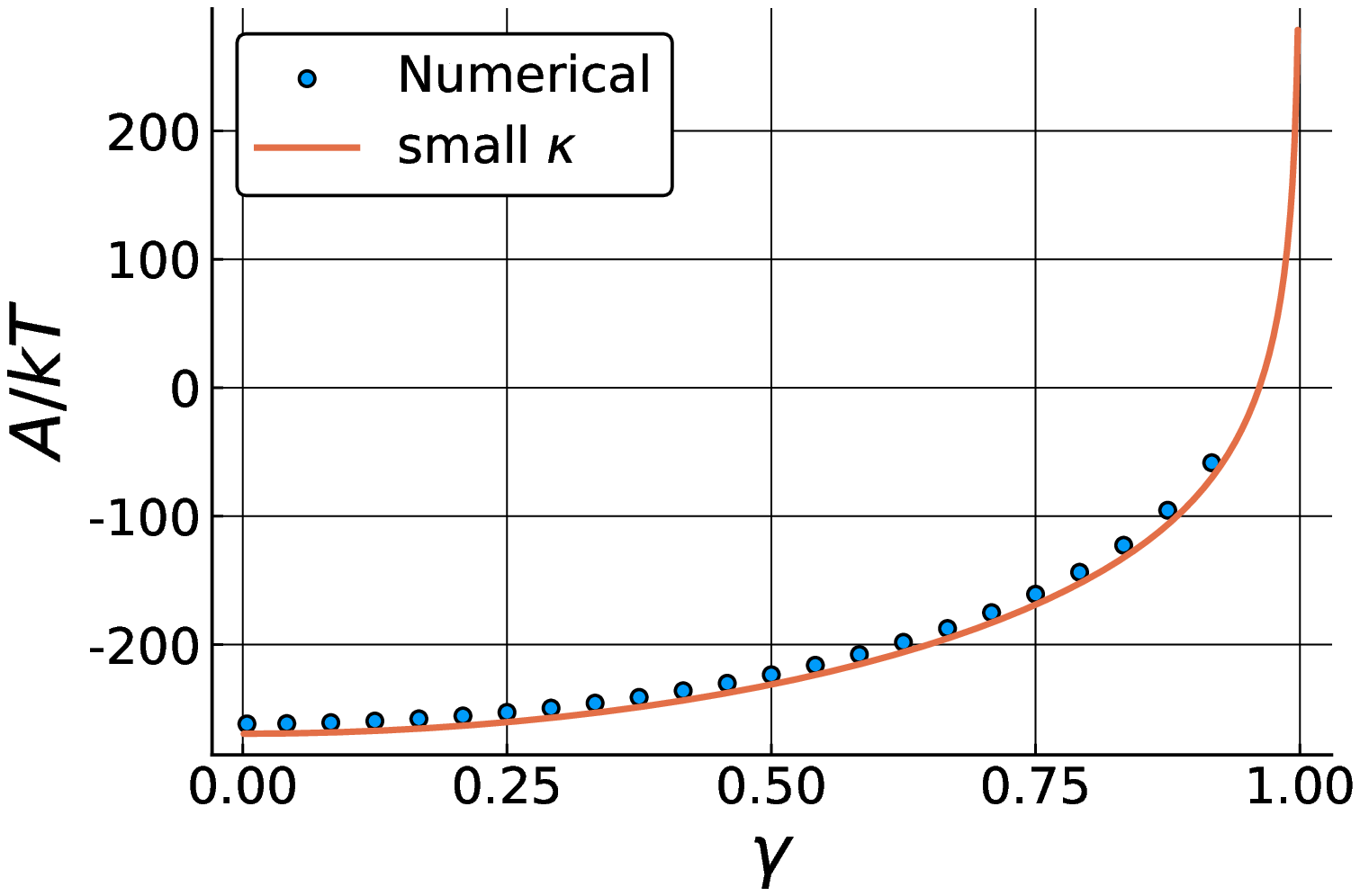}
		&
		\includegraphics[width=\FigureWidthTwoColsNumericalGraph]{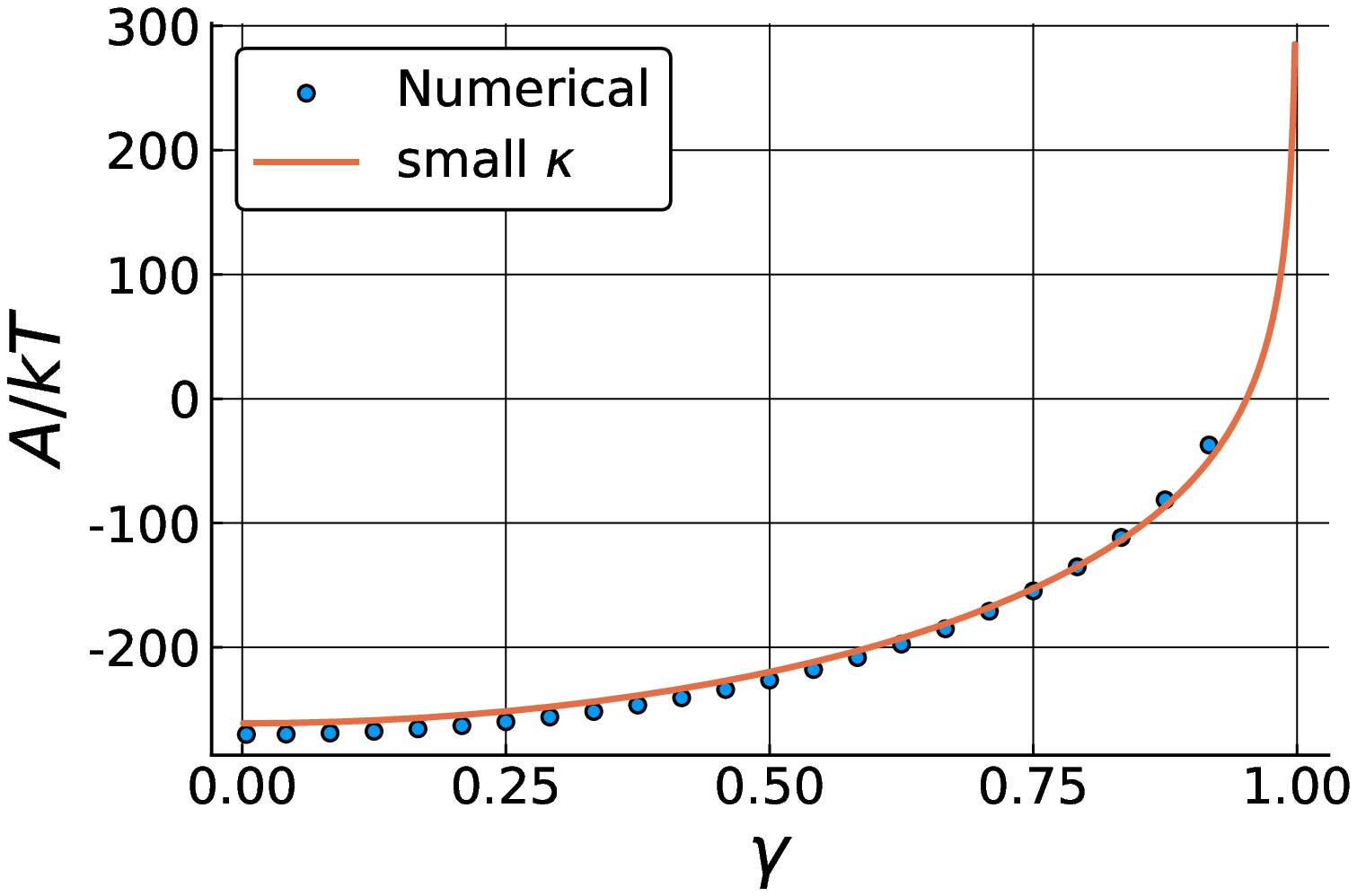}
		\\
		\includegraphics[width=\FigureWidthTwoColsNumericalGraph]{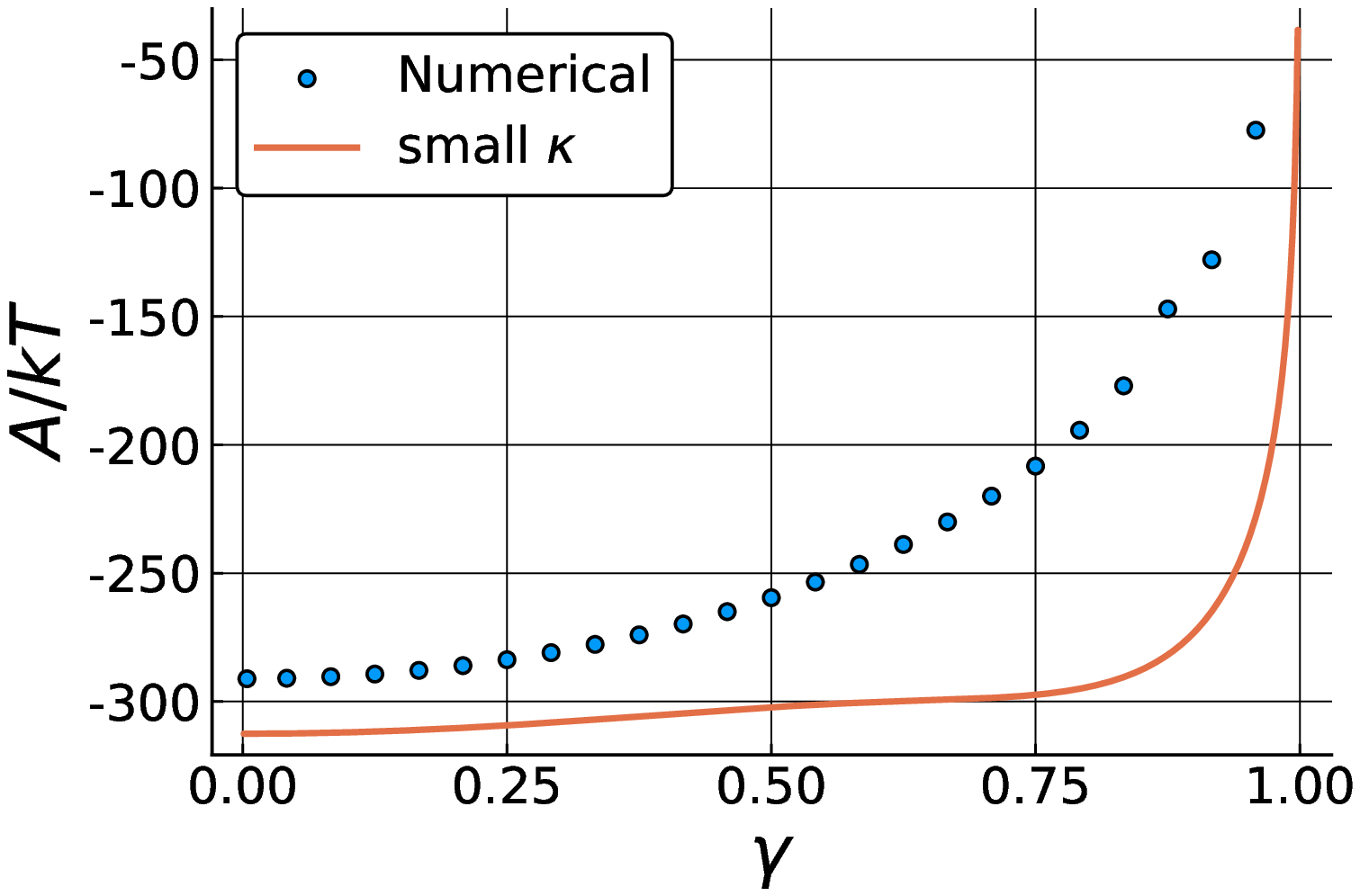}
		&
		\includegraphics[width=\FigureWidthTwoColsNumericalGraph]{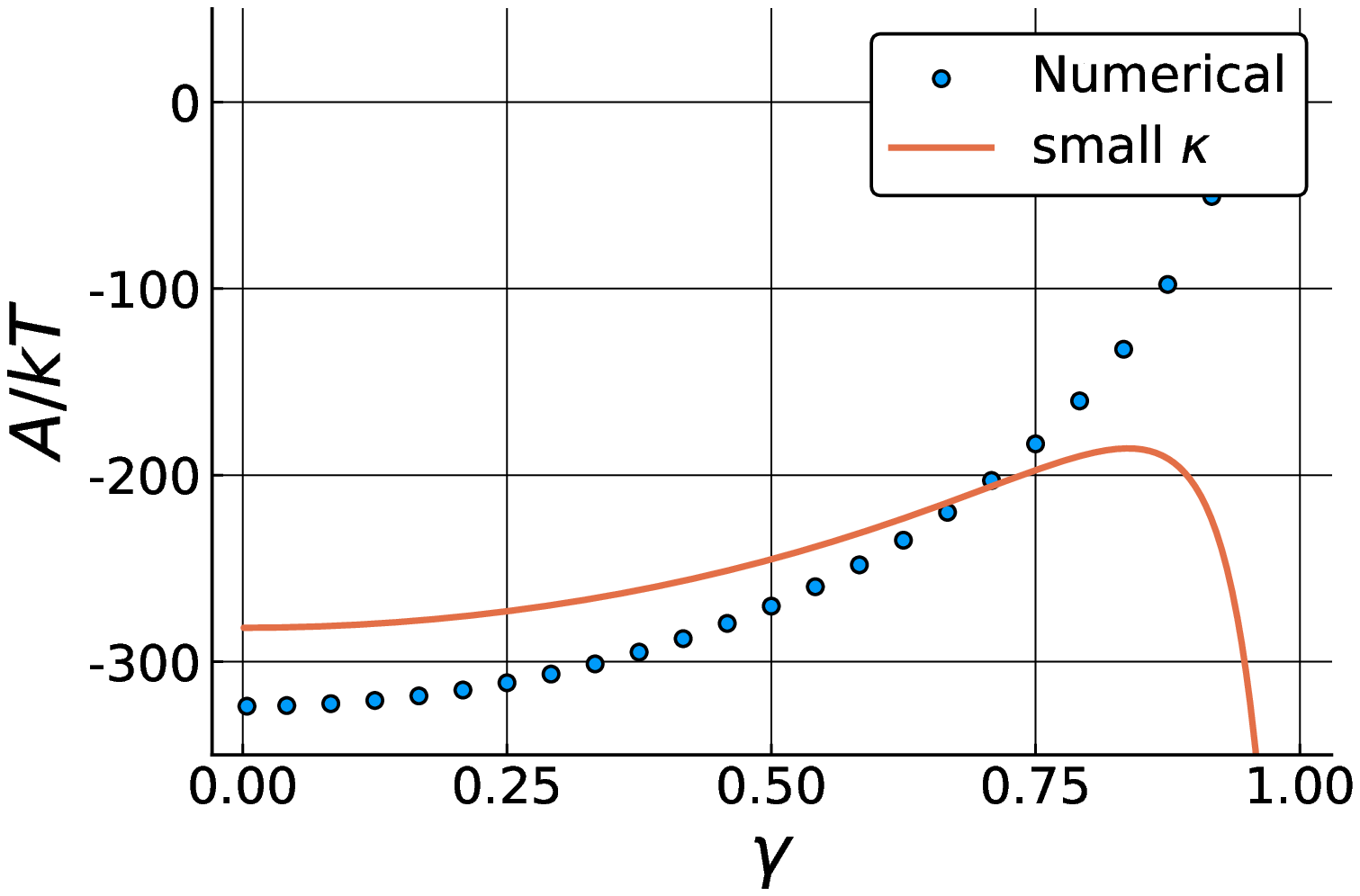}
	\end{tabulary}
	\caption{Comparison of the predicted $\A / \kB \T$ with $\stch$ relationship using the small $\unodim$ approximation and the numerical solutions.
		TI chains appear on the right and uniaxial chains on the left;
		$|\uxnodim| = 0.0625, 0.25, 1.0$ (top row, middle, bottom); $\uznodim = \uxznodim = 0.0$}
	\label{fig:so-A-Ex}
\end{figure}

\begin{figure}[htb!]
	\centering
	\begin{tabulary}{\linewidth}{c c}
		\includegraphics[width=\FigureWidthTwoColsNumericalGraph]{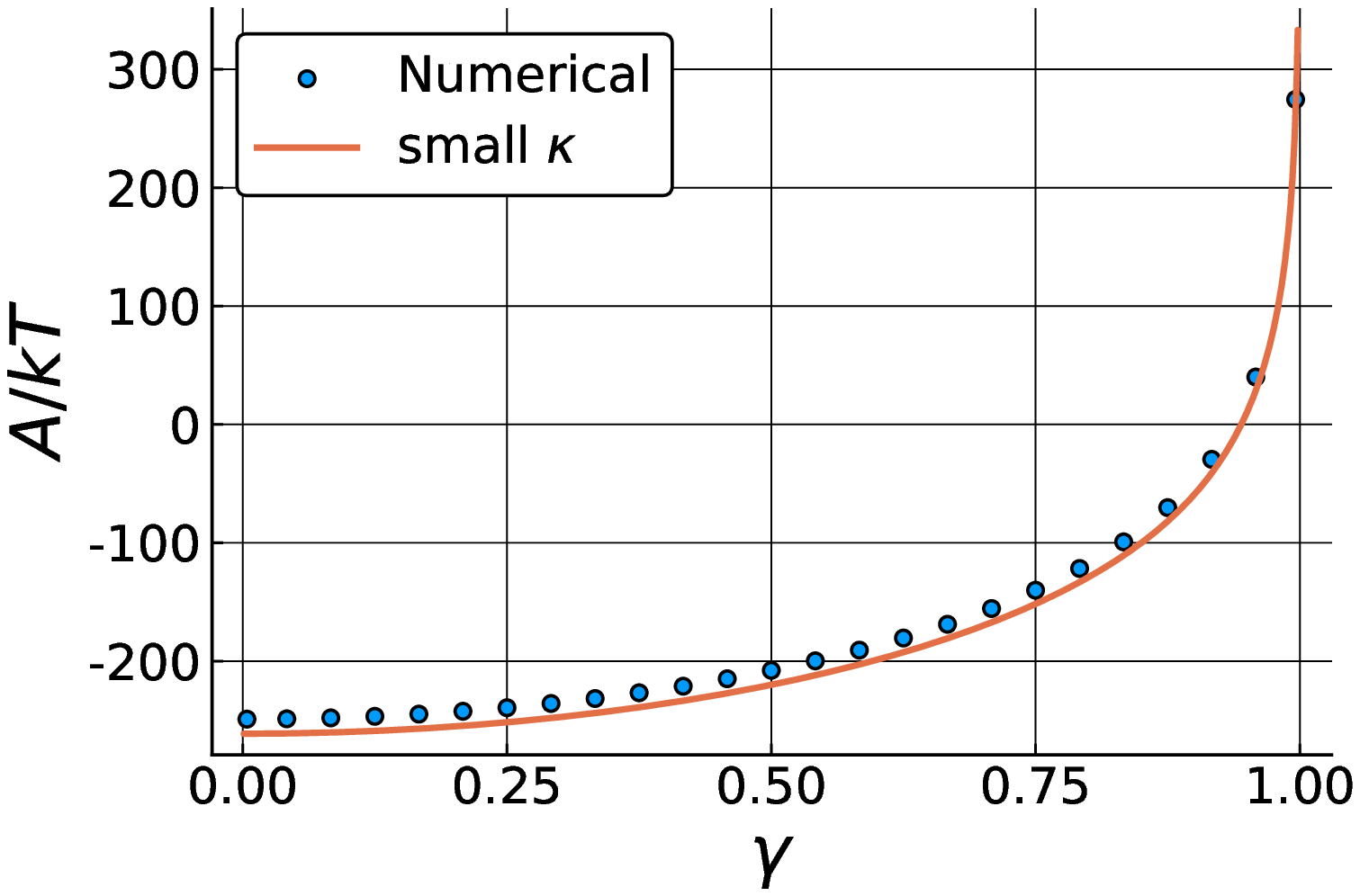}
		&
		\includegraphics[width=\FigureWidthTwoColsNumericalGraph]{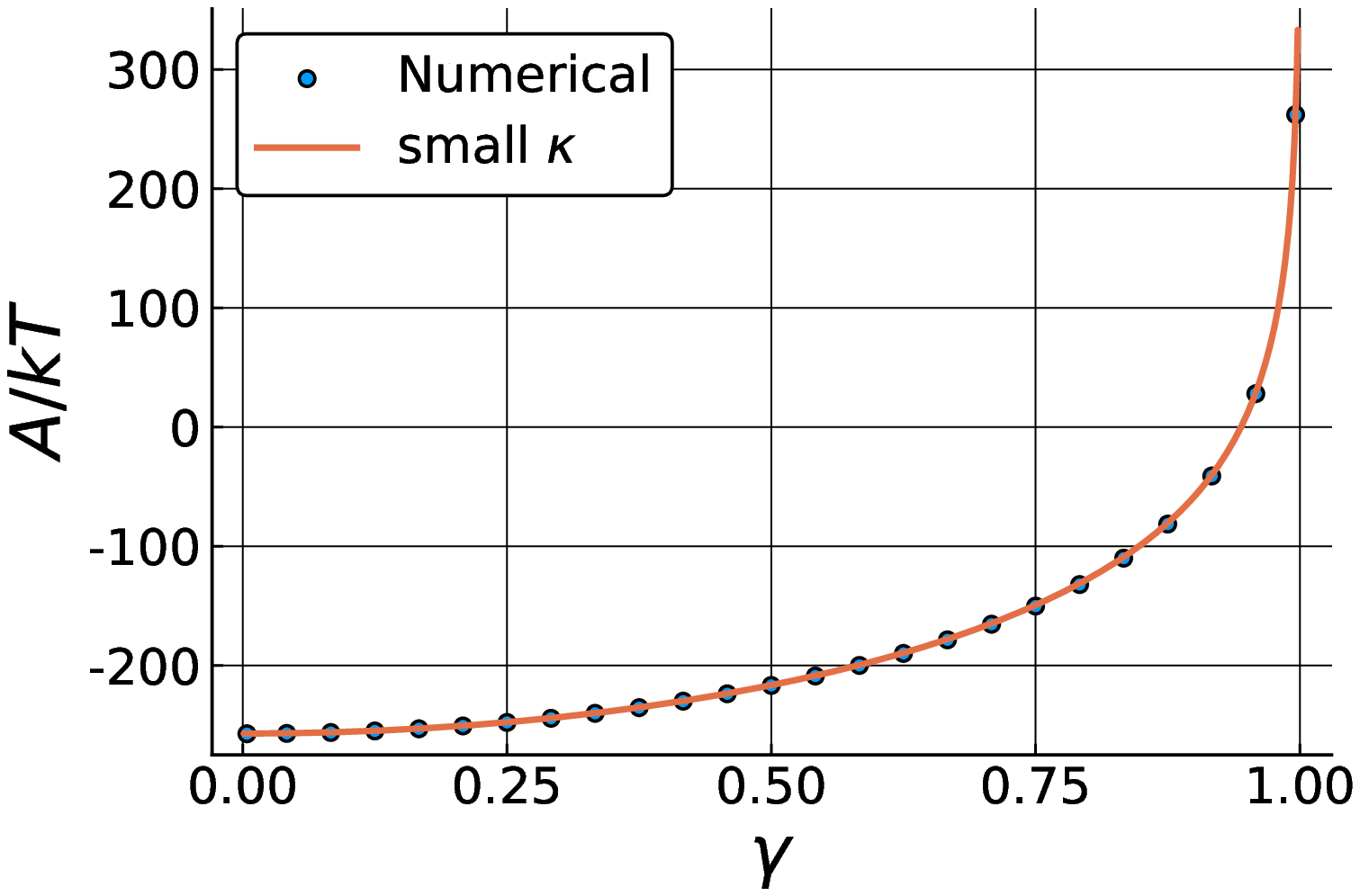}
		\\
		\includegraphics[width=\FigureWidthTwoColsNumericalGraph]{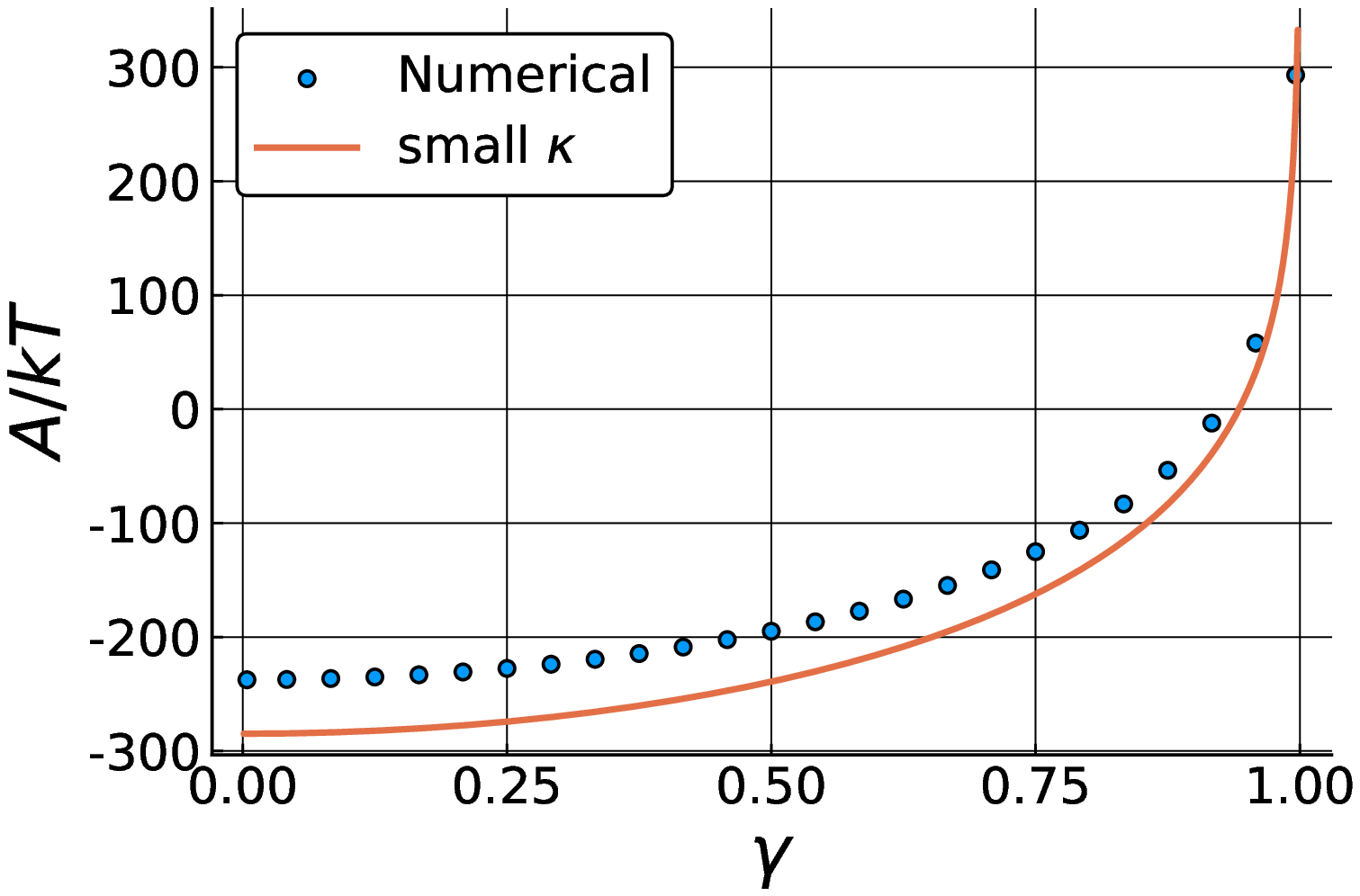}
		&
		\includegraphics[width=\FigureWidthTwoColsNumericalGraph]{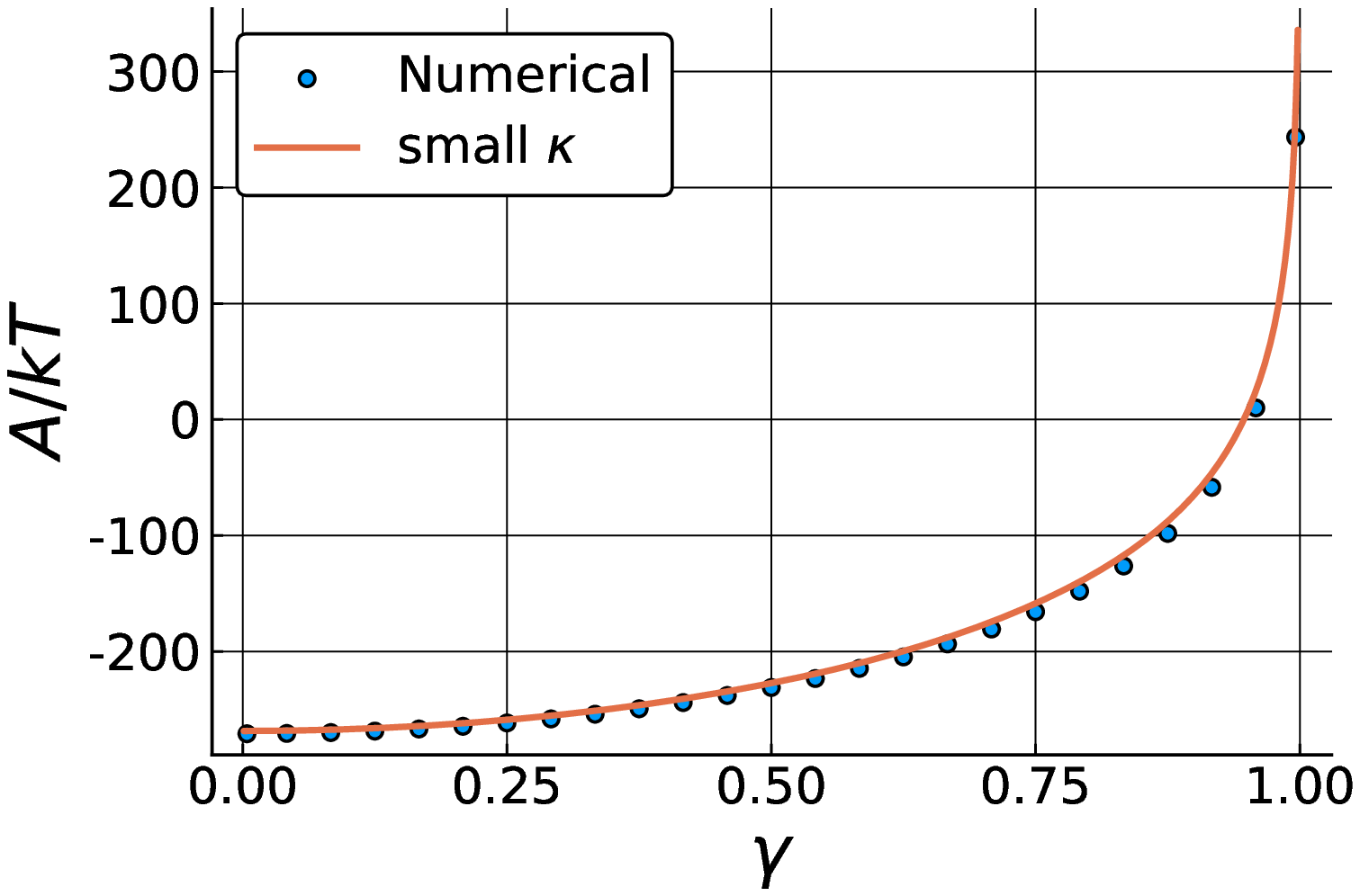}
		\\
		\includegraphics[width=\FigureWidthTwoColsNumericalGraph]{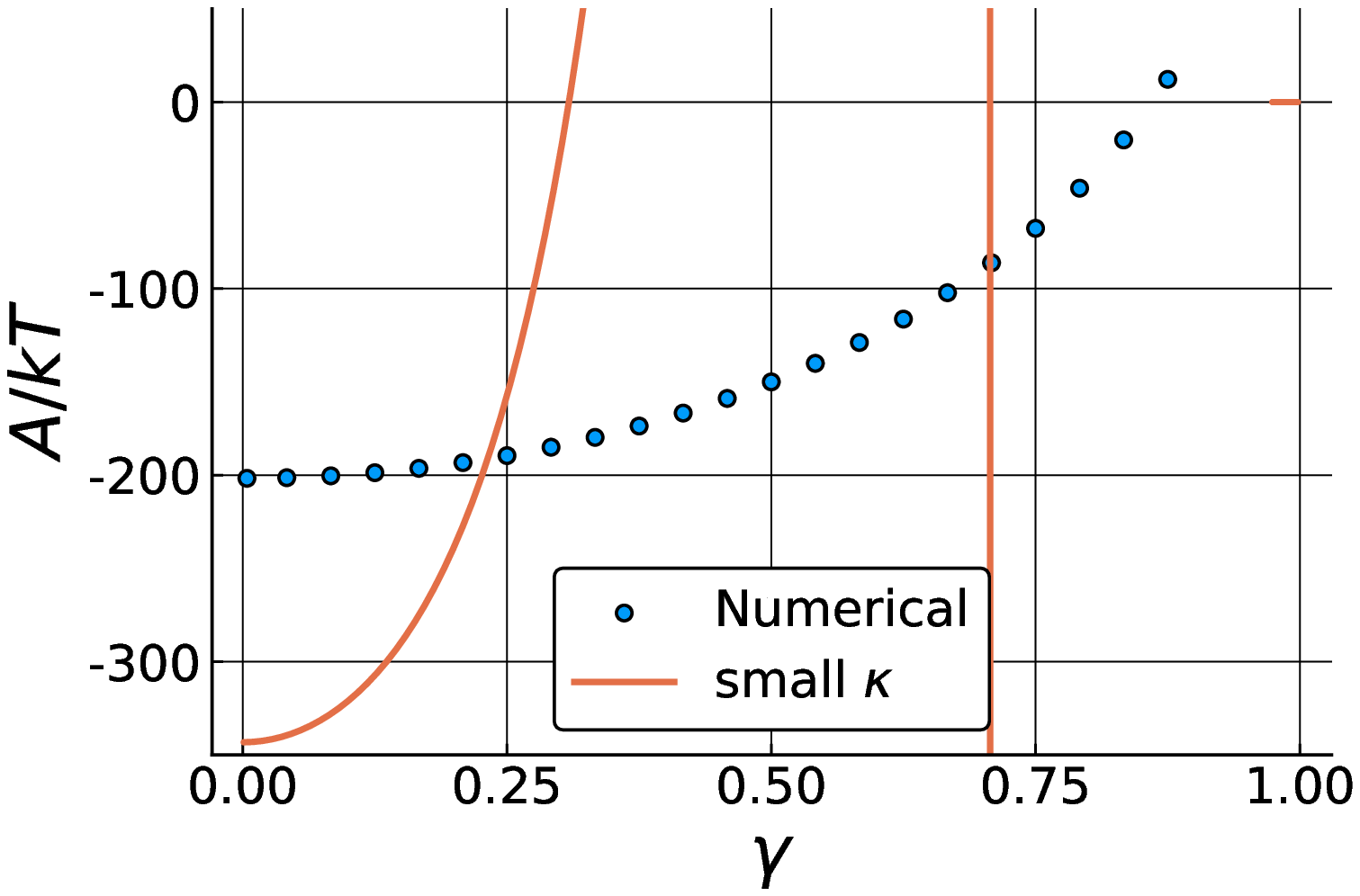}
		&
		\includegraphics[width=\FigureWidthTwoColsNumericalGraph]{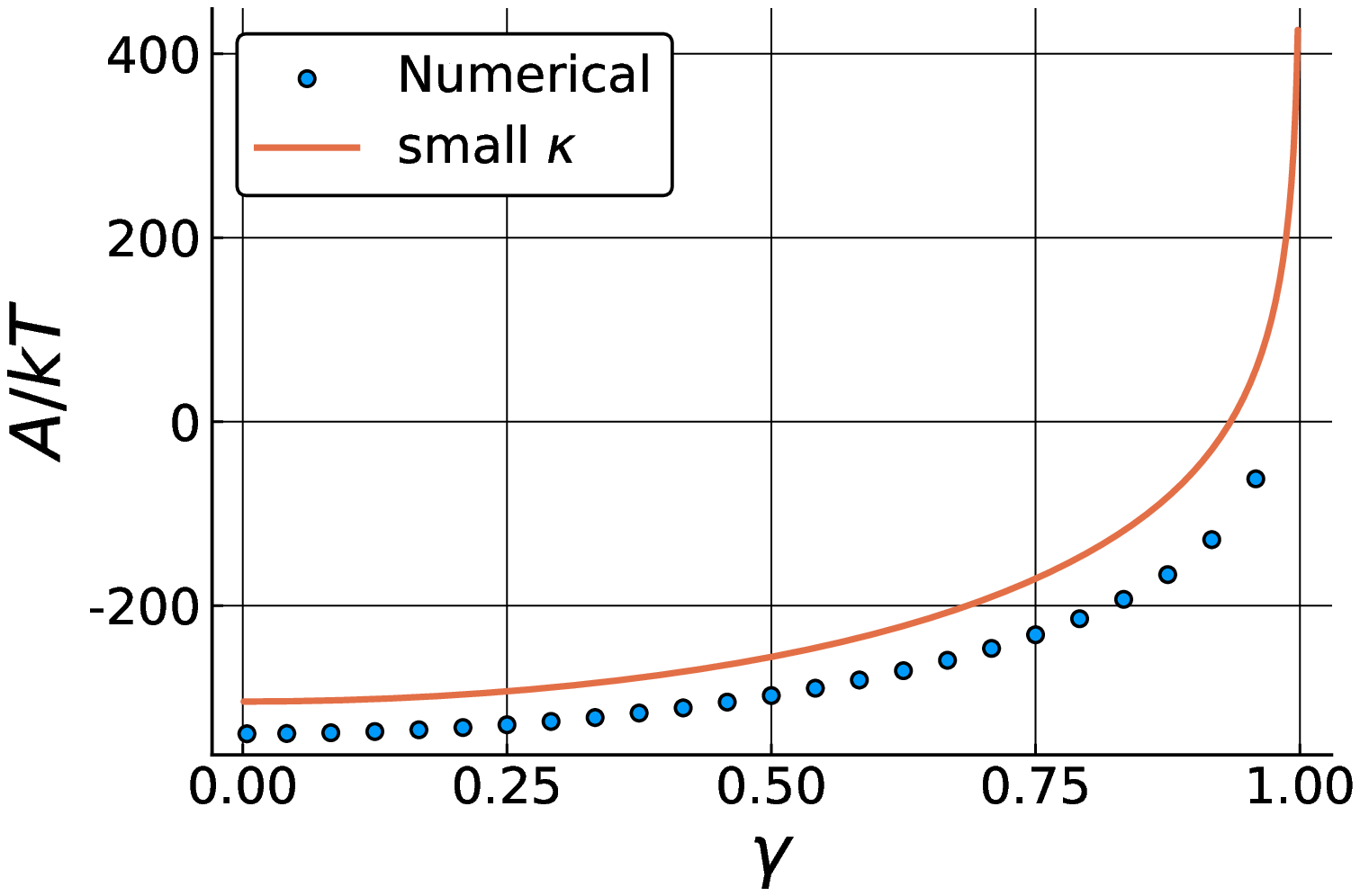}
	\end{tabulary}
	\caption{Comparison of the predicted $\A / \kB \T$ with $\stch$ relationship using the small $\unodim$ approximation and the numerical solutions.
		TI chains appear on the right and uniaxial chains on the left;
		$|\uxnodim|=|\uznodim|=|\uxznodim| = 0.0625, 0.25, 1.0$ (top row, middle, bottom)}
	\label{fig:so-A-Exz}
\end{figure}

Fig. \ref{fig:so-A-Ex} and Fig. \ref{fig:so-A-Exz} show the $\A / \kB \T$-$\stch$ relationship--comparing the small $\unodim$ approximation to the numerical solutions--for increasing $|\uxnodim|$ ($|\uxnodim| = 0.0625, 0.25, 1.0$) and $|\uxznodim|$ ($|\uxznodim| = |\uxnodim| = |\uznodim| = 0.0625, 0.25, 1.0$), respectively.
Similar to what was shown in Fig. \ref{fig:so-A-Ez}, the small $\unodim$ agrees well with the numerical solutions over the entire domain of stretch (i.e. $\stch \in [0, 1)$) when $|\uxnodim| \leq 0.25$ and $|\uxznodim| \leq 0.25$.
However, the approximation does not agree with the numerical solutions when $|\uxnodim| \geq 1.0$ and $|\uxznodim| \geq 1.0$ and can show nonphysical features.

\begin{figure}[htb!]
	\centering
	\begin{tabulary}{\linewidth}{c c}
		\includegraphics[width=\FigureWidthTwoColsNumericalGraph]{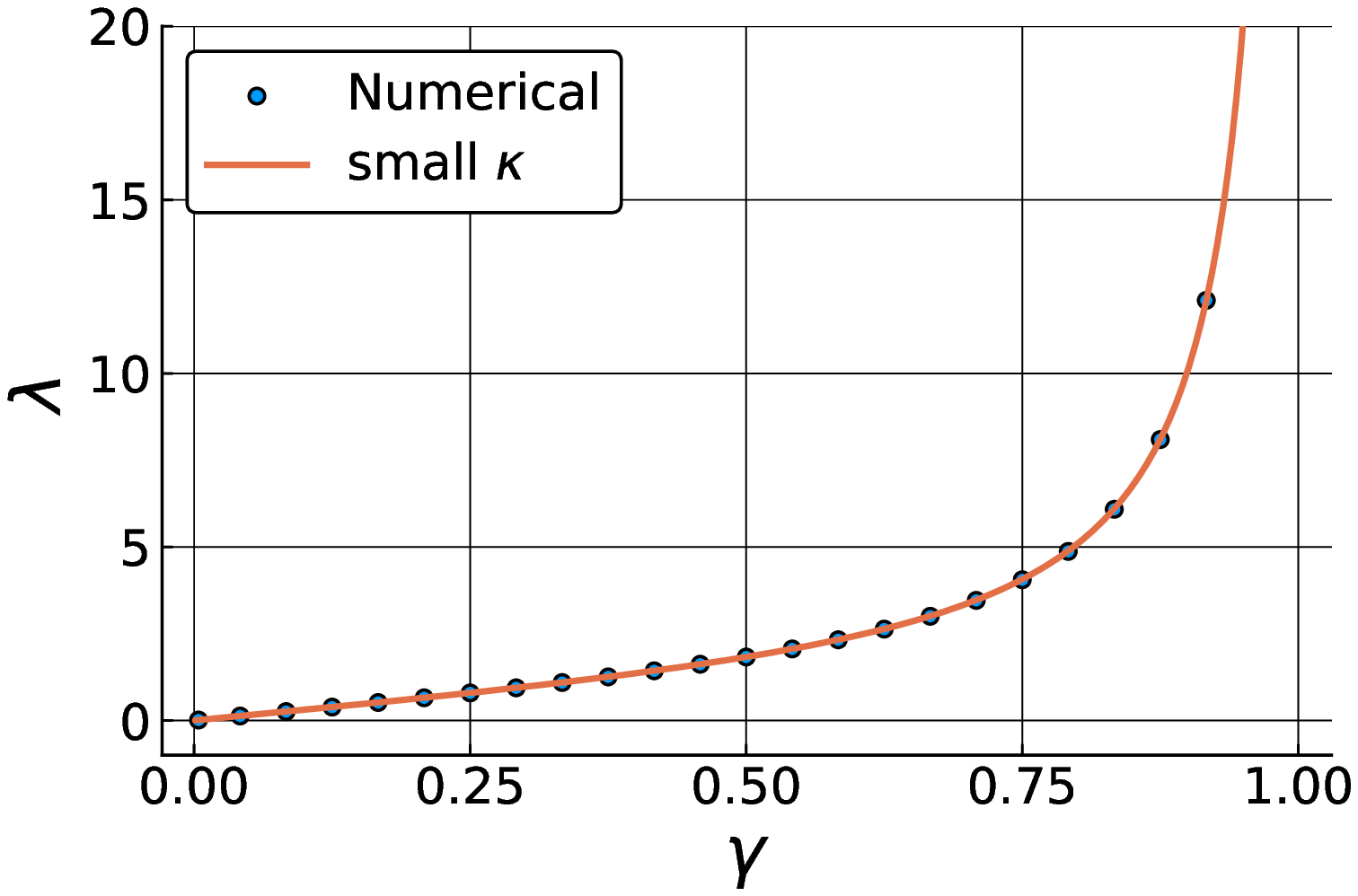}
		&
		\includegraphics[width=\FigureWidthTwoColsNumericalGraph]{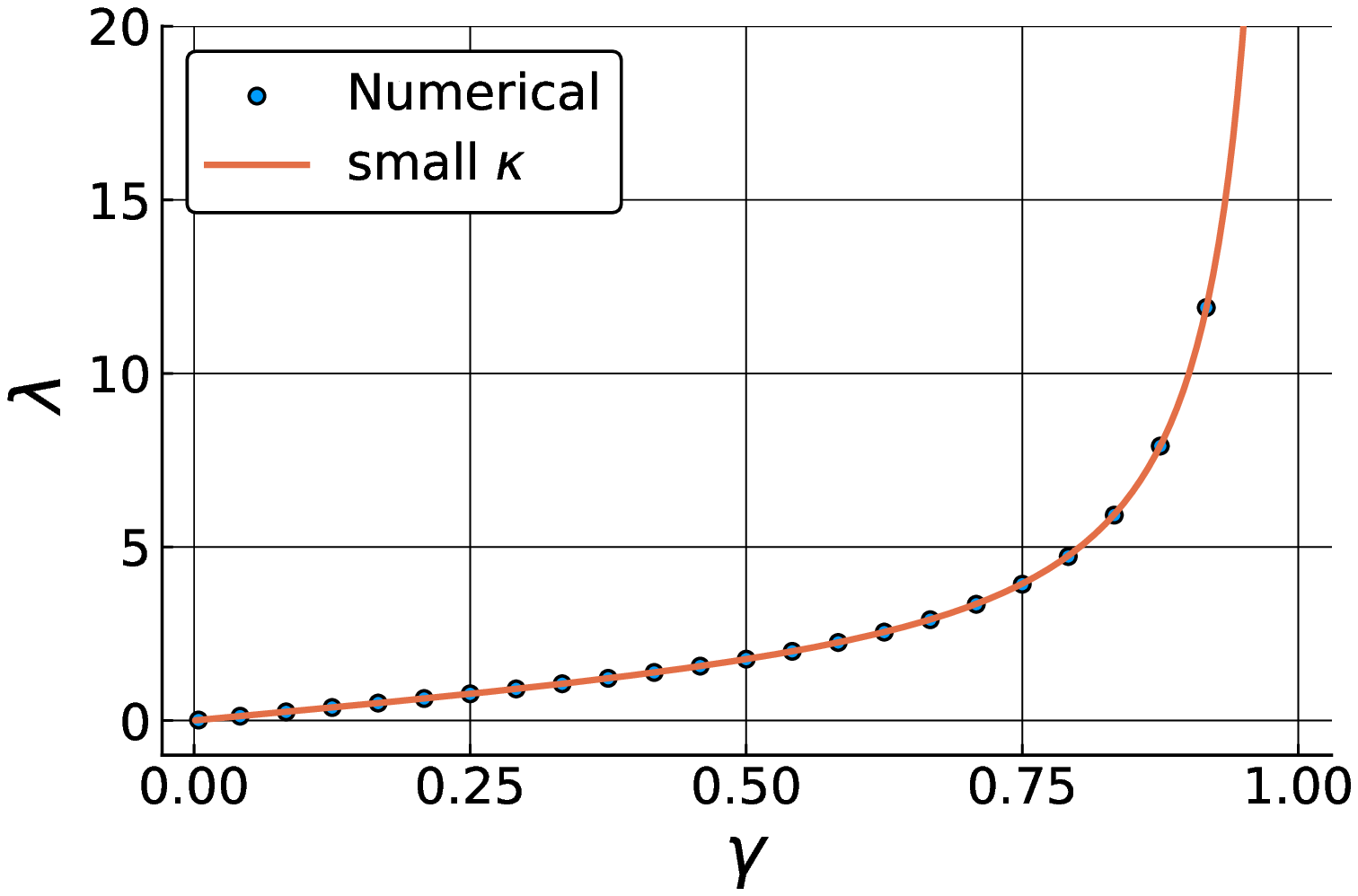}
		\\
		\includegraphics[width=\FigureWidthTwoColsNumericalGraph]{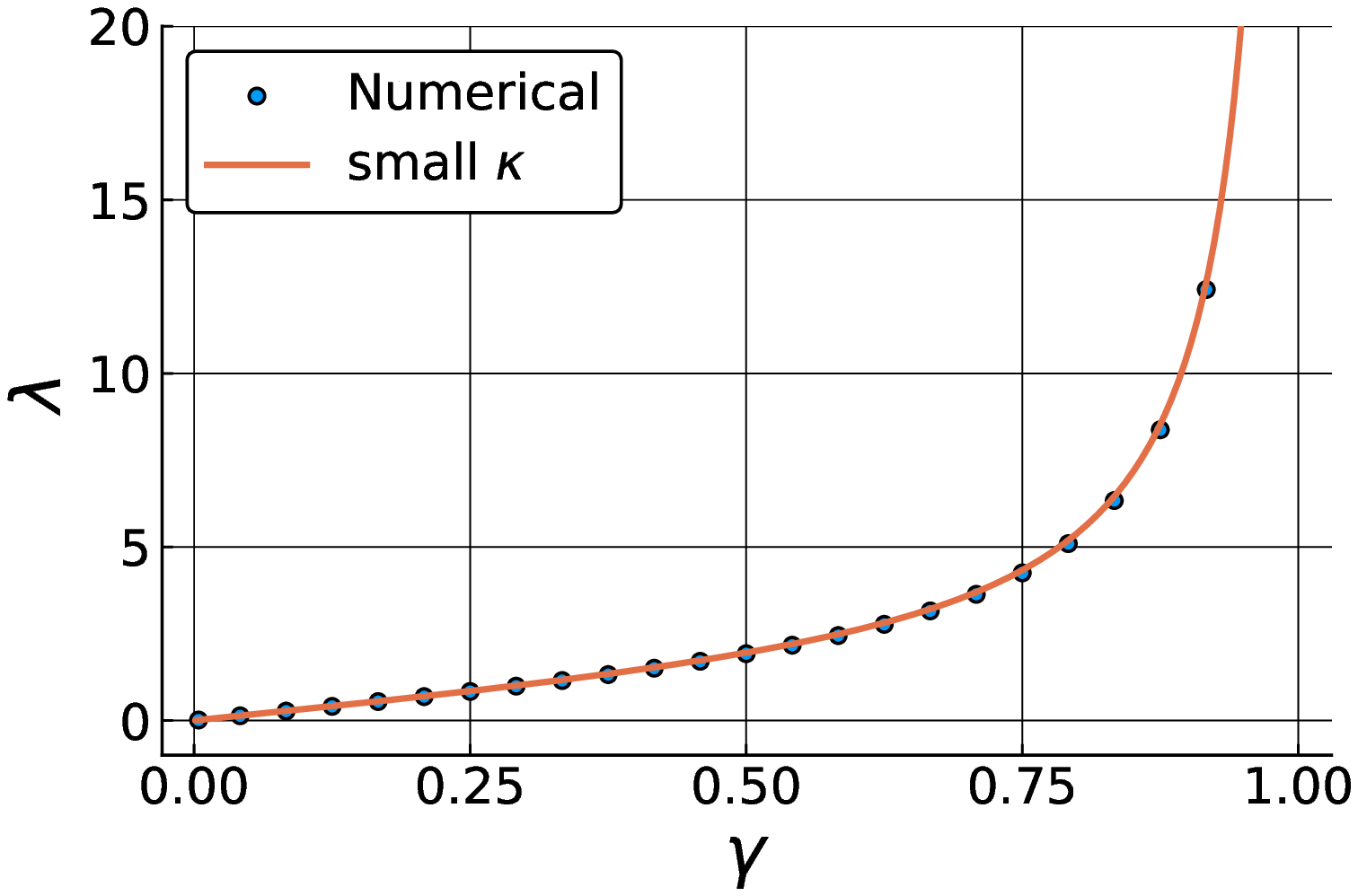}
		&
		\includegraphics[width=\FigureWidthTwoColsNumericalGraph]{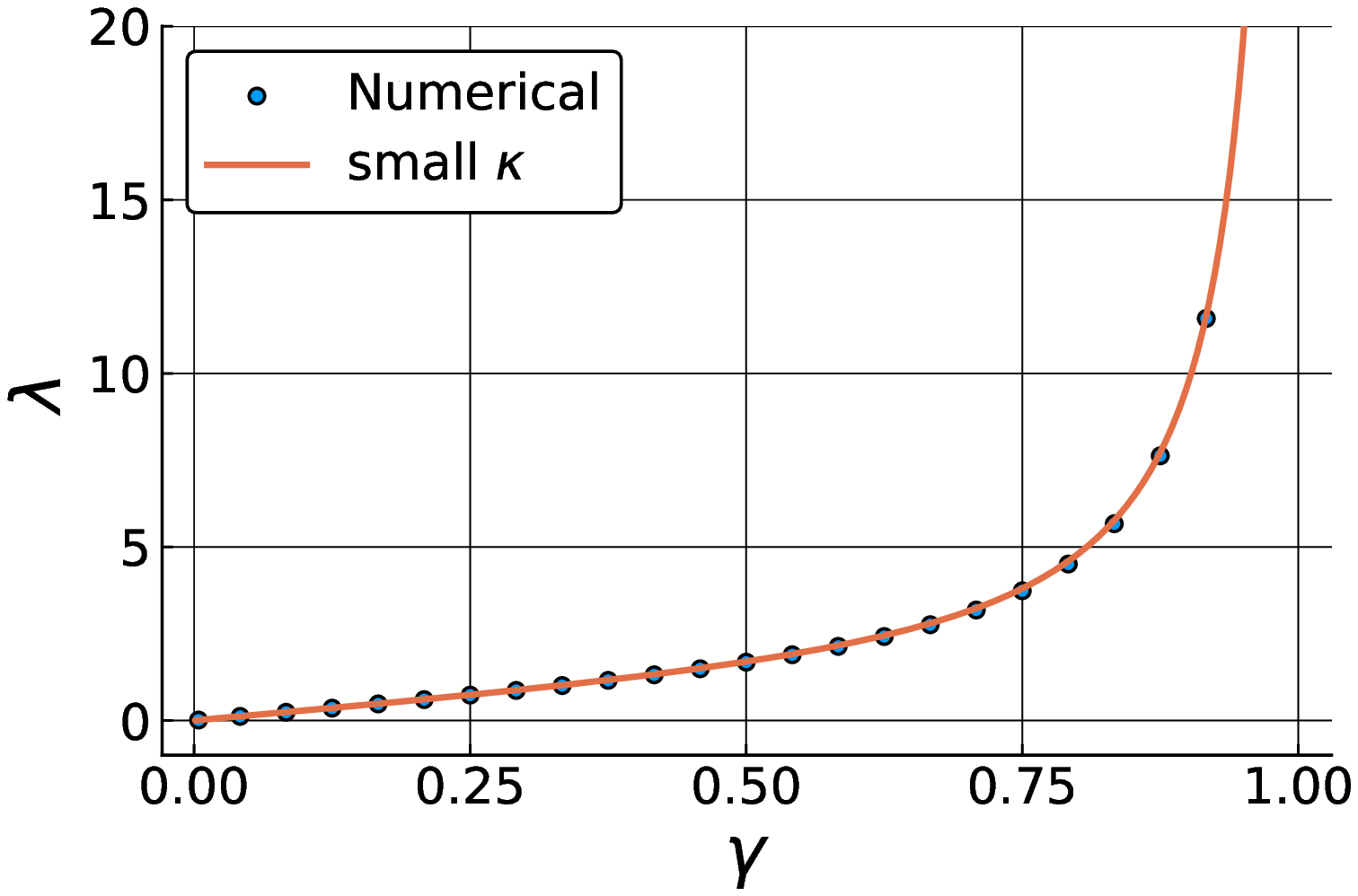}
		\\
		\includegraphics[width=\FigureWidthTwoColsNumericalGraph]{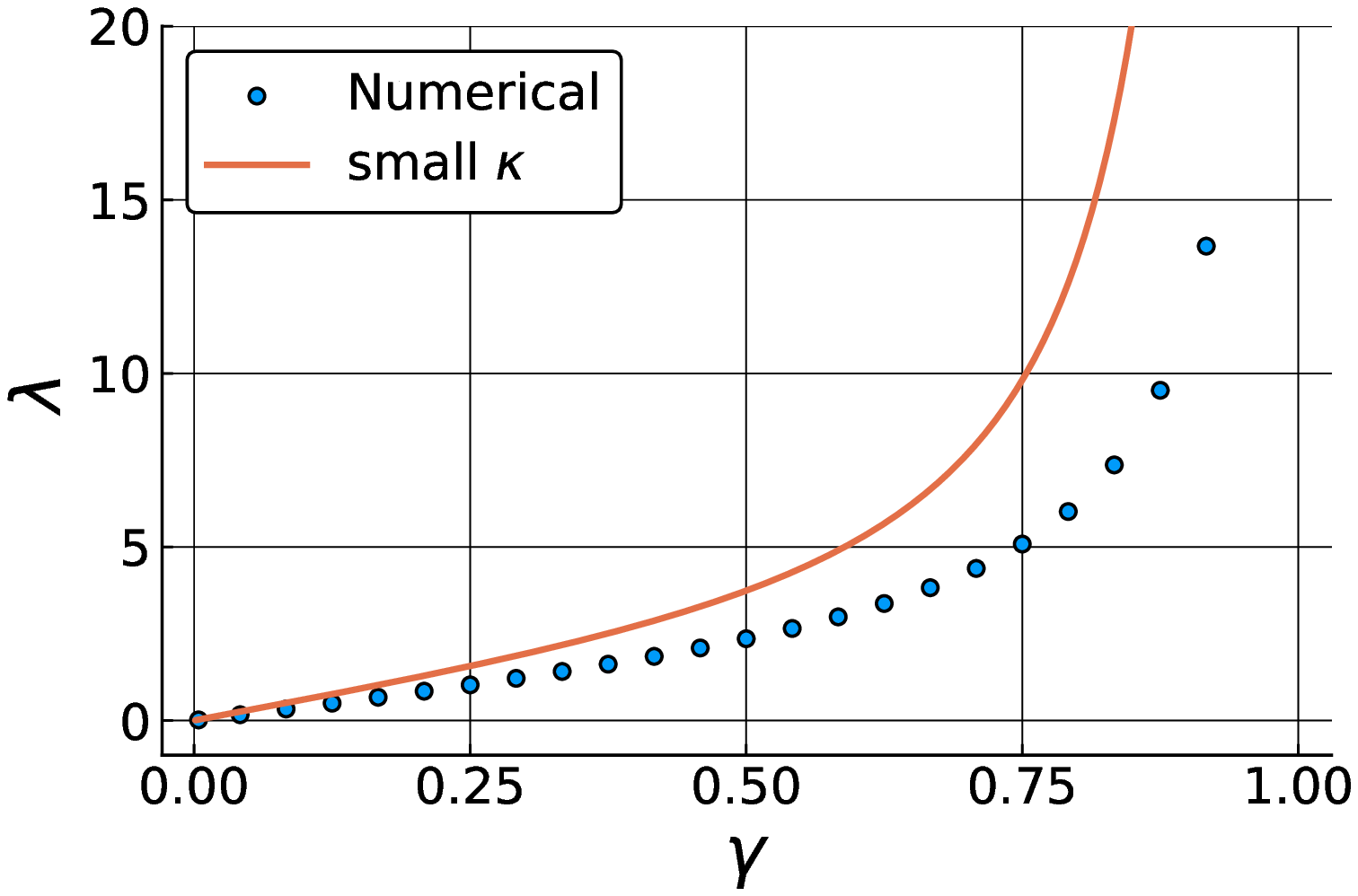}
		&
		\includegraphics[width=\FigureWidthTwoColsNumericalGraph]{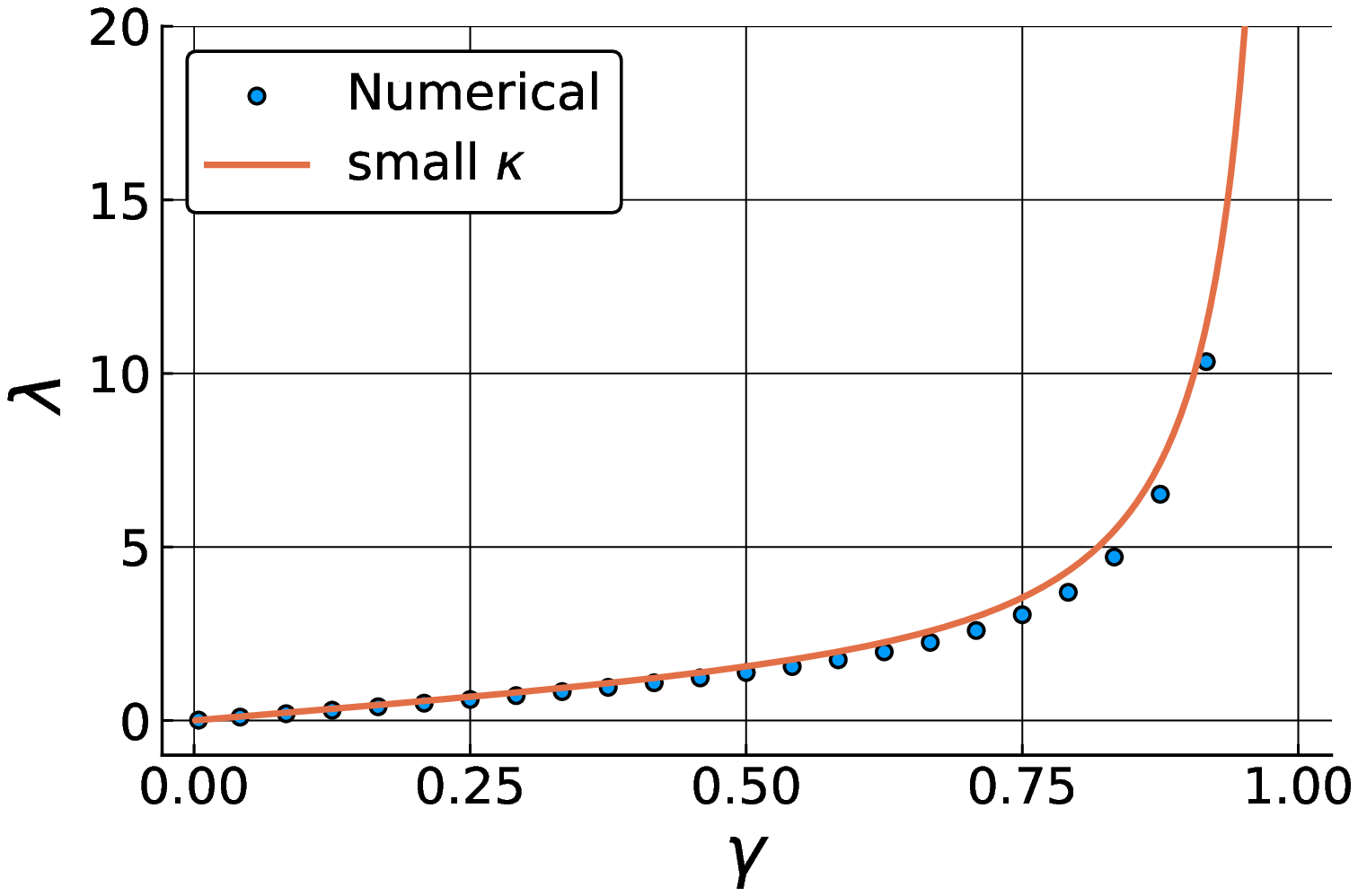}
		\\
		\includegraphics[width=\FigureWidthTwoColsNumericalGraph]{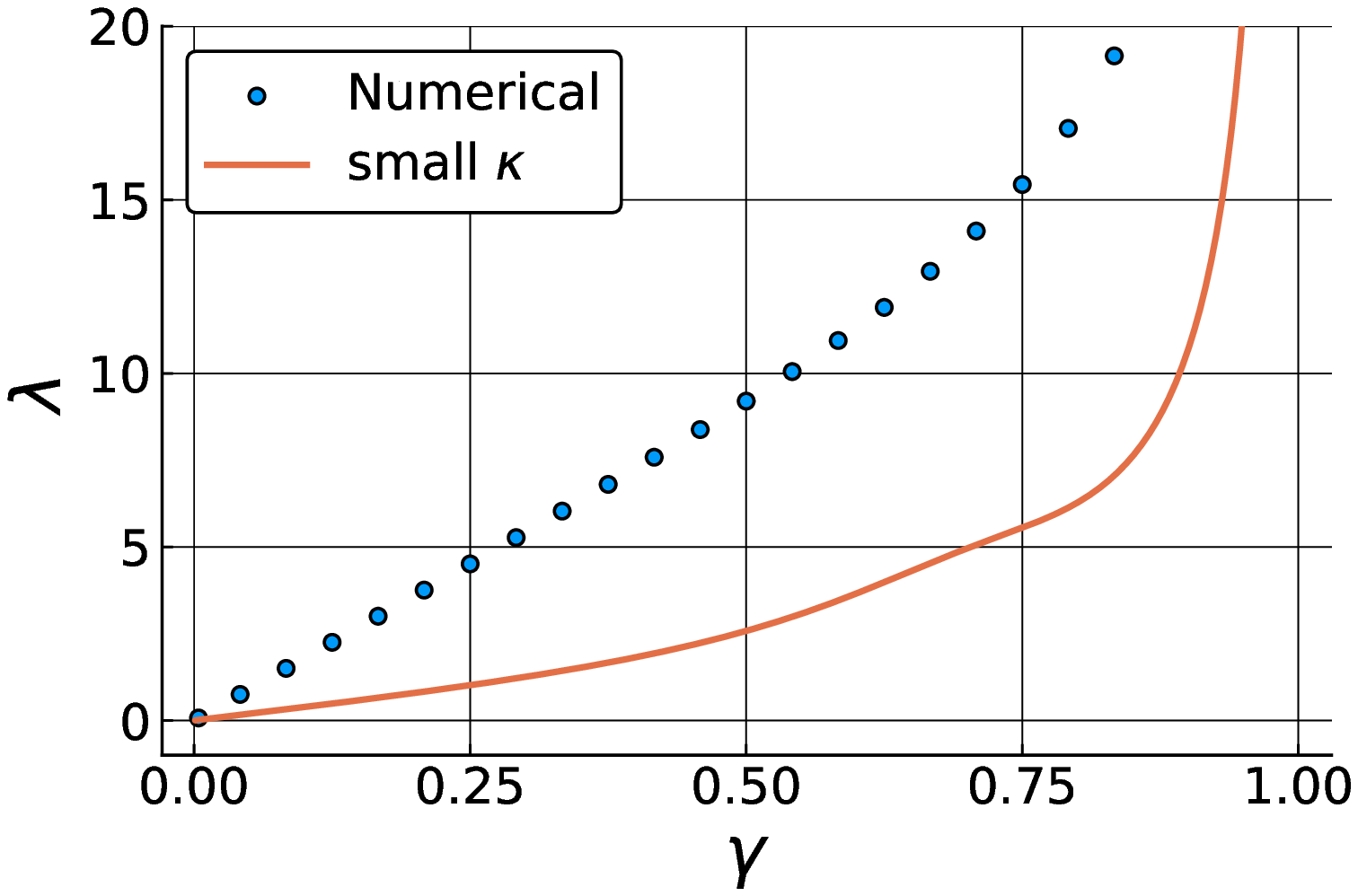}
		&
		\includegraphics[width=\FigureWidthTwoColsNumericalGraph]{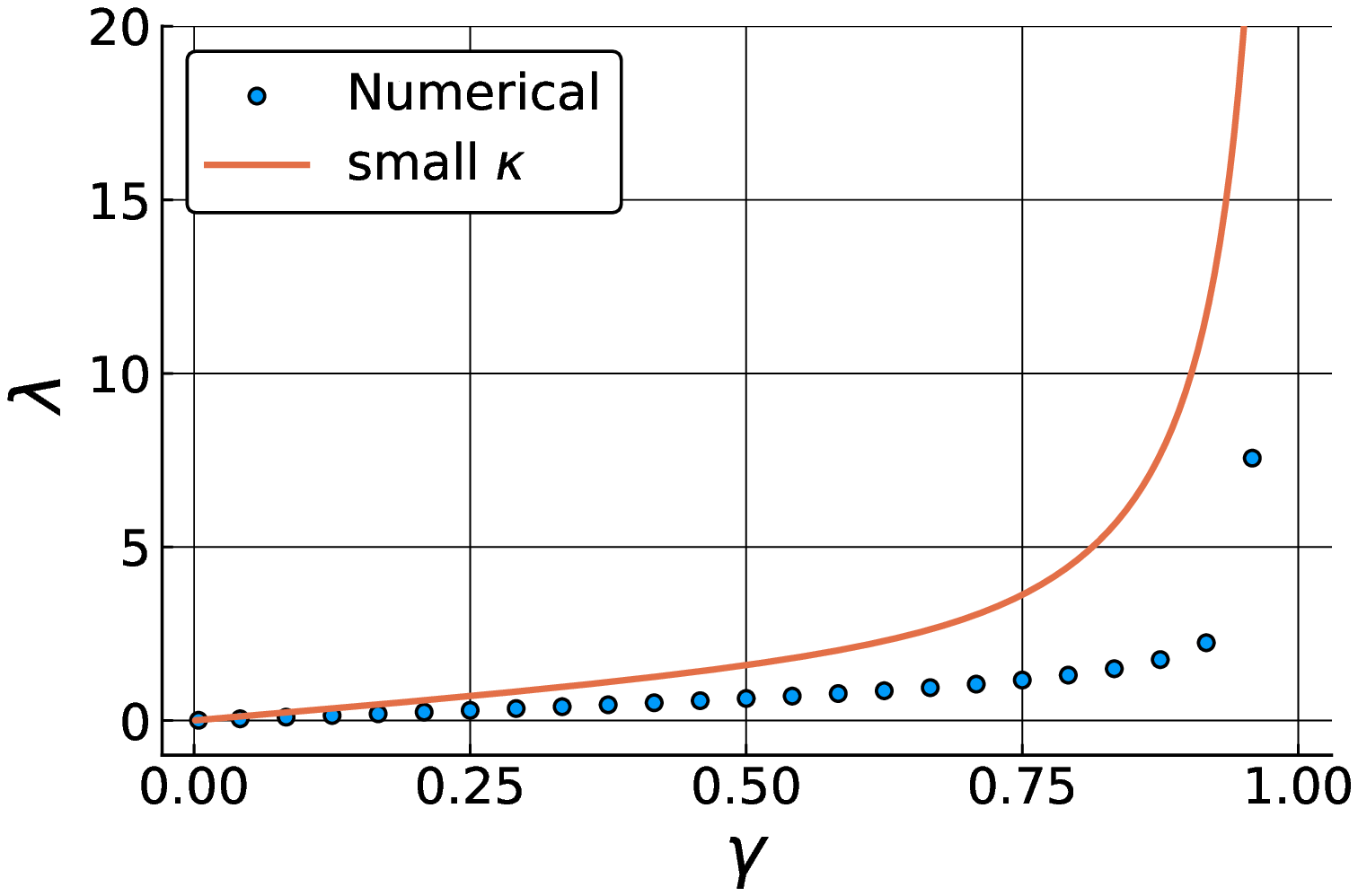}
	\end{tabulary}
	\caption{Comparison of the predicted $\zmult$ with $\stch$ relationship using the small $\unodim$ approximation and the numerical solutions.
		TI chains appear on the right and uniaxial chains on the left;
		$|\uznodim| = 0.0625, 0.25, 1.0, 9.0$ (top row, middle-top, middle-bottom, bottom); $\uxnodim = \uxznodim = 0.0$.
		Note that the small $\unodim$ approximation, like the numerical solutions, also predicts a linear regime followed by a super linear regime}
	\label{fig:so-f-Ez}
\end{figure}

In regards to electro-elasticity, we are also interested in predicting the correct force-length relationship of DE chains.
As a result, it is important to consider the accuracy of the small $\unodim$ approximate solution with respect to the numerical solutions in reproducing $\mults$.
In general, we find a similar relationship between accuracy of the small $\unodim$ approximation in reproducing $\zmult$ obtained from the numerical solution (i.e. the component of $\mults$ in the direction of stretch) as was found in the closed-form approximation's accuracy of reproducing $\A / \kB \T$; that is, the small $\unodim$ approximation agrees well for $|\uznodim|, |\uxnodim|, |\uxznodim| \leq 0.25$ but not for $|\uznodim|, |\uxnodim|, |\uxznodim| \geq 1.0$.
An example of this is shown in Fig. \ref{fig:so-f-Ez}, where $\zmult$ is plotted with respect to $\stch$ for TI (left) and uniaxial (right) chains with $|\uznodim| = 0.625$ (top), $0.25$ (middle), and $1.0$ (bottom).
Lastly, note that, even when it is inaccurate, the small $\unodim$ approximation reproduces the qualitative behavior of the numerical solutions.
Specifically, the small $\unodim$ approximation also predicts a linear regime followed by a superlinear stiffening regime.

We see the clear pattern that the accuracy of our closed-form expressions is good -- using numerical solutions as a benchmark -- for $|\uznodim|, |\uxnodim|, |\uxznodim| \leq 0.25$, but breaks down quantitatively and qualitatively for for $|\uznodim|, |\uxnodim|, |\uxznodim| \geq 1.0$.
The failure of the approximation for $|\uznodim|, |\uxnodim|, |\uxznodim| \geq 1.0$ can be attributed to a number of reasons.
First, the monomer orientation density $\density$ must be nonnegative, but we have no guarantee that the approximate density given by \eqref{eq:density-small-om} is indeed nonnegative when $\uznodim, \uxnodim, \uxznodim \geq 1.0$.
Second, the error of the Taylor series approximation in \eqref{eq:density-small-om} grows with $|\unodim|$.
Third, a Newton-Raphson iteration was used to approximate a solution to the nonlinear system of equations given by \eqref{eq:eap-so-cn}-\eqref{eq:eap-so-crx}, and the quality of this approximation is naturally expected to degrade as the initial guess -- the classical entropic polymer chain with $\unodim = 0$ -- gets further from the regime of interest.
The limitations of the approach taken in this section motivate the asymptotic matching approach taken in  \Fref{sec:patched}.

%%%%%%%%%%%%%%%%%%%%%%%%%%%%%%%%%%%%%%%%%%%%%%%%%%%%%%%%%%%%%%
%%%%%%%%%%%%%%%%%%%%%%%%%%%%%%%%%%%%%%%%%%%%%%%%%%%%%%%%%%%%%%
%%%%%%%%%%%%%%%%%%%%%%%%%%%%%%%%%%%%%%%%%%%%%%%%%%%%%%%%%%%%%%
%%%%%%%%%%%%%%%%%%%%%%%%%%%%%%%%%%%%%%%%%%%%%%%%%%%%%%%%%%%%%%

\section{Closed-form Approximation in the Limit of Small Chain Tension}
\label{sec:small-lambda}

To investigate the limit of $|\mults| \ll 1$, we use a similar process to \Fref{sec:small-omega}; that is, we (1.) use a Taylor expansion in the small quantity ($|\mults|$ in the present case) and (2.) solve the approximate system of equations (for the unknowns $\C$ and $\mults$) that result from enforcing the normalization and kinematic constraints.

We choose the coordinate system such that the polar axis is in the direction of the electric field, that is $\ethree = \ezero / |\ezero|$, and $\rvec$ lies in the $\eone,\ethree$-plane.
This choice allows one to leave the dimensionless energy term, $\uslnodim \cos^2 \polar$, inside the argument of the exponential while integrating.
Next, we write $\exp\left[-\uslnodim \cos^2 \polar + \mults \cdot \nvec\right]$ as $\exp\left[-\uslnodim \cos^2 \polar\right] \times \exp\left[\mults \cdot \nvec\right]$, then perform a Taylor expansion for the second exponential in the product up to first order to obtain the approximate density function:
\begin{equation} \label{eq:density-small-lambda}
	\densitysl\left(\azi, \polar\right) = \C\left(1 + \zmult \cos \polar + \xmult \sin \polar \cos \azi\right) \exp\left[\uslnodim \cos^2 \polar\right]
\end{equation}
Substituting \eqref{eq:density-small-lambda} into the constraint equations, namely \eqref{eq:cn-eap} and \eqref{eq:cr-eap}, and integrating gives\footnote[4]{
    When $\uslnodim$ is negative, we use:
    \begin{equation*}
        \erfw / \sqrt{\uslnodim} 
        = \erf\left(\im \sqrt{|\uslnodim|}\right) / \im \sqrt{|\uslnodim|} 
		= -\erfi\left(\sqrt{|\uslnodim|}\right) / \im^2 \sqrt{|\uslnodim|} 
		= \erfi\left(\sqrt{|\uslnodim|}\right) / \sqrt{|\uslnodim|}
    \end{equation*}
    where $\erfi$ is the imaginary error function.
    Hence, all the quantities on the left side of \eqref{eq:eap-sl-cn}--\eqref{eq:eap-sl-crx} are real.
}
\begin{equation}
    \label{eq:eap-sl-cn}
    \N = 2 \pi^{3/2} \C \erfw / \sqrt{\uslnodim} 
\end{equation}
\begin{equation}
    \label{eq:eap-sl-crz}
    \frac{\rz}{\mlen} = \pi \C \zmult \left(\frac{\sqrt{\pi} \erfw}{\uslnodim^{3/2}} - \frac{2 e^{-\uslnodim}}{\uslnodim}\right) 
\end{equation}
\begin{equation}
    \label{eq:eap-sl-crx}
    \frac{\rx}{\mlen} = \pi \C \xmult \left(\frac{\sqrt{\pi} \left(2\uslnodim-1\right) \erfw}{2 \uslnodim^{3/2}} + \frac{ e^{-\uslnodim}}{\uslnodim}\right)
\end{equation}
Notice that \eqref{eq:eap-sl-cn}--\eqref{eq:eap-sl-crx} are linear in the unknowns and can be readily solved to obtain:
\begin{align} \label{eq:eap-sl-c}
    \C &= \Csl \\ \label{eq:eap-sl-z}
    \zmult &= \zmultsl \\ \label{eq:eap-sl-x}
    \xmult &= \xmultsl 
\end{align}
where $\stchz = \rz / \N \mlen$ and $\stchx = \rx / \N \mlen$.

\subsection{Free Energy}

Having obtained an approximate solution for the unknowns $\C$ and $\mults$, we turn our attention to the free energy.
Substituting the approximate density function \eqref{eq:density-small-lambda} into \eqref{eq:A-approx}, and using the Taylor expansion
\begin{equation*}
    \log\left(1 + \zmult \cos \polar + \xmult \sin \polar \cos \azi\right) \approx \zmult \cos \polar + \xmult \sin \polar \cos \azi
\end{equation*}
and integrating results in
\begin{equation} \label{eq:A-approx-sl}
\begin{split}
    A = \kB \T \left[\frac{\pi^{3/2} \C \erf\left(\sqrt{\uslnodim}\right)}{2 \uslnodim^{3/2}} \left(\xmult^2\left(2\uslnodim - 1\right) + 2\left(\zmult^2 - 2\uOnodim \uslnodim \right) + 4\uslnodim \log \C\right) \right. +\\
    \left. \frac{\pi \C \exp\left(-\uslnodim\right) \left(\xmult^2 - 2 \zmult^2\right)}{\uslnodim} - \N \log \N\right]
\end{split}
\end{equation}

\begin{figure}[htb!]
	\centering
	\begin{tabulary}{\linewidth}{c c}
		\includegraphics[width=\FigureWidthTwoColsNumericalGraph]{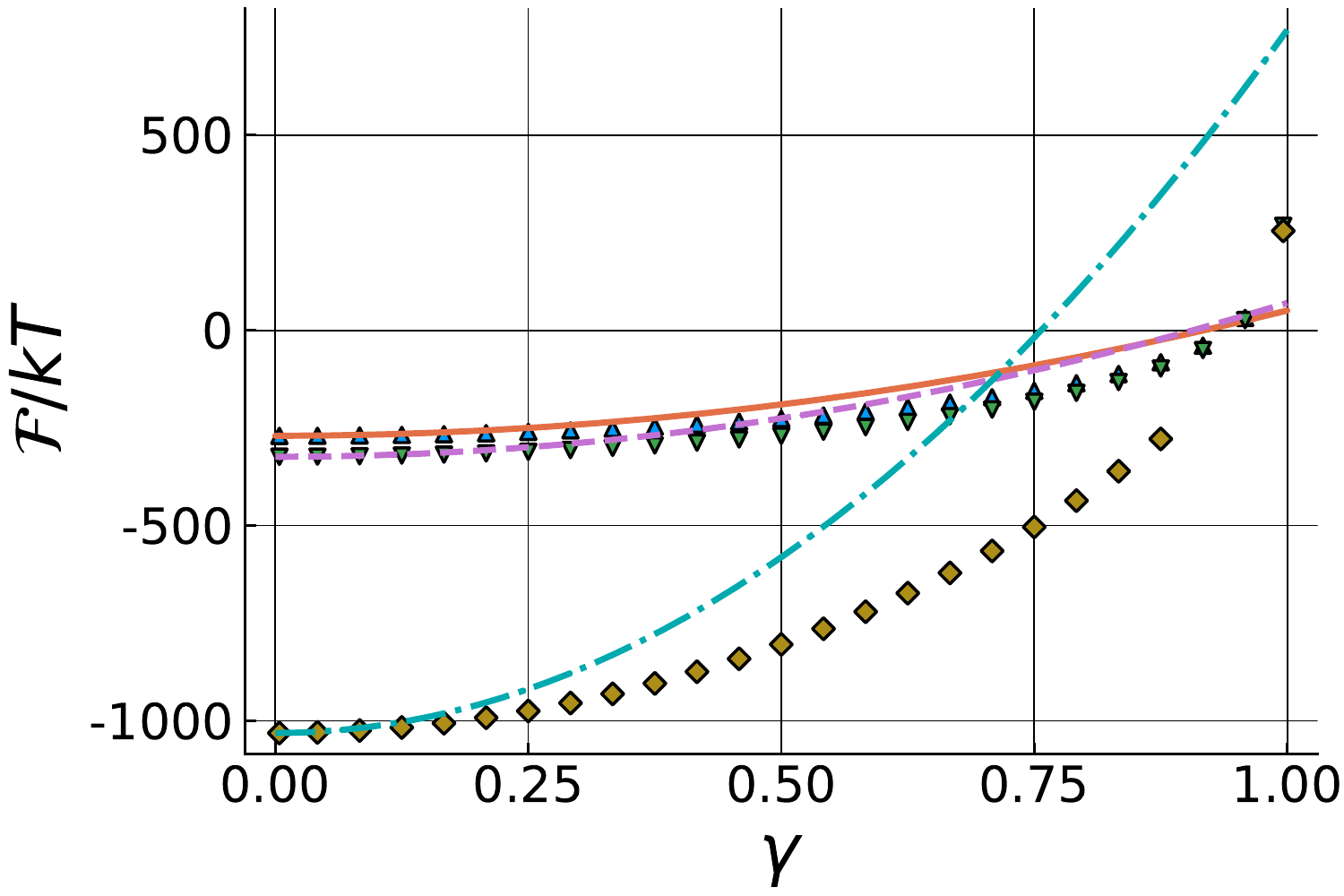}
		&
		\includegraphics[width=\FigureWidthTwoColsNumericalGraph]{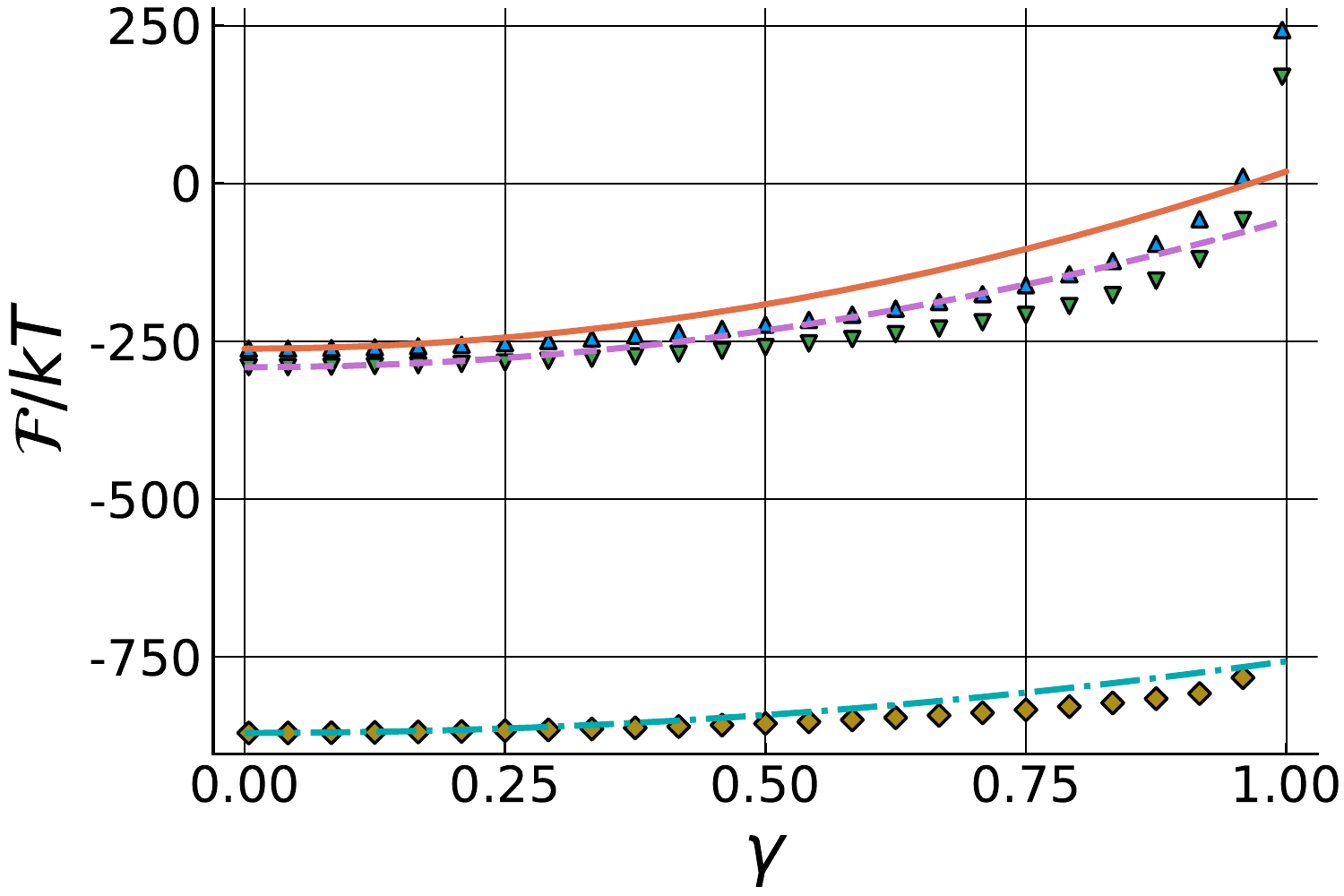}
		\\
		\includegraphics[width=\FigureWidthTwoColsNumericalGraph]{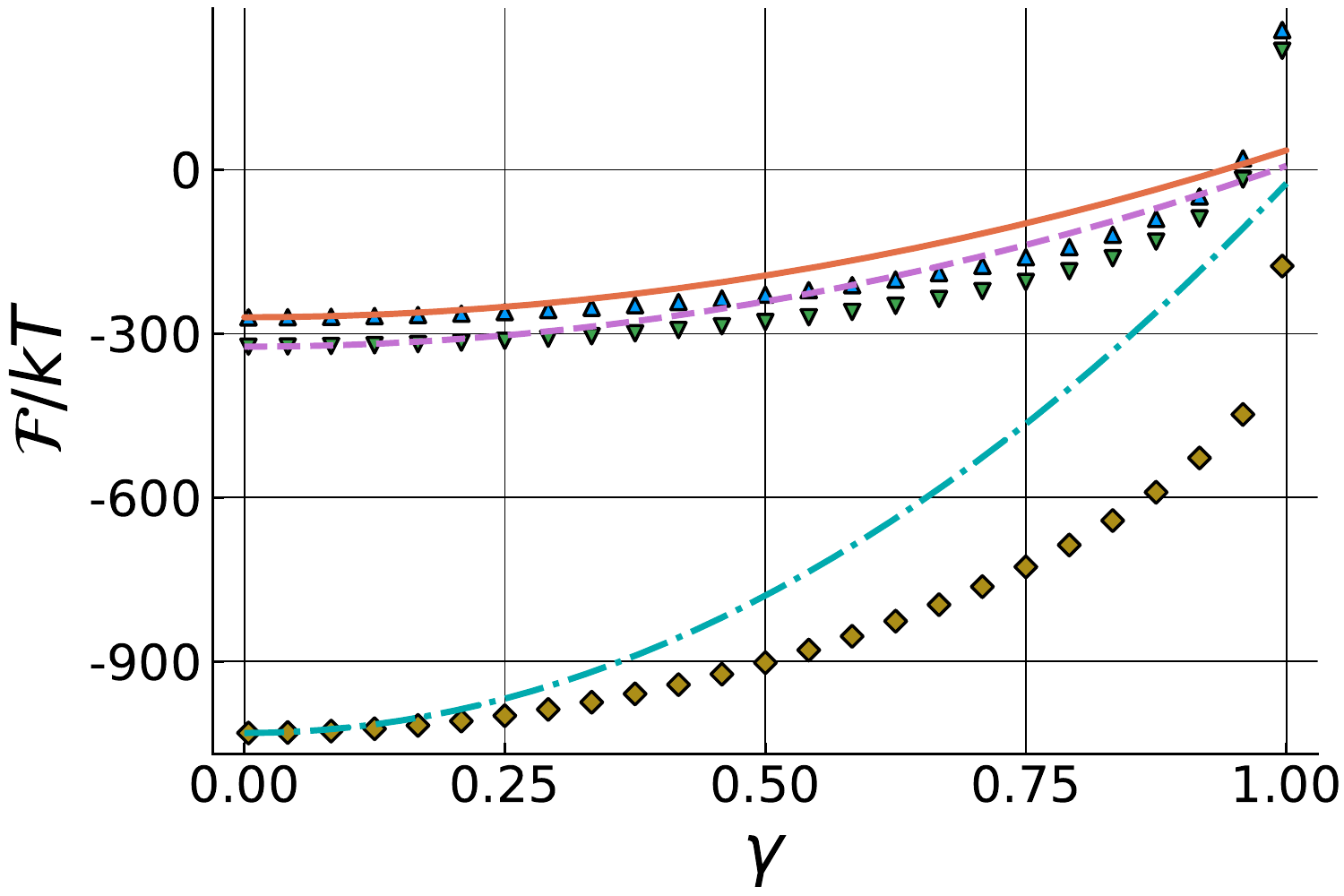}
		&
		\includegraphics[width=\FigureWidthTwoColsNumericalGraph]{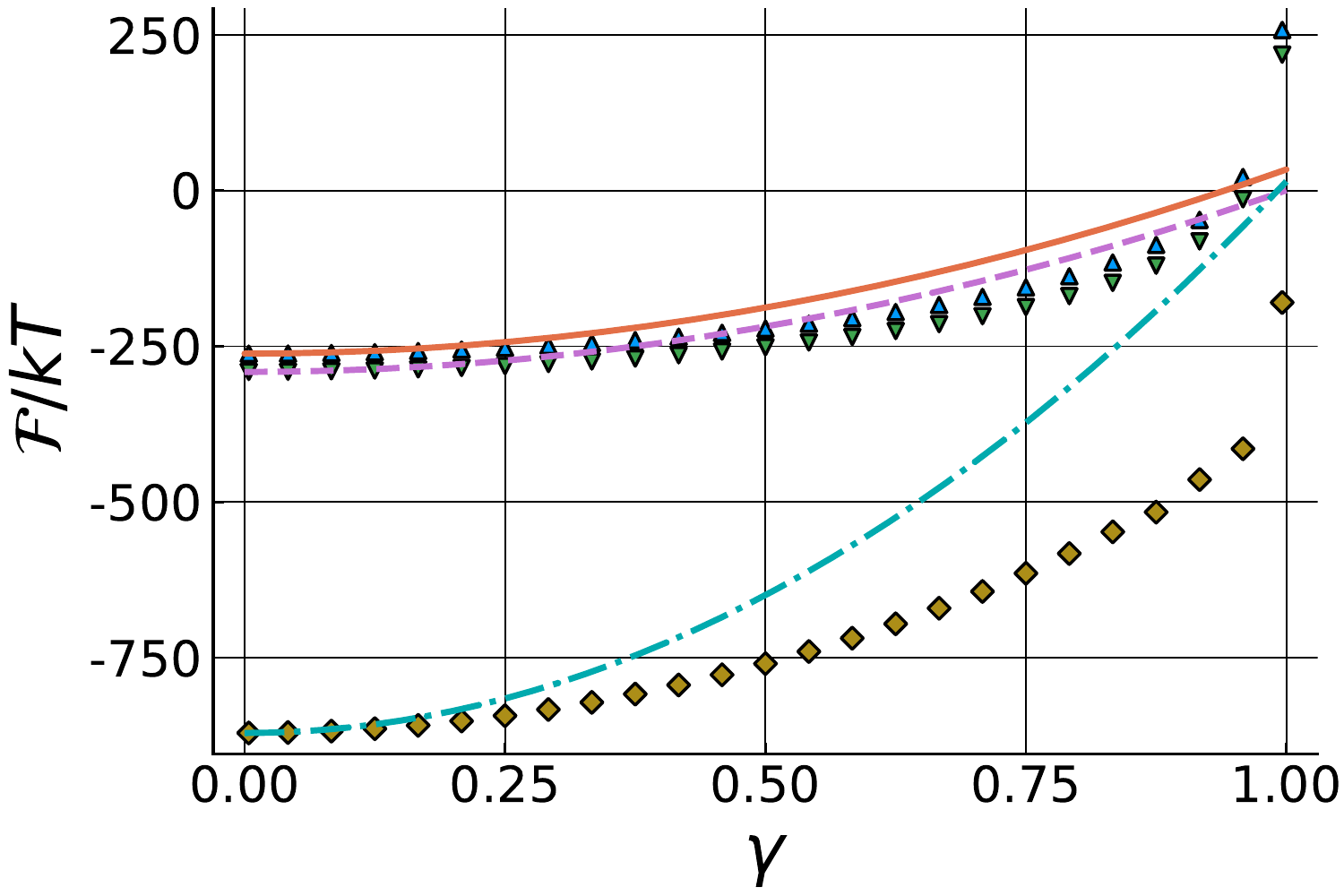}
		\\
		\includegraphics[width=\FigureWidthTwoColsNumericalGraph]{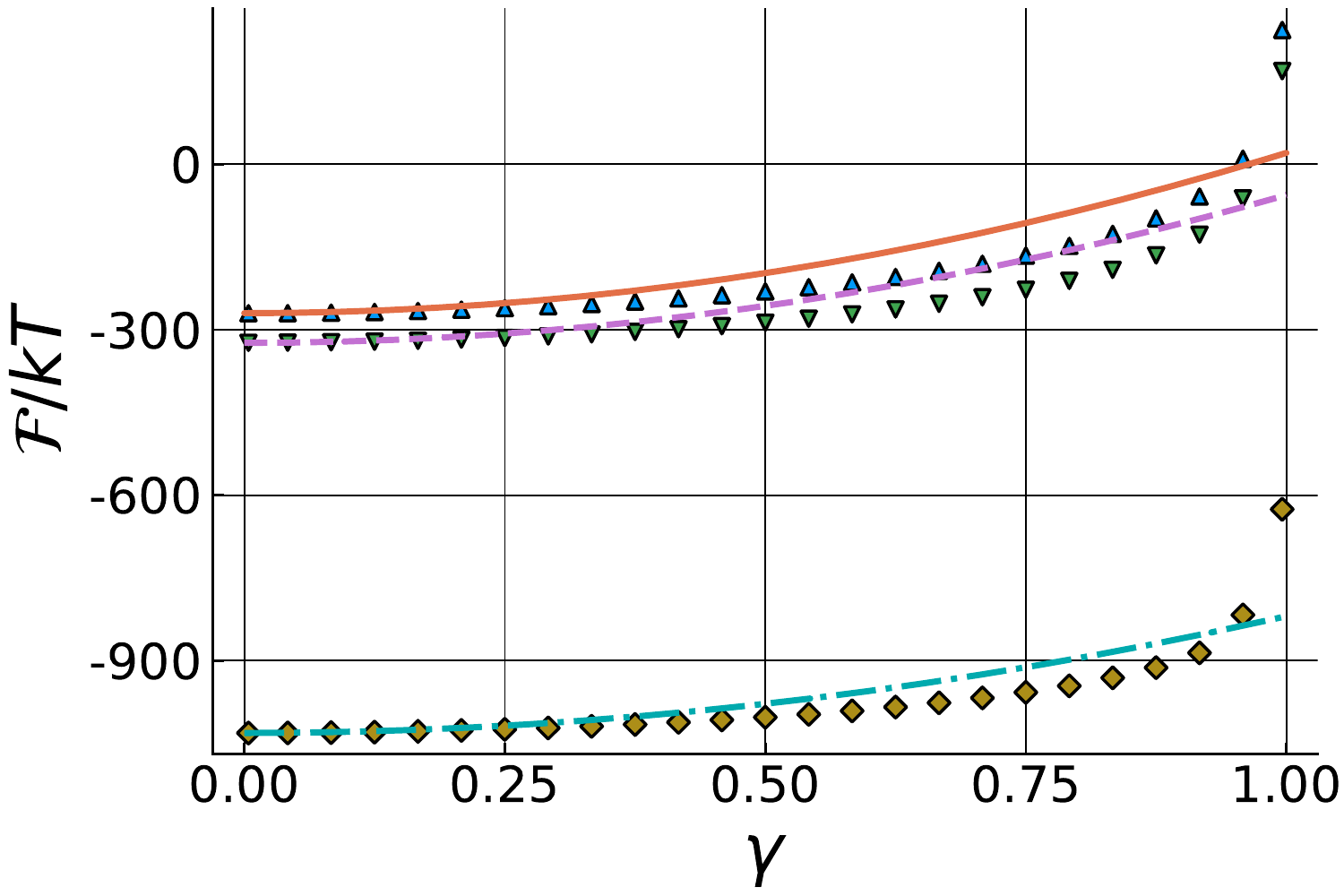}
		&
		\includegraphics[width=\FigureWidthTwoColsNumericalGraph]{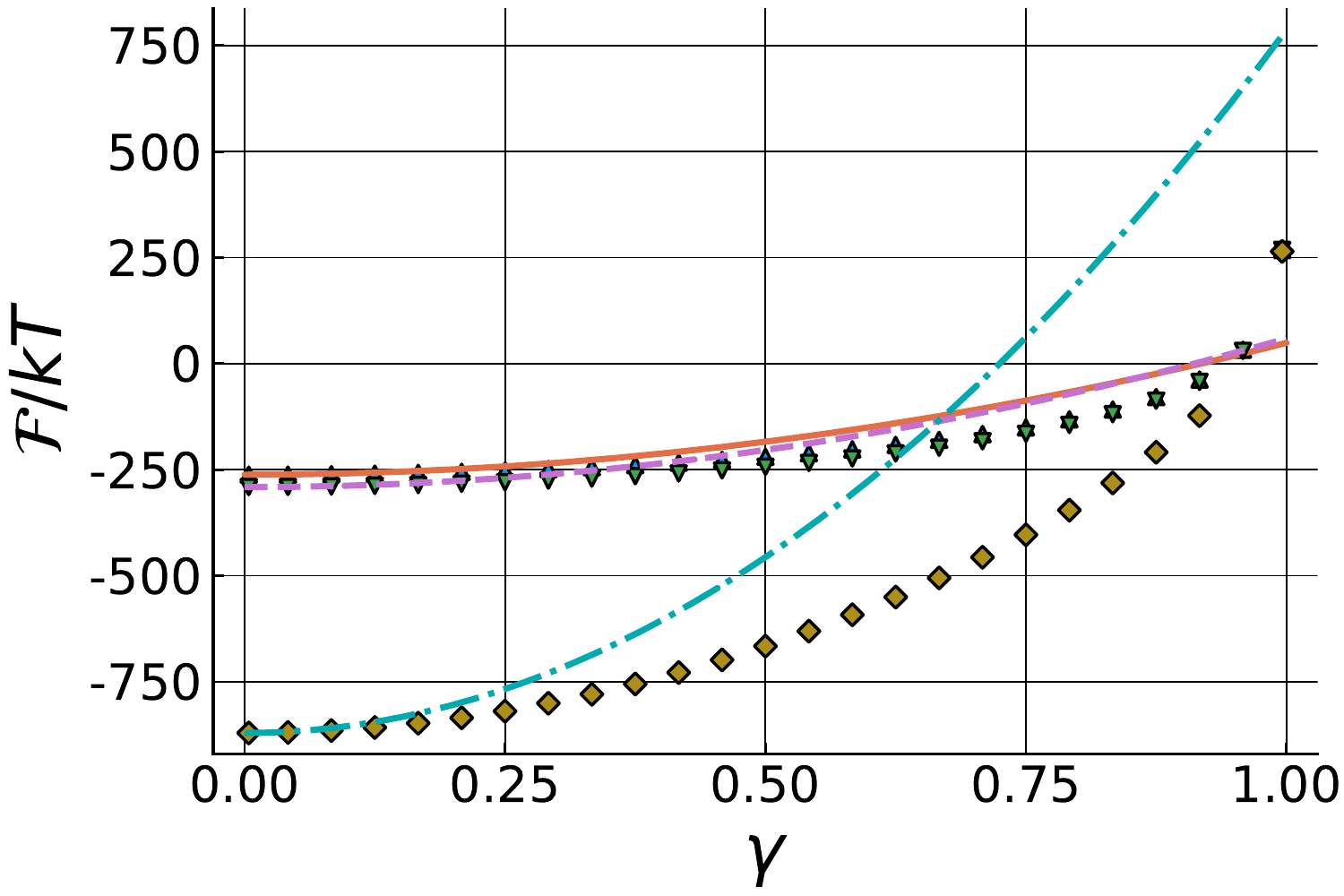}
		\\
		\multicolumn{2}{c}{
			\includegraphics[scale=0.3]{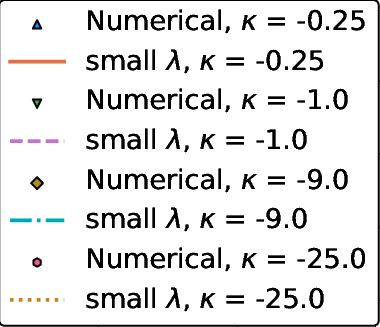}
		}
	\end{tabulary}
	\caption{Comparison of the predicted $\A / \kB \T$ with $\stch$ relationship using the small $|\mults|$ approximation and the numerical solutions.
		TI chains appear on the right and uniaxial chains on the left;
		$\etorangle = 0, \frac{\pi}{4}, \frac{\pi}{2}$ (top row, middle, bottom).}
	\label{fig:sl-A}
\end{figure}

Again, as in \Fref{sec:small-omega}, having obtained an approximate solution for the unknowns $\C$ and $\mults$ and an approximate expression for the free energy, we test its accuracy by comparing it with numerical solutions obtained in \Fref{sec:numerical}.
Fig. \ref{fig:sl-A} shows the $\A / \kB \T$--$\stch$ relationship as approximated by both the small $|\mults|$ approximate solution and the numerical solutions.
The plots are shown for TI (right) and uniaxial (left) DE chains, oriented at $\etorangle = 0$ (top), $\etorangle = \frac{\pi}{4}$ (middle) and $\etorangle = \frac{\pi}{2}$ (bottom).
In all cases, the zero and small stretch ($\stch \leq 0.1$) agree nearly exactly.
In addition, in contrast to the small $|\unodim|$ closed-form approximation (\Fref{sec:small-omega}), all of the curves are convex -- as is desired, for reasons discussed in more detail in \Fref{sec:numerical} and  \Fref{sec:small-omega}.

However, none of the small $|\mults|$ curves have finite extensibility, i.e., $\A / \kB \T$ does not approach infinity as $\stch \rightarrow 1$.
Lastly, note that the accuracy of the approximation in the regime of moderate to large stretch ($\stch > 0.25$) depends on $|\uslnodim|$, the type of monomers (TI or uniaxial) the chain is composed of, and the orientation of the chain end-to-end vector with respect to the electric field (i.e. $\etorangle$).
More specifically, the TI chains with $\etorangle = 0$ and $\etorangle = \frac{\pi}{4}$ and the uniaxial chains with $\etorangle = \frac{\pi}{4}$ and $\etorangle = \frac{\pi}{2}$ predict $\A / \kB \T$ to be larger than obtained numerically, consequently predicting overly stiff chains, for $\stch$ in about the interval $(0.25, 0.99)$.

Recall that these chain end-to-end vector orientations are such that monomers are kinematically constrained to high-energy states as $\stch$ increases.

\subsection{Force-Stretch Relation} \label{sec:sl-fs}

\begin{figure}[htb!]
	\centering
	\begin{tabulary}{\linewidth}{c c}
		\includegraphics[width=\FigureWidthTwoColsNumericalGraph]{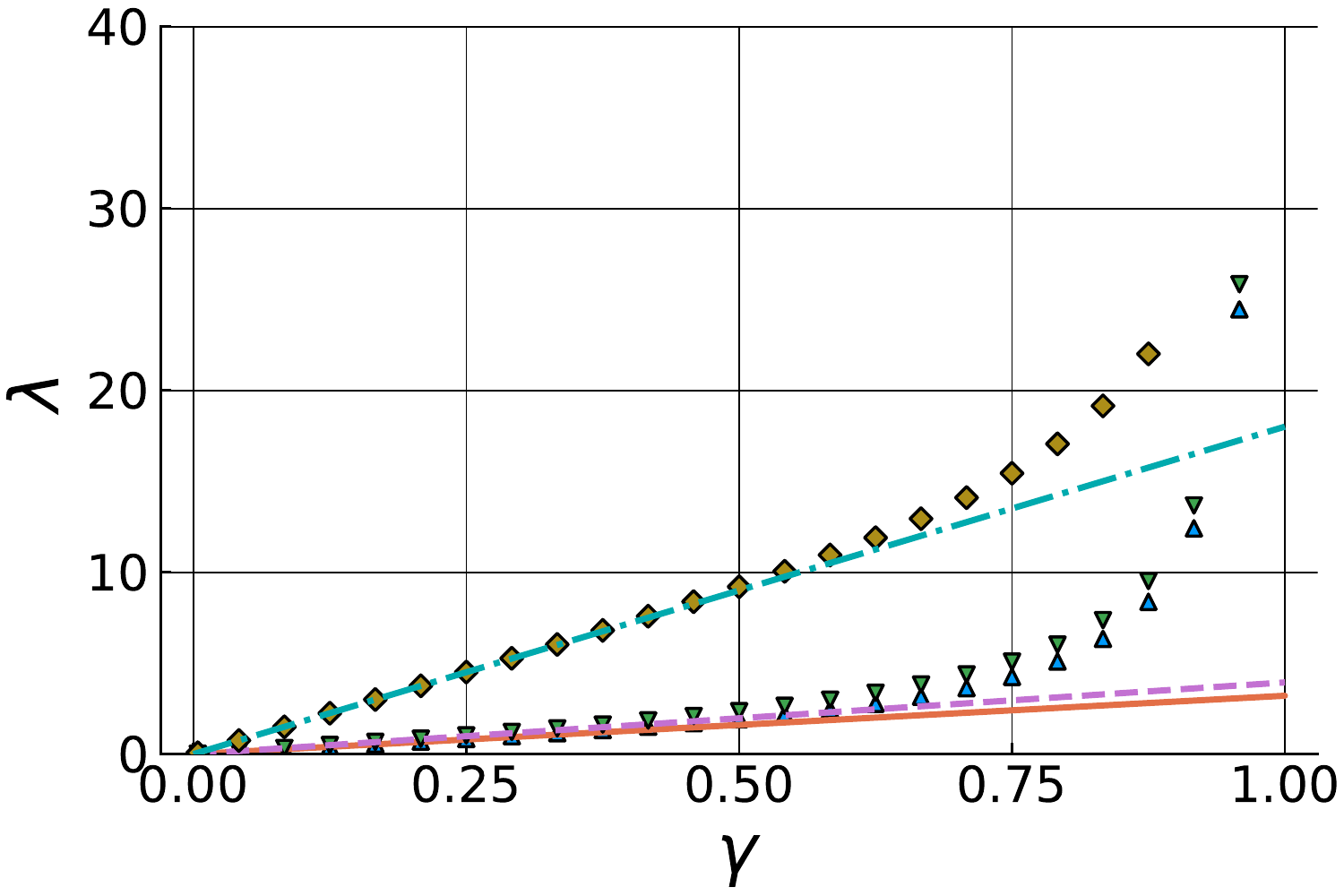}
		&
		\includegraphics[width=\FigureWidthTwoColsNumericalGraph]{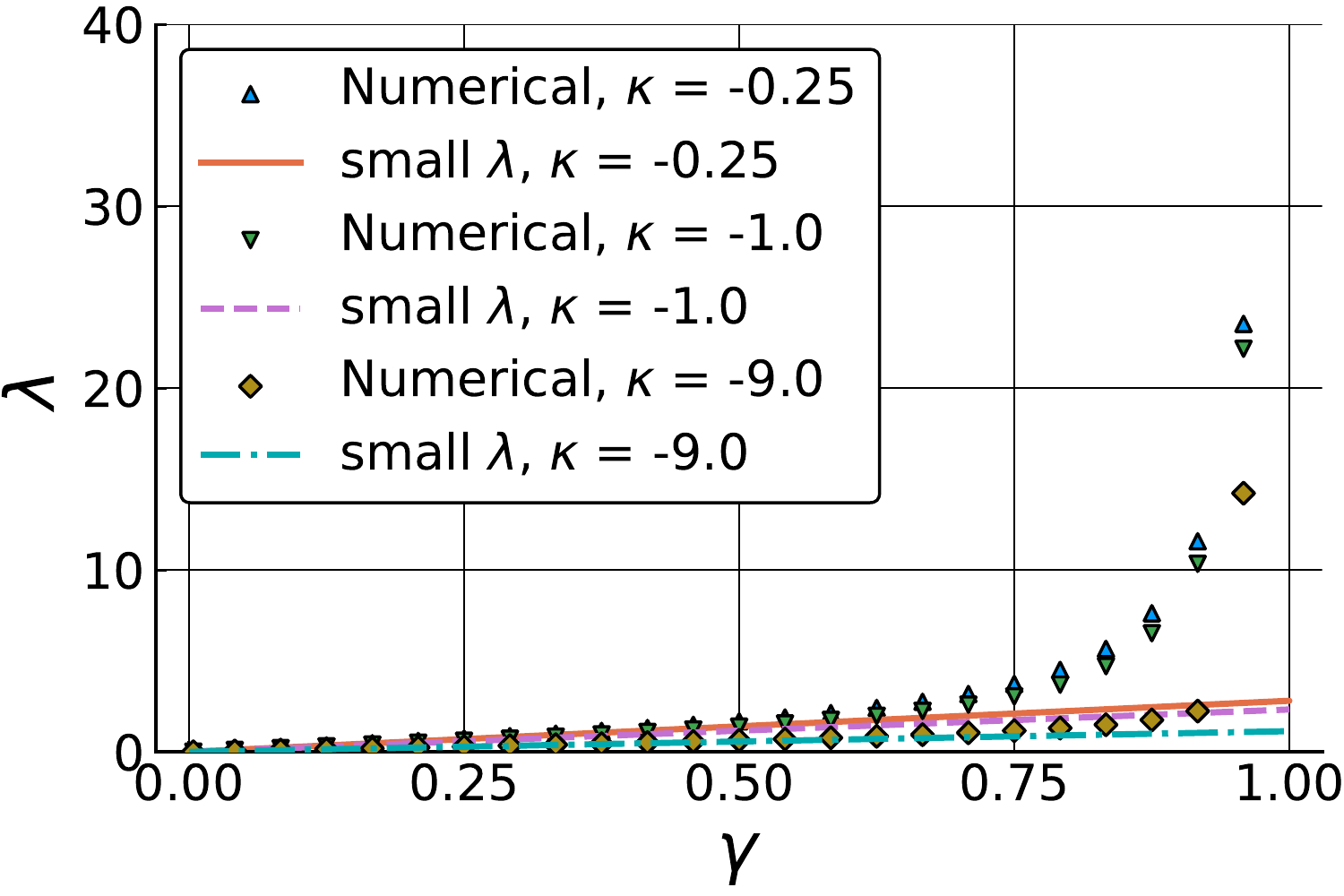}
		\\
		\includegraphics[width=\FigureWidthTwoColsNumericalGraph]{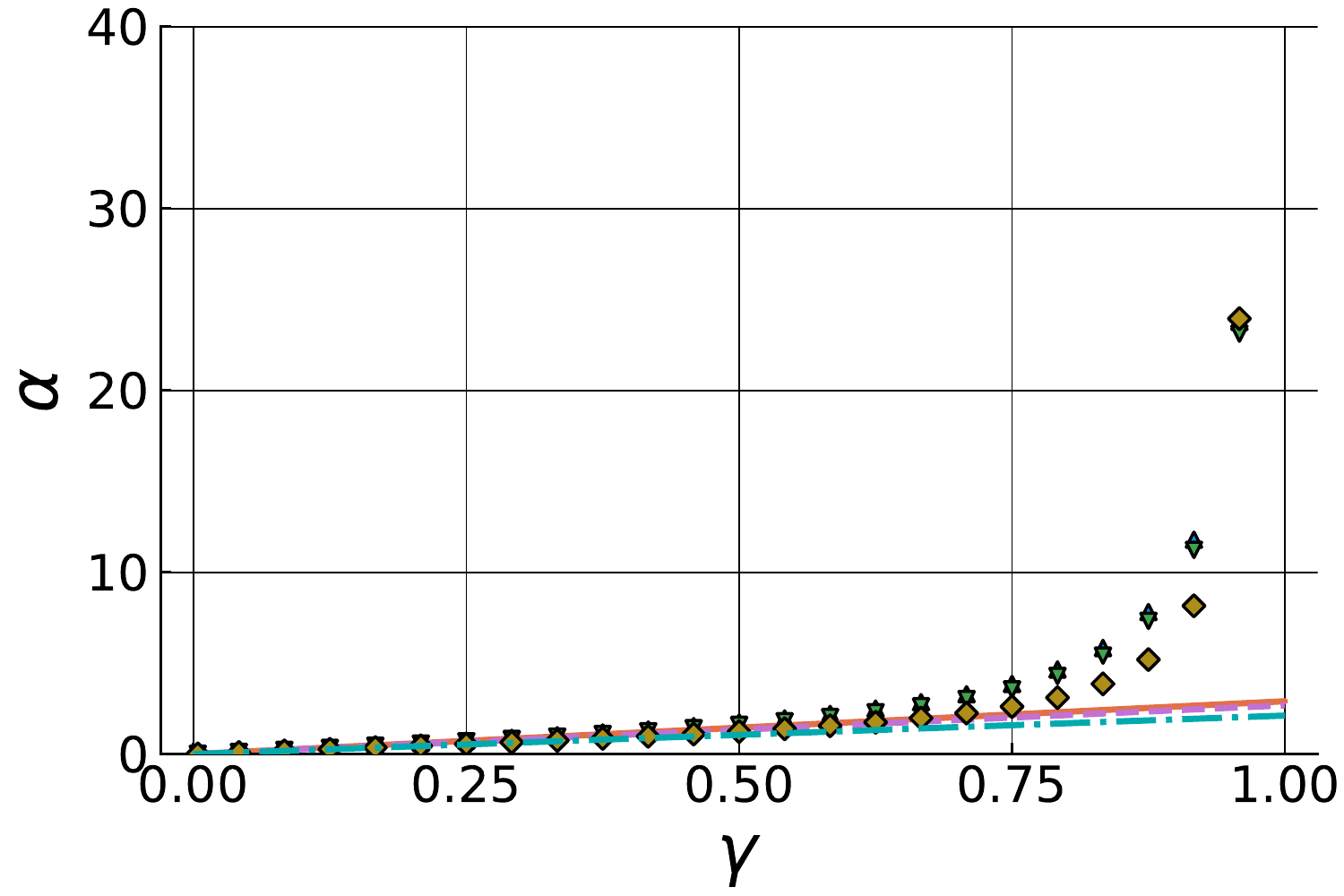}
		&
		\includegraphics[width=\FigureWidthTwoColsNumericalGraph]{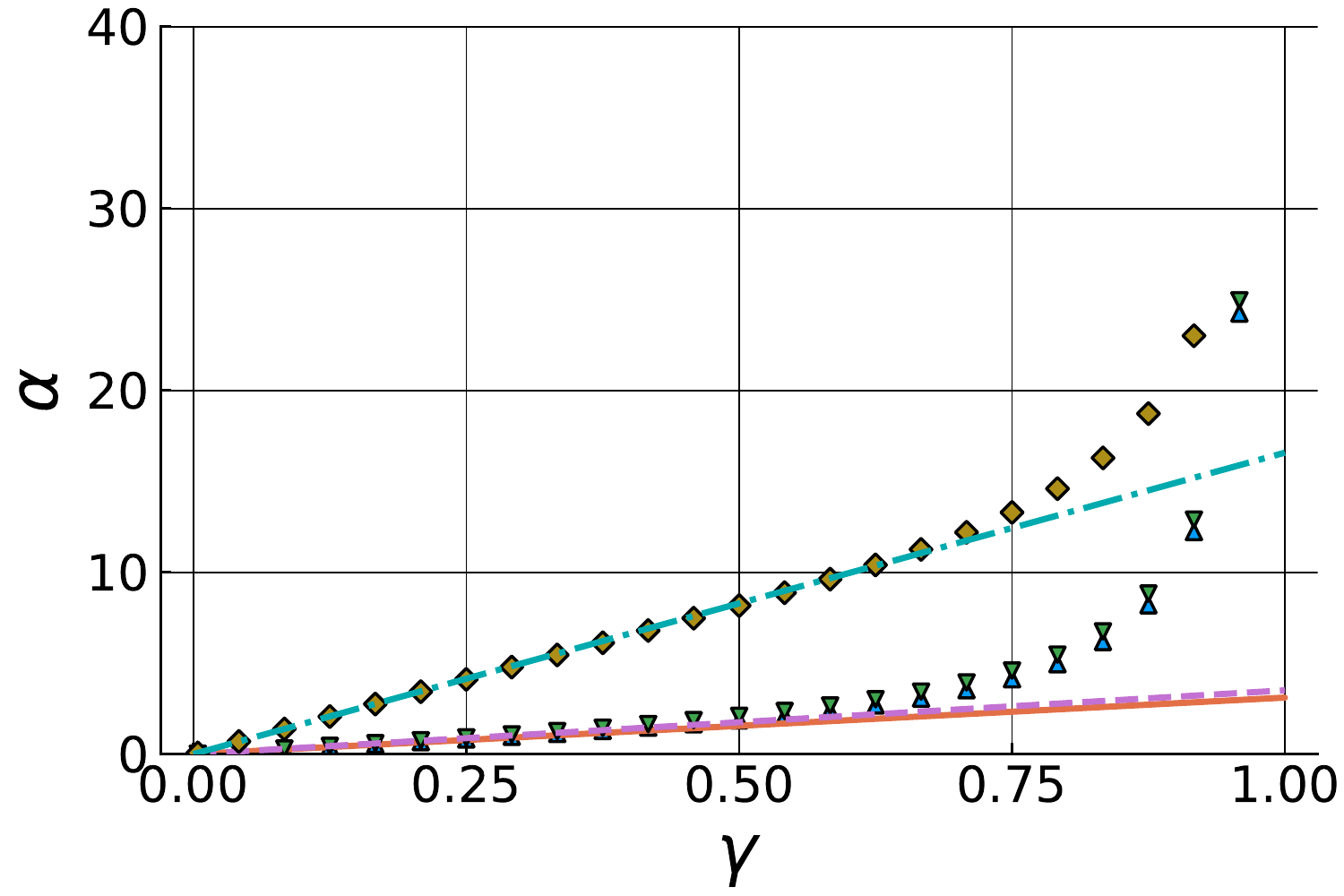}
	\end{tabulary}
	\caption{Comparison of the predicted component of $\mults$ in the direction of stretch with $\stch$ relationship using the small $|\mults|$ approximation and the numerical solutions.
		TI chains appear on the right and uniaxial chains on the left;
		$\etorangle = 0, \frac{\pi}{4}, \frac{\pi}{2}$ (top row, middle, bottom).
		The small $|\mults|$ approximation matches the linear regime almost exactly, but does not capture the super linear regime.}
	\label{fig:sl-f}
\end{figure}

In addition to investigating the accuracy of the $\A / \kB \T$ approximation, we also consider the $|\mults|$--$\stch$ relationship.
Fig. \ref{fig:sl-f} shows the component of $\mults$ in the direction of stretch for TI (left) and uniaxial (right) chains oriented at $\etorangle = 0$ (top) and $\etorangle = \frac{\pi}{2}$.
Clearly, in regards to the numerical solutions, there are two regimes in the force--stretch relationship: a linear regime (generally $\stch \in (0, 0.5)$) and a superlinear regime.
From \eqref{eq:eap-sl-z} and \eqref{eq:eap-sl-x}, it can be seen that the small $|\mults|$ closed-form approximation predicts a linear relationship and in Fig. \ref{fig:sl-f} we see that, with respect to the numerical solutions, the small $|\mults|$ captures the linear regime almost exactly.
The error with respect to the numerical solutions does not occur until the superlinear regime $\stch > 0.5$.
Thus, the error of the small $|\mults|$ approximation in the $\A / \kB \T$ curves over $\stch \in (0.25, 0.5)$ is likely due to error in the normalization constant, $\C$.
Indeed, \eqref{eq:eap-sl-c} shows that this approximation predicts a normalization constant that does not change with stretch.
Consequently, the normalization condition is only satisfied at $\stch = 0$.

\subsection{Net Chain Dipole} \label{sec:sl-chain-polarization}

Lastly, we consider the accuracy of the approximation of the net chain dipole.
Using \eqref{eq:eap-sl-c}-\eqref{eq:eap-sl-z} in \eqref{eq:chain-polarization}, we obtain $\chainpolarmag_1 = \chainpolarmag_2 = 0$ and
\begin{equation} \label{eq:sl-chain-polarization-z}
    \chainpolarmag_3 = \frac{\ezeromag \N}{2 \uslnodim}\left[\sus{1} + \sus{2}\left(2\uslnodim - 1\right)\right] + \frac{e^{-\uslnodim} \ezeromag \N \dsus}{\sqrt{\pi \uslnodim} \erfw}
\end{equation}
Note that the above expressions for the net chain dipole do not have a dependence on the chain end-to-end vector which,
from \Fref{sec:numerical-dipole}, we know is incorrect.
However, the above expressions are exact when $\rvec = \mathbf{0}$.
%\todo[inline]{Please fix $p_x$ and so on to the correct coordinate system.}

%%%%%%%%%%%%%%%%%%%%%%%%%%%%%%%%%%%%%%%%%%%%%%%%%%%%%%%%%%%%%%%%%%%%%%%%%%%%
%%%%%%%%%%%%%%%%%%%%%%%%%%%%%%%%%%%%%%%%%%%%%%%%%%%%%%%%%%%%%%%%%%%%%%%%%%%%
%%%%%%%%%%%%%%%%%%%%%%%%%%%%%%%%%%%%%%%%%%%%%%%%%%%%%%%%%%%%%%%%%%%%%%%%%%%%
%%%%%%%%%%%%%%%%%%%%%%%%%%%%%%%%%%%%%%%%%%%%%%%%%%%%%%%%%%%%%%%%%%%%%%%%%%%%

\section{Asymptotic Matching} \label{sec:patched}

Obtaining an approximate, closed-form solution that is both accurate for $|\unodim| > 1$ and moderate stretches ($\stch > 0.25$) has proved difficult.
%However, here we take a different approach.
%Instead of assuming some parameter is small \emph{a priori}, we can use what we have learned thus far to guide our thinking in developing a new solution.
Although determining the monomer density function is difficult for general $\stch$, we do know the exact function at $\stch = 0$, which is obtained by recognizing that at $\stch = 0$, $\mults = \mathbf{0}$\footnote[5]{
  It is evident that $\mults \rightarrow \mathbf{0}$ as $\rvec \rightarrow \mathbf{0}$ because the Boltzmann term in the exponential (i.e. $\unodim \left(\edir \cdot \nvec\right)^2$) is invariant with respect to $\nvec \rightarrow -\nvec$.
  Because of this symmetry, for \eqref{eq:cr-eap} to be satisfied at $\rvec = \mathbf{0}$, we require $\mults = \mathbf{0}$.
}; and is $\density = \C \exp \left(\uslnodim \left(\ezerodir \cdot \nvec\right)^2 \right)$ where $\C$ is given by \eqref{eq:eap-sl-c}.
In addition, we know that at $\stch = 1$, the kinematic constraint dictates that $\density = \N \dirac\left(\rdir - \nvec\right)$.
We expect the density to transition from the $\stch = 0$ density to the $\stch = 1$ density as the chain is stretched.

Similarly, we can expect the small $|\mults|$ closed-form approximation (\Fref{sec:small-lambda}) to be accurate in the neighborhood of $\stch = 0$ and recovers the exact solution as $\stch \rightarrow 0$.
Also, we know that the Kuhn and Gr\"{u}n solution (\Fref{sec:kuhn-and-grun}) not only recovers $\density = \N \dirac\left(\rdir - \nvec\right)$ in the limit of $\stch \rightarrow 1$, but also is nearly exact in the neighborhood of $\stch = 1$.
Although the Kuhn and Gr\"{u}n solution is derived under the assumption that all chain configurations have the same potential energy and that is not true when electrical interactions are present, as $\stch \rightarrow 1$, the kinematic constraint dominates and all of the admissible chain configurations have approximately the same energy since the individual monomers in each chain must be oriented close to $\rdir$.

The strategy that we use here is to interpolate between these two solutions to generate a closed-form approximation that is reasonably accurate for $|\unodim| \ge 1$, over the entire range of stretch.
Ideally, we would take this approach with the monomer density function; however, this leads to a difficulty in calculating the entropy in \eqref{eq:A-approx-eap} (i.e. $\intoverSns{\density \log \density}$).
To see this, let $\densitysm$ and $\densitykg$ denote the small $|\mults|$ density and Kuhn and Gr\"{u}n density, respectively.
Then, ideally, we would use an approximation of the form
\begin{equation*}
    \density \approx \wsm \densitysm + \wkg \densitykg
\end{equation*}
where $\wsm$ and $\wkg$ are the weights of each respective monomer density, and are functions of the macroscopic parameters (i.e. $\wsm = \wsm\left(\unodim, \stch, \ezerodir, \rdir\right)$, $\wkg = \wkg\left(\unodim, \stch, \ezerodir, \rdir\right)$).
% Since $\wsm$ and $\wkg$ are weights, we require
% \begin{align*}
% 	\wsm + \wkg &= 1.0 \\
% 	0.0 \leq \wsm &\leq 1.0 \\
% 	0.0 \leq \wkg &\leq 1.0
% \end{align*}
However, the integral in \eqref{eq:A-approx-eap} would include a term that would be the logarithm of a sum of two exponentials, which we are unable to evaluate or approximately evaluate in closed-form.

\subsection{Free Energy} \label{sec:asymptotic-A}

To avoid the difficulty illustrated just above, we instead aim to approximate $\A$ as a weighted average.
Let $\Asm$ denote the free energy approximation \eqref{eq:A-approx-sl} derived in \Fref{sec:small-lambda}.
As mentioned previously, the approximation is accurate near $\stch = 0$.
Additionally, substituting $\densitykg$ into \eqref{eq:A-approx-eap} results in a free energy approximation that is exact in the limit $\stch \rightarrow 1$:
\begin{align}
	\Akg &= \Ukg - T \Skg \\
	\Ukg &= \intoverSns{\densitykg \um} \\
		 &= \N \kB \T \left[\uznodim - \uOnodim + \frac{\stch}{\zmultzero}\left(\uxnodim - 2\uznodim\right)\right]
\end{align}
and $\Skg$ was derived in \Fref{sec:kuhn-and-grun} in \eqref{eq:entropy-length}.

Finally, we make the approximation
\begin{equation} \label{eq:asymptotic-A}
	\begin{split}
	\A &\approx \Aas \\
	&= \Akg + \left(1 - \stch^2\right)\left( \lim_{\stch \rightarrow 0} \Asm - \lim_{\stch \rightarrow 0} \Akg \right) \\
	&= \N \kB \T \left\{ -\uOnodim + \stch \zmultzero + \log \left(\frac{\zmultzero \csch \left[\zmultzero\right]}{4 \pi}\right) + \right. \\
	& \qquad \qquad \left. \frac{\stch \unodim}{\zmultzero}  + \left(1 - \stch^2 \right) \left[-\frac{\unodim}{3} + \log \left(\frac{2 \sqrt{\unodim}}{\sqrt{\pi} \erfw}\right)\right]  + \right. \\
	& \qquad \qquad \left. \left(\unodim - 3 \frac{\stch \unodim}{\zmultzero}\right) \left(\edir \cdot \rdir\right)^2\right\}
	\end{split}
\end{equation}
Notice that \eqref{eq:asymptotic-A} has the following features:
(1.) it recovers the exact solution when $\unodim = 0$, and (2.) it is exact in the limits of zero stretch and full stretch.
In principle, the (stretch) limiting behavior would be recovered with any choice of exponent on $\stch$; however, the stretch term was chosen to be quadratic because of additional physical considerations.
It was discovered in \Fref{sec:sl-fs} that there are generally two regimes to the force-length relation of an DE chain: a linear regime at small to moderate stretches followed by a superlinear regime.
By choosing the stretch term to be quadratic, we reproduce the linear regime while the $\Akg$ term recovers the super linear regime.
We do {\em not} add a term that is linear in $\stch$ because it would result in a constant contribution to the chain force which is not observed.
%Consider that the end-to-end vector, $\rvec$, is formed by joining monomers end to end; thus, $\rvec \rightarrow -\rvec$ would have the effect of reversing the direction of each of the monomers in the chain.
%Also, (1.) the energy of an electro-responsive monomer is invariant with respect to $\nvec \rightarrow -\nvec$ and (2.) entropic elasticity is rotationally invariant (e.g. Fig. \ref{fig:fe-surfaces}).
%Consequently, the free energy of a DE chain is invariant with respect to $\rvec \rightarrow -\rvec$.
%This means that, physically, \emph{while it is important where the two ends of the chain are fixed, there is no difference between calling one the beginning and the other the end, or vice versa}.
%Since there is no physical difference between $\rvec$ and $-\rvec$, the free energy should not have a linear contribution in $\stch$.

\begin{figure}[htb!]
	\centering
	\begin{tabulary}{\linewidth}{c c}
		\includegraphics[width=\FigureWidthTwoColsNumericalGraph]{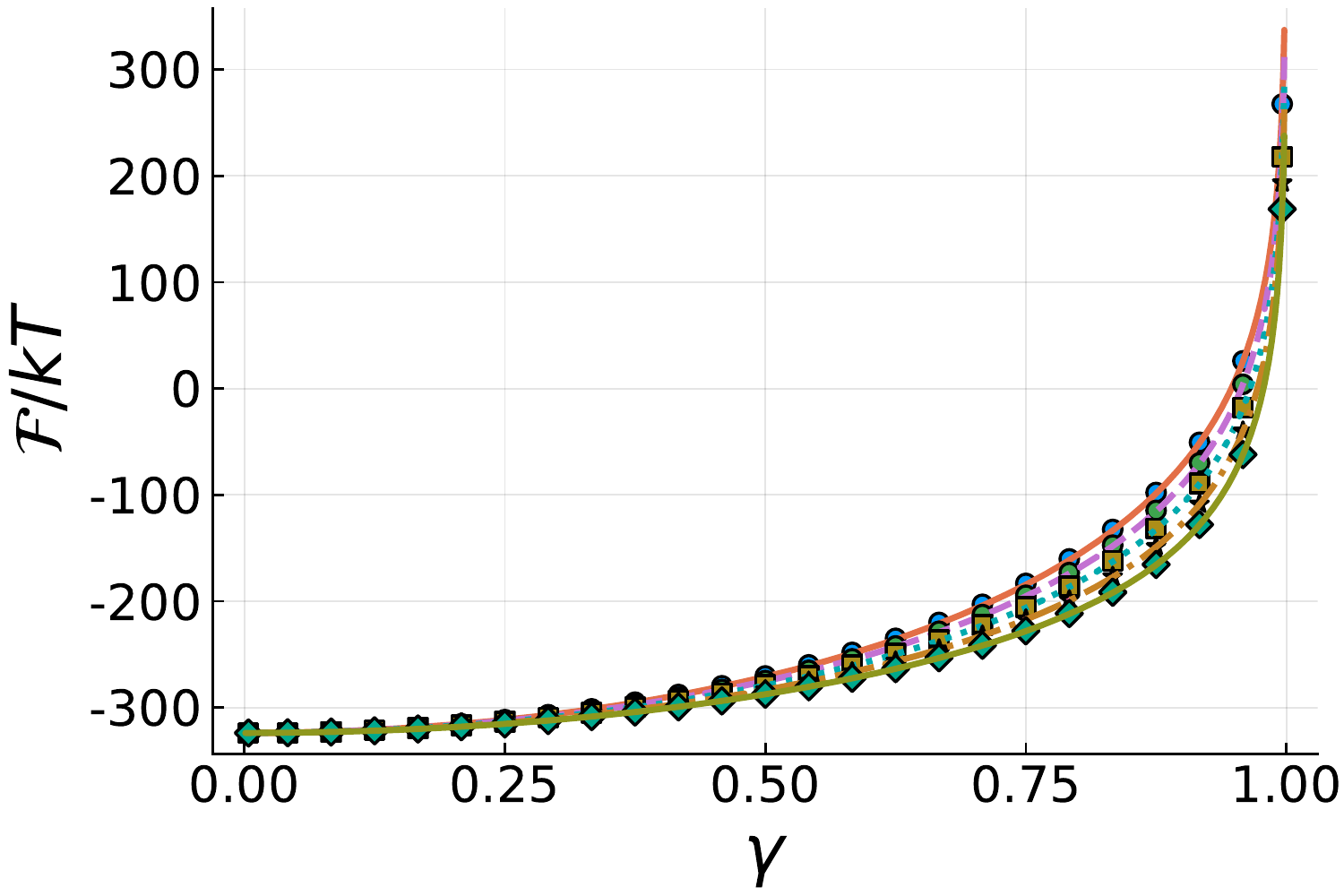}
		&
		\includegraphics[width=\FigureWidthTwoColsNumericalGraph]{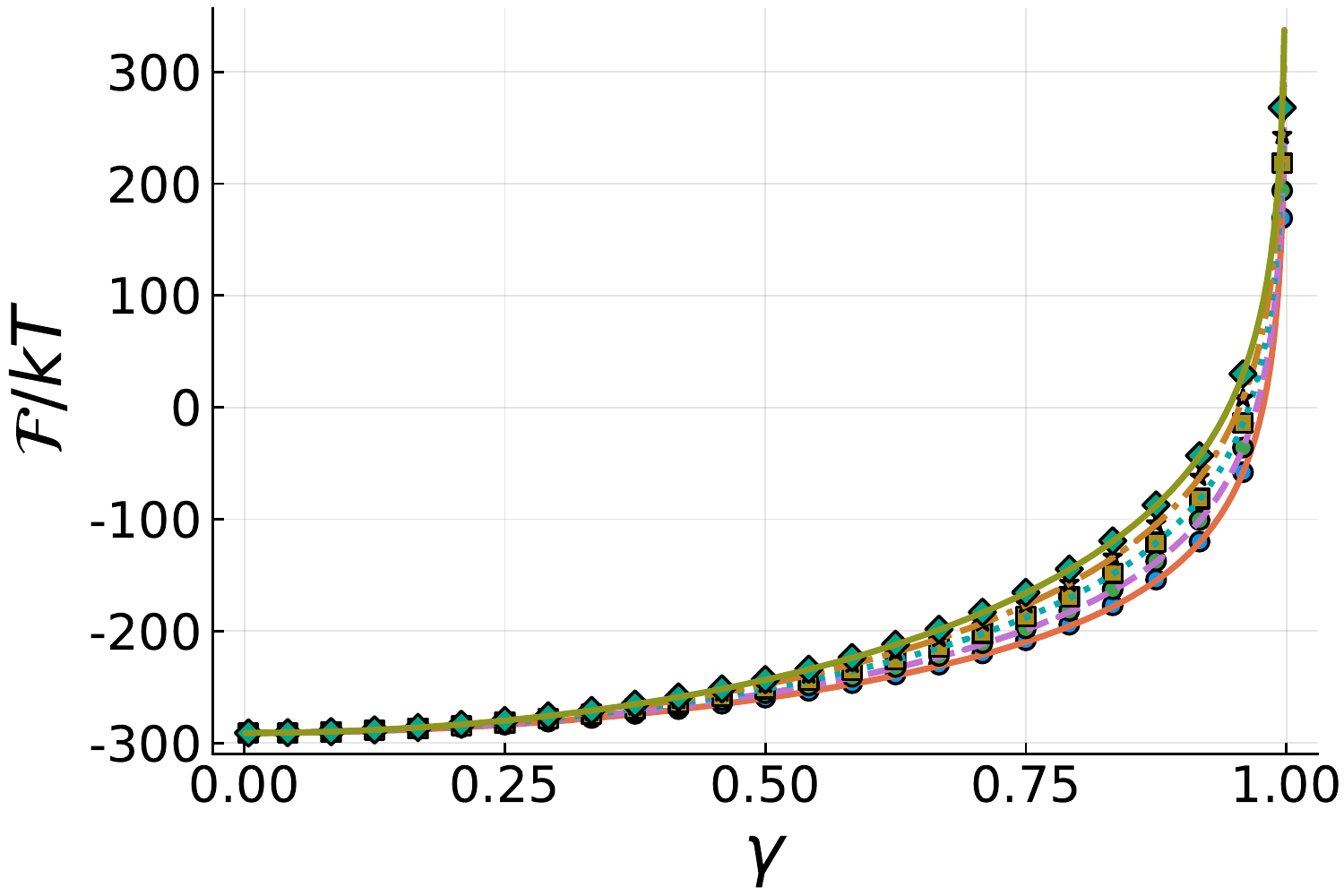}
		\\
		\includegraphics[width=\FigureWidthTwoColsNumericalGraph]{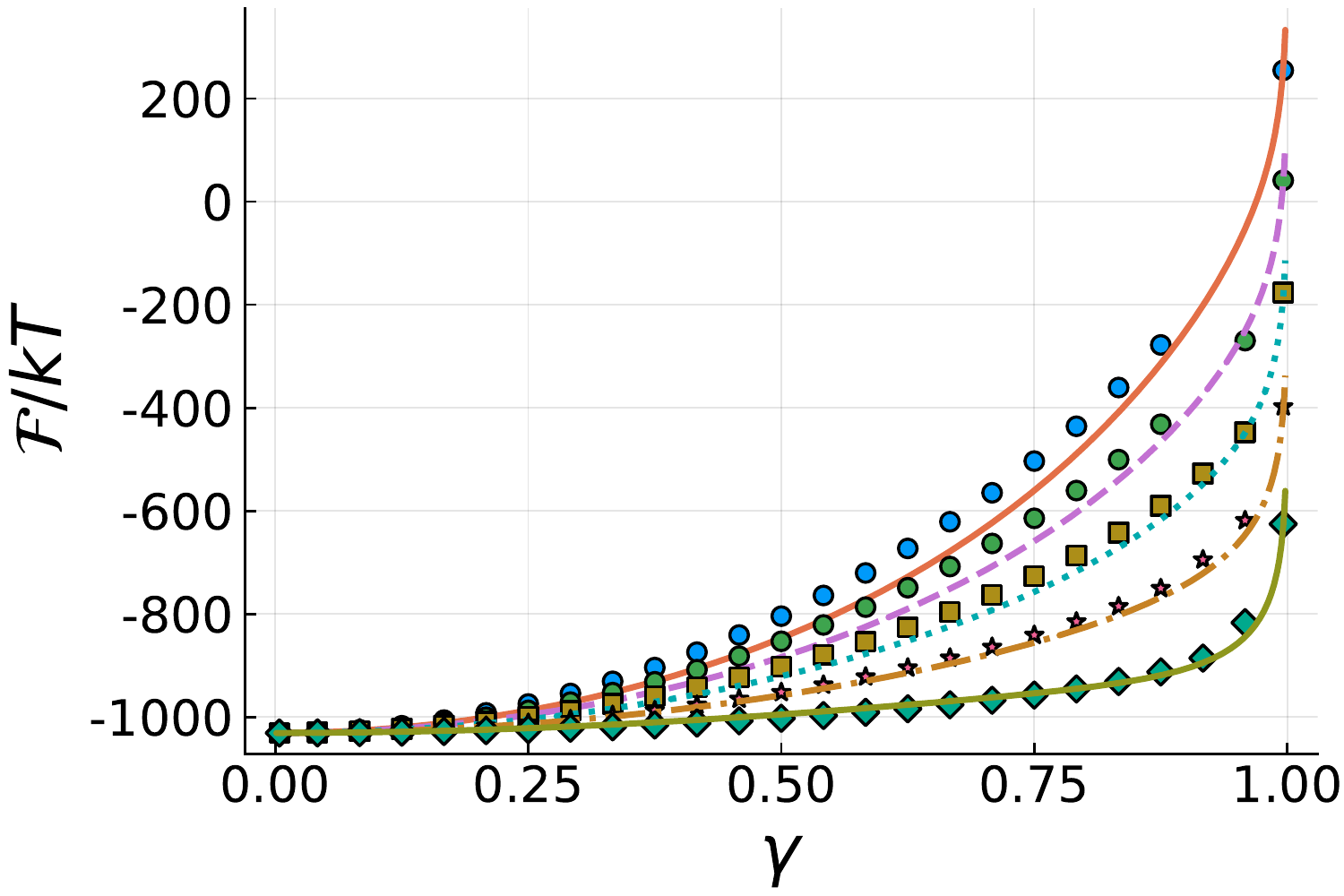}
		&
		\includegraphics[width=\FigureWidthTwoColsNumericalGraph]{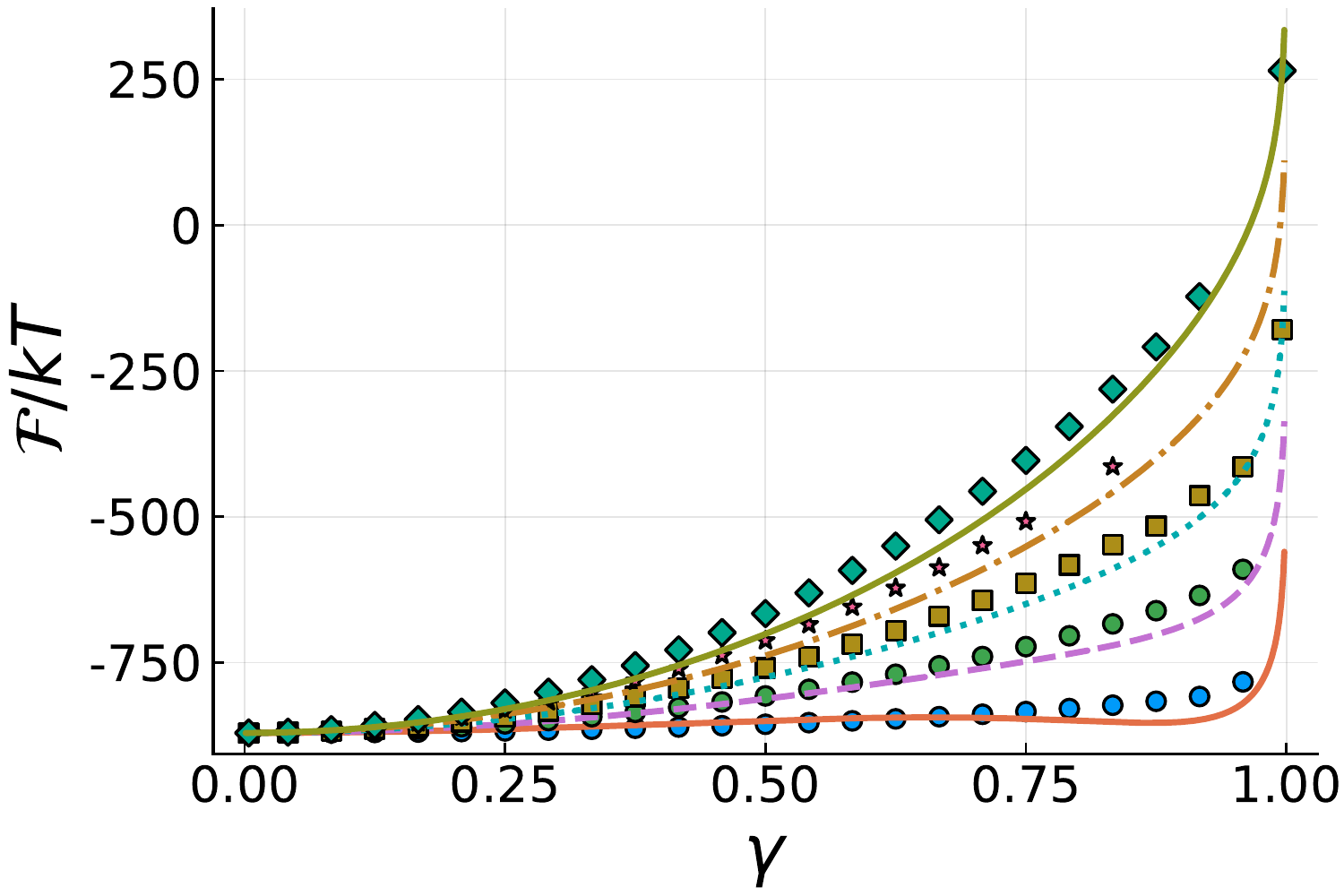}
		\\
		\includegraphics[width=\FigureWidthTwoColsNumericalGraph]{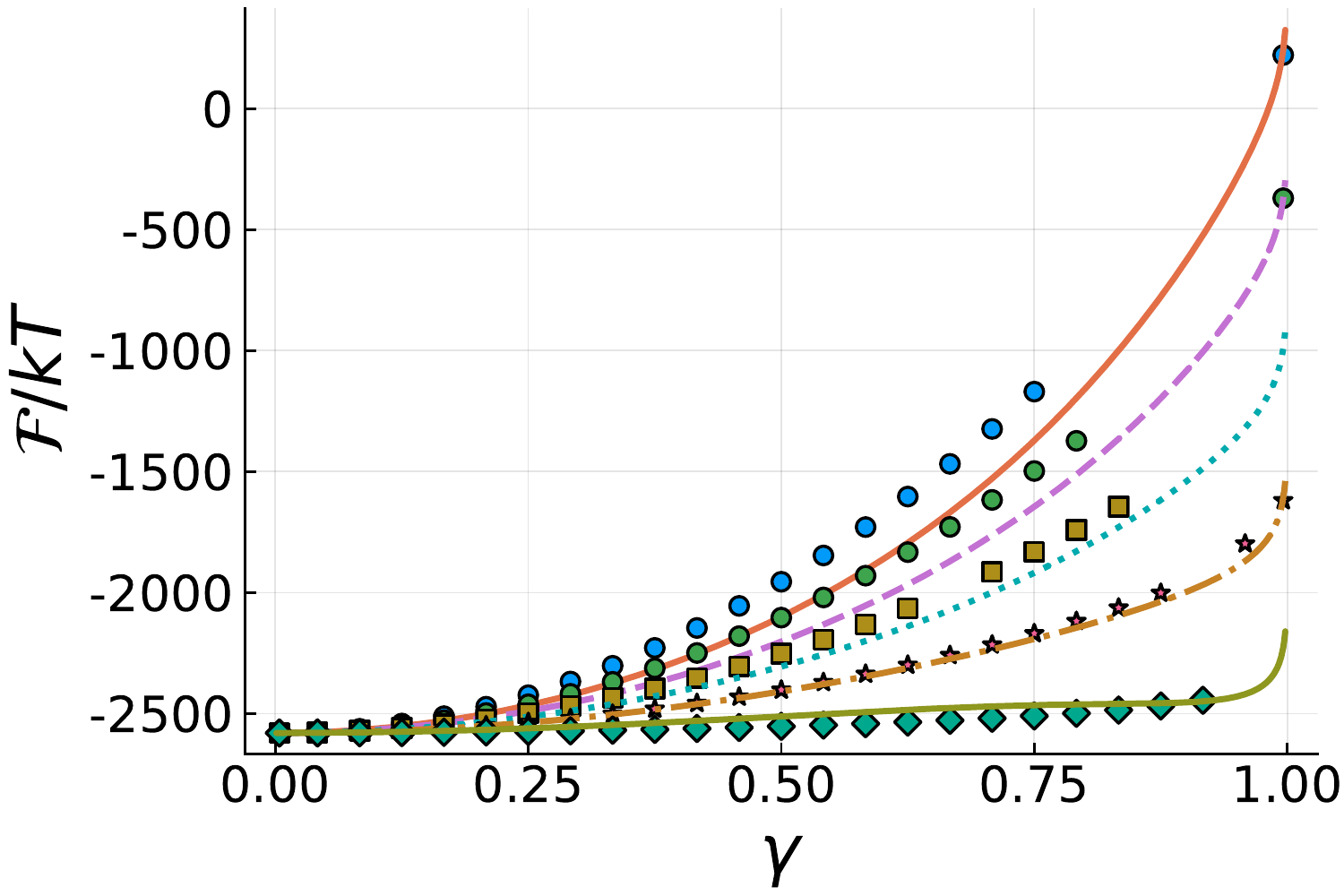}
		&
		\includegraphics[width=\FigureWidthTwoColsNumericalGraph]{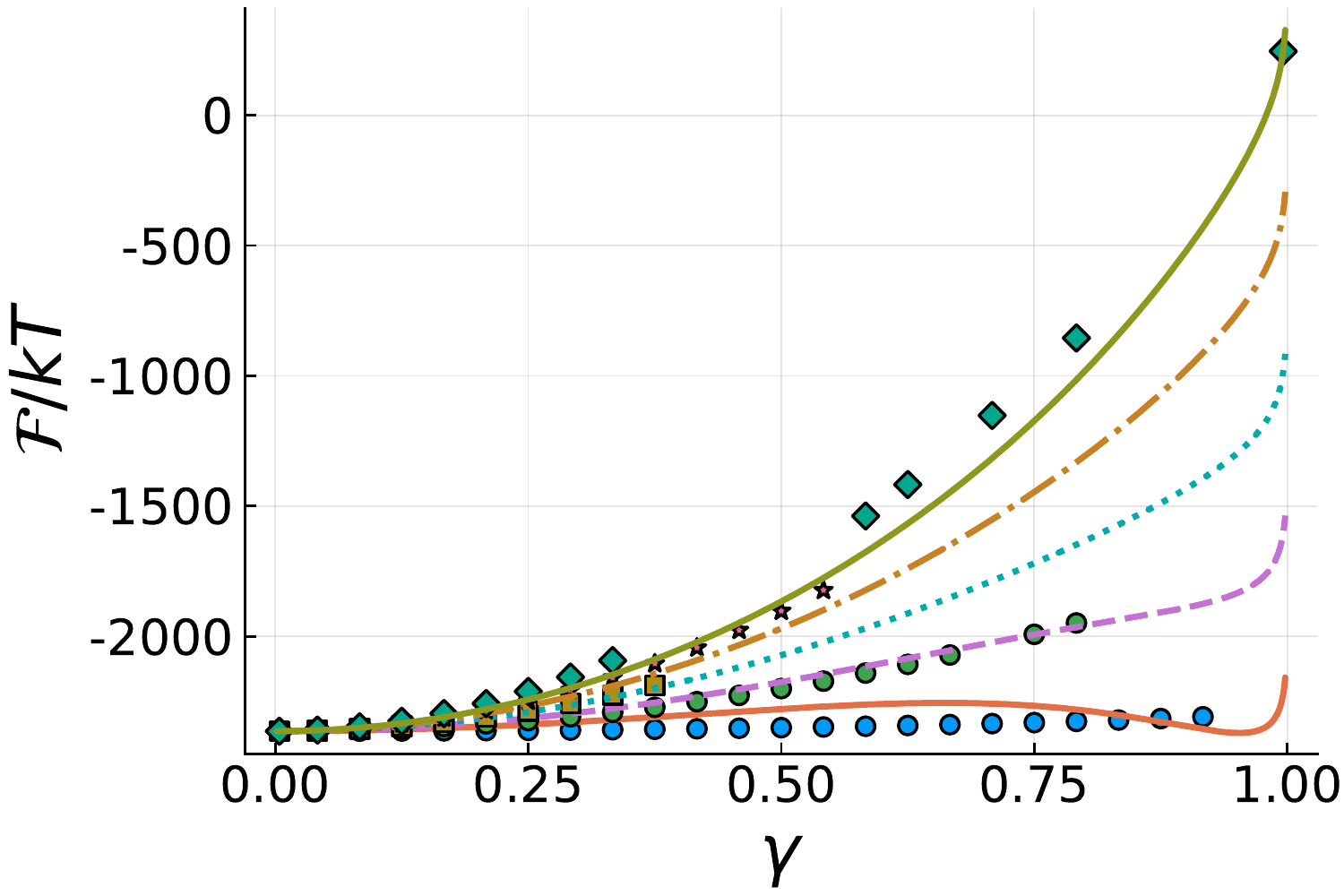}
		\\
		\multicolumn{2}{c}{\fbox{\includegraphics[height=1.7in]{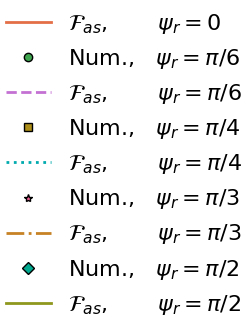}}}
	\end{tabulary}
	\caption{Comparison of the predicted $\A / \kB \T$ with $\stch$ relationship using the asymptotic matching approximation and the numerical solutions.
		TI chains appear on the right and uniaxial chains on the left;
		$\unodim = 1.0, -1.0, 9.0, -9.0, 25.0$, and $-25.0$, (top left to bottom right, respectively).}
	\label{fig:asymptotic-A}
\end{figure}

Fig. \ref{fig:asymptotic-A} shows the predicted free energy-stretch relation of the $\Aas$ approximation compared to the numerical solution for $\unodim = 1.0, -1.0, 9.0, -9.0, 25.0$, and $-25.0$, (top left to bottom right, respectively).
The comparison is shown for chain orientations with respect to the electric field of $\etorangle = 0, \pi / 6, \pi / 4, \pi / 3$, and $\pi / 2$.
It can be seen from Fig. \ref{fig:asymptotic-A} that the approximation developed using asymptotic matching and motivated by the numerical results, $\Aas$, agrees well with the numerical solutions for a wide range of chain stretch, chain orientation, and $\unodim$.

\subsection{Net Chain Dipole}

\begin{figure}[htb!]
	\centering
	\begin{tabulary}{\linewidth}{c c}
		\includegraphics[width=\FigureWidthTwoColsNumericalGraph]{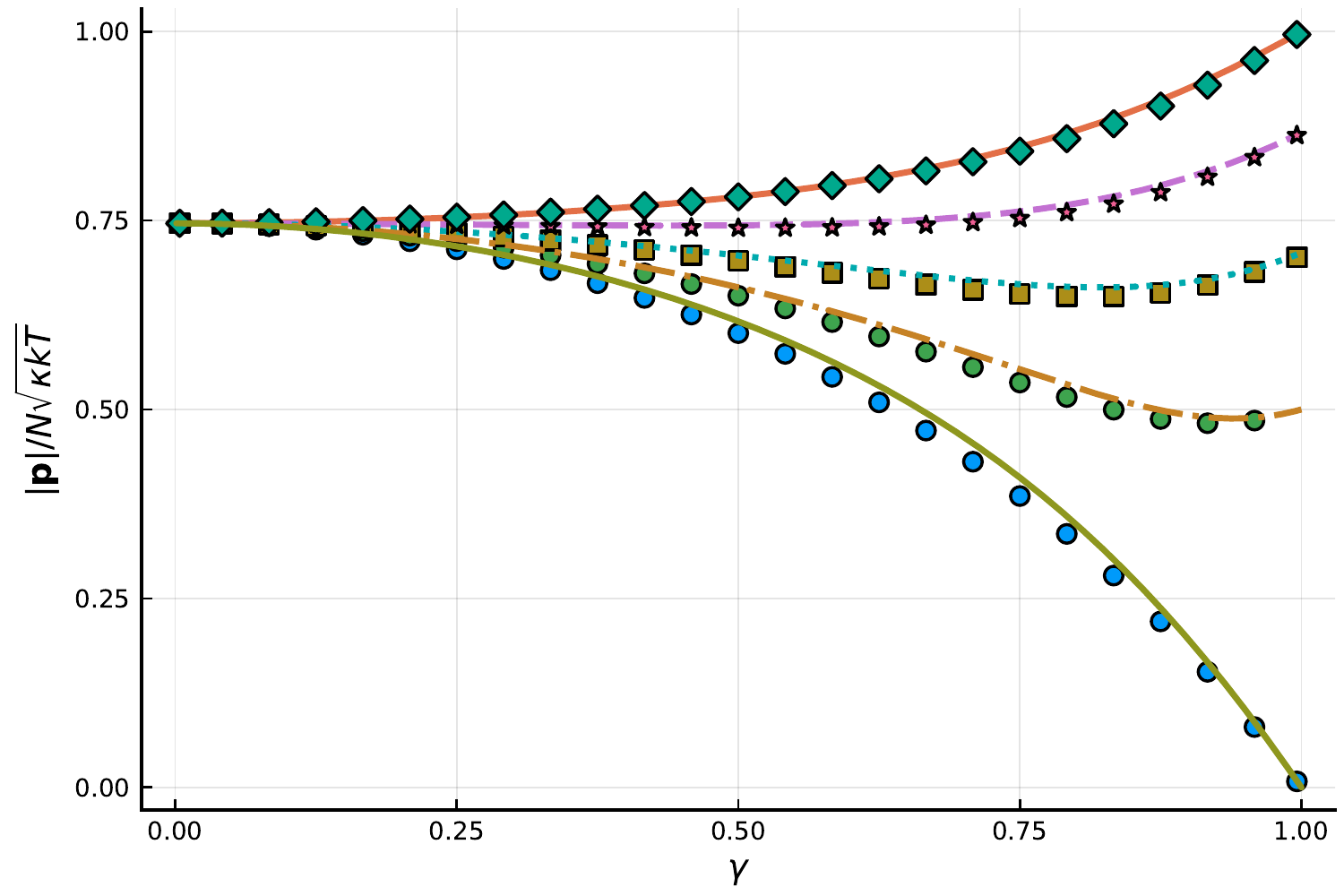}
		&
		\includegraphics[width=\FigureWidthTwoColsNumericalGraph]{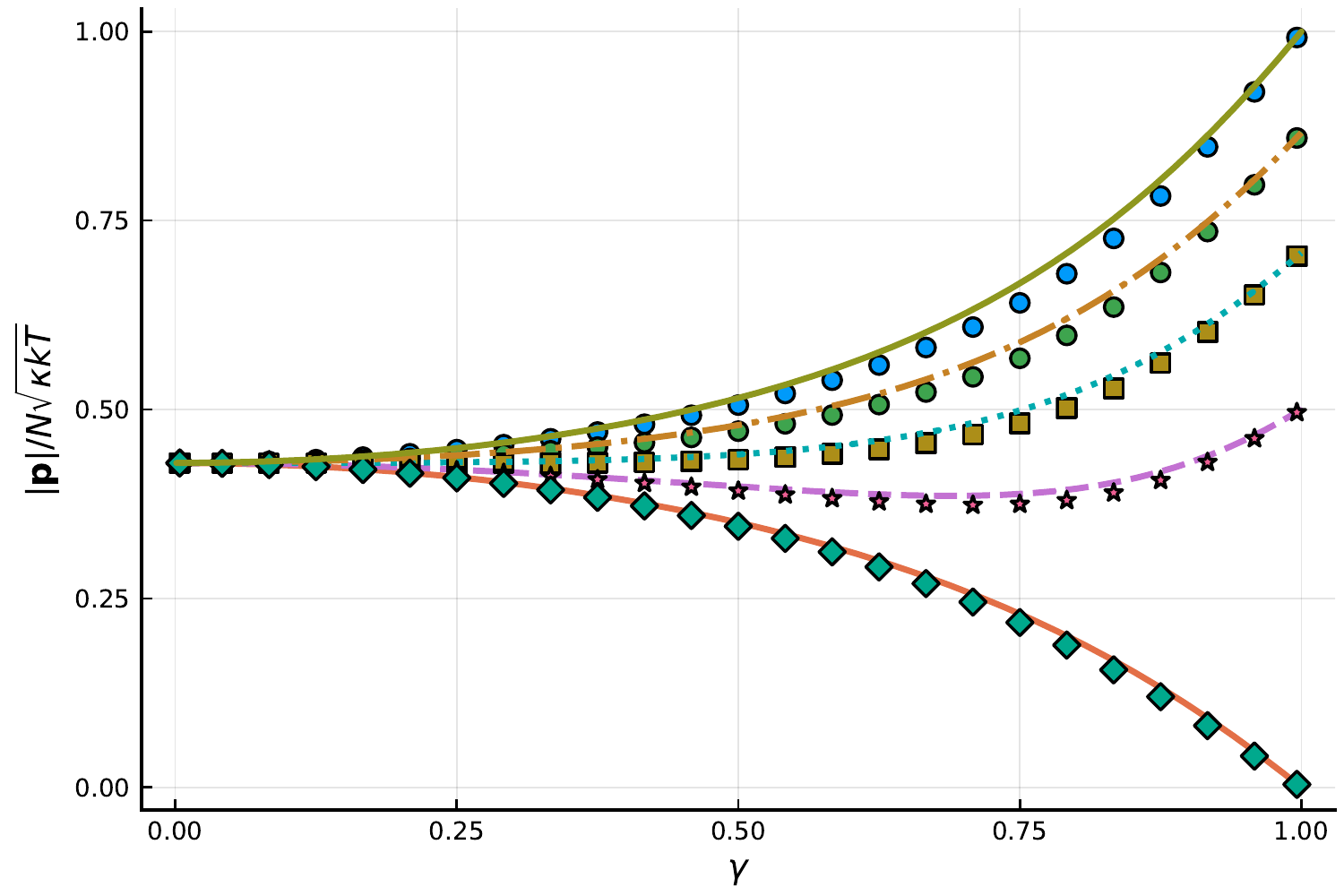}
		\\
		\includegraphics[width=\FigureWidthTwoColsNumericalGraph]{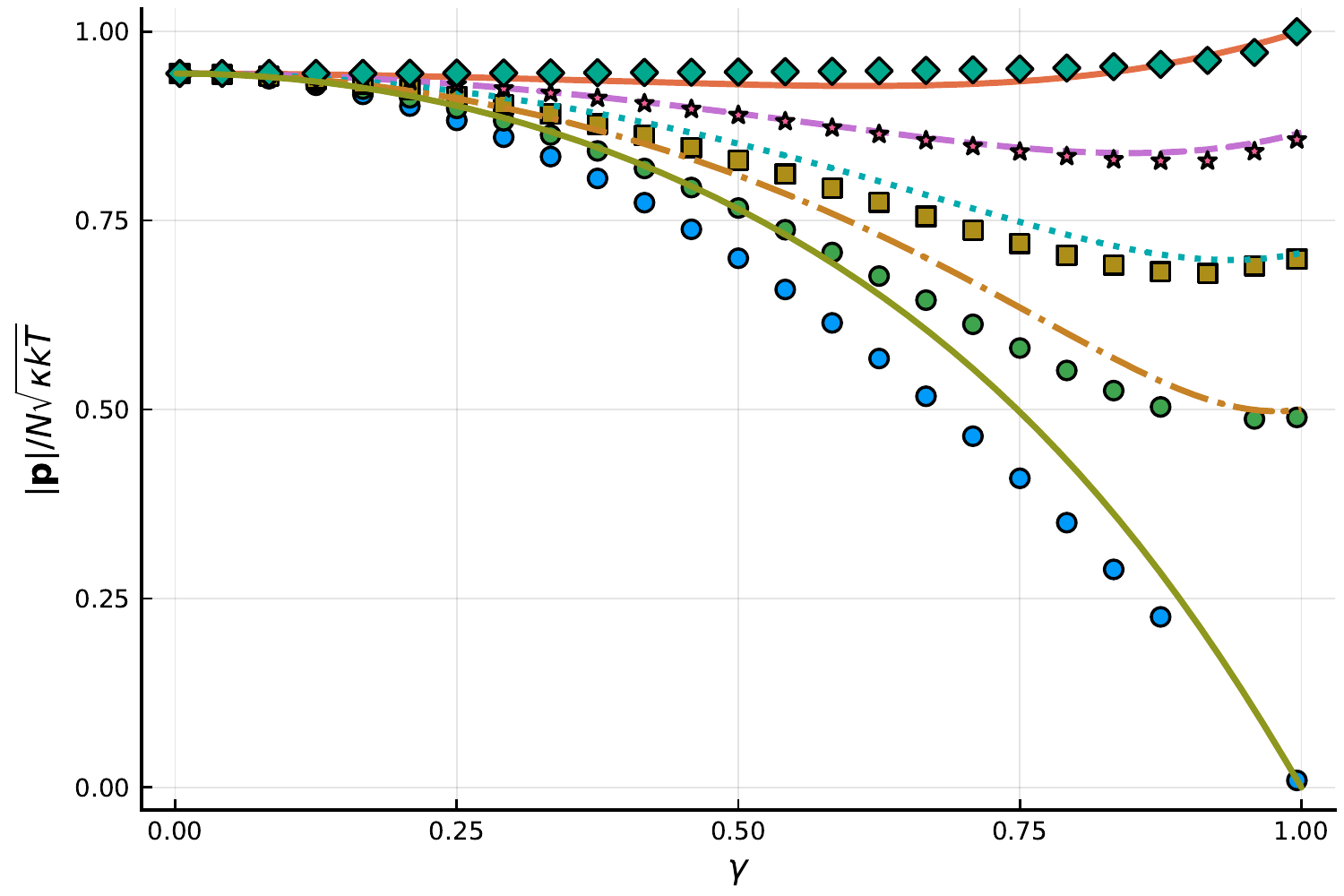}
		&
		\includegraphics[width=\FigureWidthTwoColsNumericalGraph]{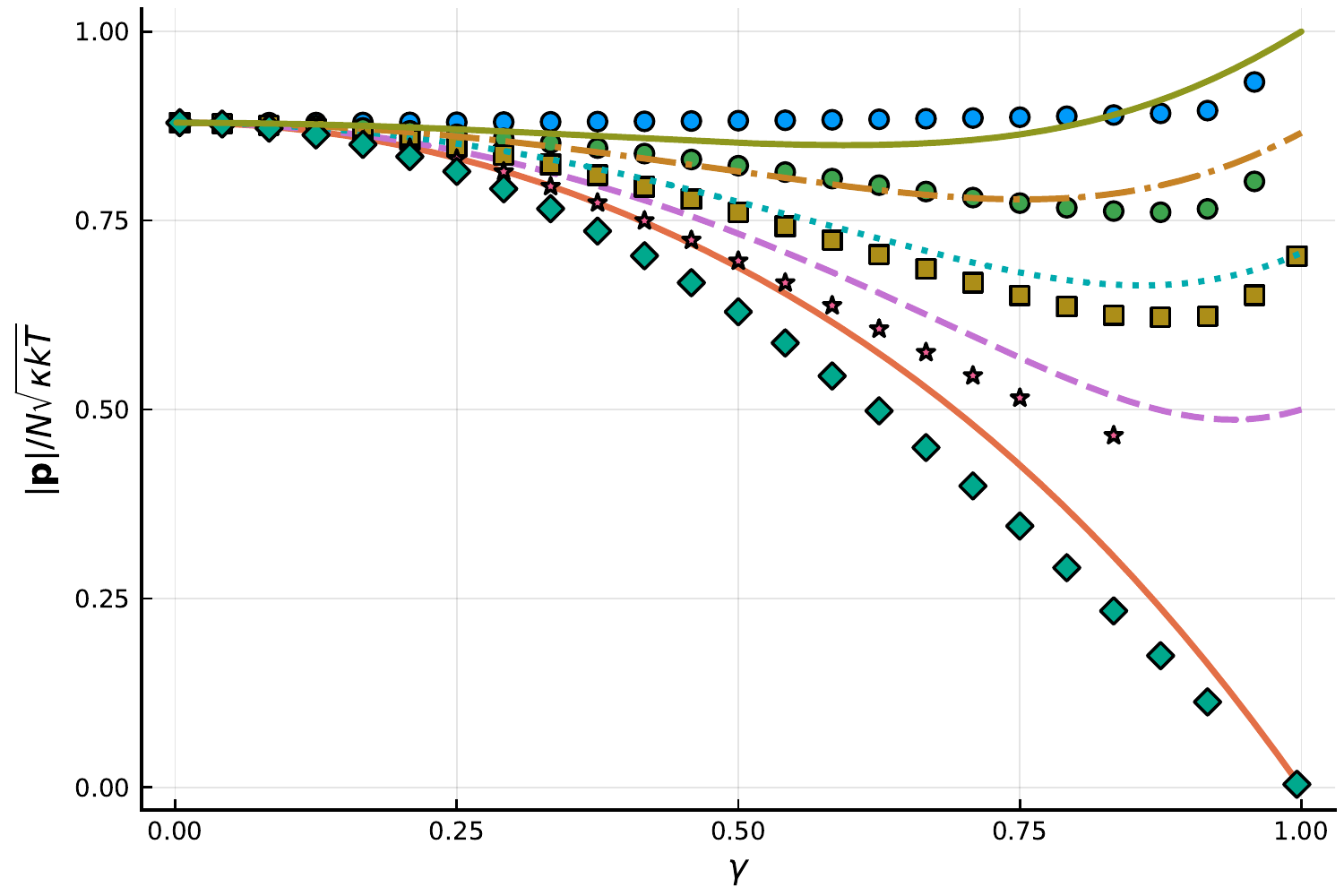}
		\\
		\multicolumn{2}{c}{\includegraphics[height=1.7in]{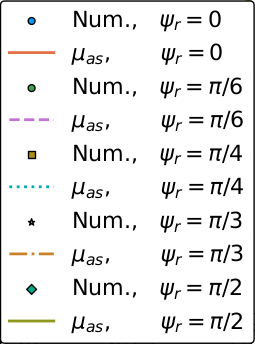}}
		\\
	\end{tabulary}
	\caption{Comparison of the predicted $|\chainpolar| / \sqrt{|\unodim| \kB \T}$ with $\stch$ relationship using the asymptotic matching approximation and the numerical solutions.
		TI chains appear on the right and uniaxial chains on the left;
		$\unodim = 1.0, -1.0, 9.0$, and $-9.0$, (top left to bottom right, respectively).}
	\label{fig:asymptotic-chain-polarization}
\end{figure}

We now make a similar approximation to the net chain dipole.
Let $\chainpolarsm$ be the net dipole derived in  \Fref{sec:sl-chain-polarization} in  \eqref{eq:sl-chain-polarization-z}.
Next, let the coordinate system be such that the polar axis (i.e. $\ethree$) is taken in the direction of chain stretch.
The approximate net chain dipole using $\densitykg$ is
\begin{equation}
    \chainpolarkg = \N \begin{Bmatrix}
    \ezerox \left[\sus{2} - \dsus \stch / \zmultzero\right] \\
    0 \\
    \ezeroz \left[\sus{1} + 2 \dsus \stch / \zmultzero\right]
    \end{Bmatrix}
\end{equation}
Lastly, the asymptotic approximation is taken to be
\begin{equation}
    \chainpolaras = \chainpolarkg + \left(1 - \stch^2\right)\left( \bfR \chainpolarsm - \lim_{\stch \rightarrow 0} \chainpolarkg \right)
\end{equation}
where $\bfR$ is a proper orthogonal matrix that rotates the coordinate system used to derived $\chainpolarsm$, i.e. the polar axis taken in the direction of the electric field, to the coordinate system in which $\chainpolarkg$ was derived\footnote[6]{
    Alternatively, we could have arrived at an approximation of the net chain dipole by utilizing the relation $-\takepartialflat{\A}{\ezero} = \chainpolar$.
}.
Fig. \ref{fig:asymptotic-chain-polarization} shows the predicted magnitude of the net chain dipole of the $\chainpolaras$ approximation compared to the numerical solution for $\unodim = 1.0, -1.0, 9.0$, and $-9.0$, (top left to bottom right, respectively).
The comparison is shown for chain orientations with respect to the electric field of $\etorangle = 0, \pi / 6, \pi / 4, \pi / 3$, and $\pi / 2$.
It can be seen from Fig. \ref{fig:asymptotic-chain-polarization} that, once again, the approximation developed using asymptotic matching and physical intuition agrees well with the numerical solutions for a wide range of chain stretch, chain orientation, and $\unodim$.

%%%%%%%%%%%%%%%%%%%%%%%%%%%%%%%%%%%%%%%%%%%%%%%%%%%%%%
%%%%%%%%%%%%%%%%%%%%%%%%%%%%%%%%%%%%%%%%%%%%%%%%%%%%%%
%%%%%%%%%%%%%%%%%%%%%%%%%%%%%%%%%%%%%%%%%%%%%%%%%%%%%%
%%%%%%%%%%%%%%%%%%%%%%%%%%%%%%%%%%%%%%%%%%%%%%%%%%%%%%

\section{Discussion} \label{sec:conclusion}

The aim of this paper was to investigate the electro-elasticity of DE chains using statistical mechanics.
Following a broadly similar approach to the classical work of \citet{kuhn1942beziehungen}, we derived equations for the most-likely monomer density function of a DE chain.
Given the complexity of the resulting equations, we used numerical methods to obtain insights into the monomer orientation distribution, free energy, net chain dipole and other physically-important quantities.
In particular, we consider the physical implications of the resulting solutions -- emphasizing the interplay between the electrostatic energy, the thermal energy, and the kinematic constraints of the chain.
We then derived closed-form approximations in the regime where the electrical contributions are small and when the chain tension is small; these approximations have a limited regime of validity when compared to the numerical solutions.
Therefore, we then interpolate between these limit solutions using insight obtained from the numerical solution to obtain a closed-form approximation for the free energy and the net chain dipole that agreed well with the numerical solution over a large range of parameter values and regimes of interest.

As a part of our study, we examined the mechanical and electrical torque acting on a chain.
There is a competition between the kinematic end-to-end vector constraint, the entropically-driven tendency for uniform orientation distribution, and the electrostatically-driven tendency to align monomer dipoles with the electric field.
The interplay between these effects result in a balance that can be identified as a mechanical torque and an electrical torque.
At the continuum scale, there are close analogies in the homogenization of electrically- and magnetically- responsive anisotropic inclusions \cite{siboni2014fiber,castaneda2011homogenization,galipeau2013finite}.

For future research, in the context of the bigger picture of predicting the electromechanical constitutive response of electro-responsive elastomers, the next step is to use the response of a single chain obtained in this paper and average over large numbers of chains in various directions, following e.g. References \citenum{verron2015questioning,arruda1993threee,miehe2004micro,davidson2013nonaffine}, to predict the macroscopic response of a network.
Alternatively, concurrent approaches to multiscale modeling, following e.g. References \citenum{ghareeb2020adaptive,marshall2014atomistic}, could also provide important insights into settings where the deformation varies at lengthscales on order of the chain lengthscale.

\section*{Conflicts of interest}
There are no conflicts to declare.

\section*{Software Availability}
A version of the code developed for this work is available at \url{github.com/grasingerm/polymer-stats}.
Links to the integration package~\cite{integration} and optimization package~\cite{NLopt} are available in the references.

\section*{Acknowledgments}
This paper draws from the doctoral dissertation of Matthew Grasinger at Carnegie Mellon University.
We thank Michael Bockstaller, Timothy Breitzman, Gal deBotton, Richard James, Carmel Majidi, Pedro Ponte Casta\~{n}eda, Prashant Purohit, M. Ravi Shankar, and Pradeep Sharma for useful discussions.
We acknowledge financial support from NSF (1635407), ARO (W911NF-17-1-0084), ONR (N00014-18-1-2528, N00014-18-1-2856), BSF (2018183), and AFOSR (MURI FA9550-18-1-0095).
We acknowledge NSF for computing resources provided by Pittsburgh Supercomputing Center.

\appendix

\section{The Constant Electric Field Ensemble}
\label{sec:constant-E-ensemble}

In the constant electric field ensemble, we begin with the assumption that the macroscopic electric field $\bfE$ varies over length scales much largers than the polymer chain length scale.
Therefore, at the chain scale, the microscopic picture is that the electric field $-\nabla\phi$ is periodic such that the average value of $-\nabla\phi$ matches the local value $\bfE_0$ of the macroscopic field.

Consider a periodic setting with the periodicity defined by the lattice vectors $\{\bff_1, \bff_2, \bff_3\}$.
The periodic unit cell $\Omega$ is defined as $\nu_1\bff_1 + \nu_2\bff_2 + \nu_3\bff_3, 0\leq\nu_1,\nu_2,\nu_3 < 1$.

We can now define the constant electric field ensemble as subject to a periodic electric field with given mean value $\bfE_0$:
\begin{equation}
    \phi(\bfx + \bff_i) = \phi(\bfx) - \bfE_0\cdot\bff_i \Rightarrow \nabla\phi|_{\bfx+\bff_i} = \nabla\phi|_{\bfx}, i = 1,2,3
\end{equation}
To fix the gauge, we assume $\int_\Omega \phi \dm\Omega = 0$.

We next show that the average value of $\nabla\phi$ over $\Omega$ is $-\bfE_0$.
\begin{equation}
\label{eqn:average1}
\begin{split}
    & \int_\Omega \nabla\phi \dm\Omega
     =
    \int_{\partial\Omega} \hat\bfm \phi \dm\partial\Omega
    \\
    & =
    \int\limits_{\substack{\nu_1=1\\0\leq\nu_2,\nu_3<1}} \hat\bfm\phi \dm A
    +
    \int\limits_{\substack{\nu_1=0\\0\leq\nu_2,\nu_3<1}} \hat\bfm\phi \dm A
    +
    \int\limits_{\substack{\nu_2=1\\0\leq\nu_1,\nu_3<1}} \hat\bfm\phi \dm A
    +
    \int\limits_{\substack{\nu_2=0\\0\leq\nu_1,\nu_3<1}} \hat\bfm\phi \dm A
    +
    \ldots
%    \int\limits_{\substack{\nu_3=1\\0\leq\nu_2,\nu_1<1}} \hat\bfm\phi \dm A
%    +
%    \int\limits_{\substack{\nu_3=0\\0\leq\nu_2,\nu_1<1}} \hat\bfm\phi \dm A
    \\
    & =
    -\bfE_0\cdot\bff_1 \int\limits_{\substack{\nu_1=1\\0\leq\nu_2,\nu_3<1}} \hat\bfm \dm A
    -
    \bfE_0\cdot\bff_2 \int\limits_{\substack{\nu_2=1\\0\leq\nu_1,\nu_3<1}} \hat\bfm \dm A
    -
    \bfE_0\cdot\bff_3     \int\limits_{\substack{\nu_3=1\\0\leq\nu_1,\nu_2<1}} \hat\bfm \dm A
    \\
    & =
    - (\bff_2 \times \bff_3 \otimes \bff_1 
    + \bff_3 \times \bff_1 \otimes \bff_2 
    +\bff_1 \times \bff_2 \otimes \bff_3) \bfE_0
\end{split}
\end{equation}
To arrive at the last step, we have used that $\hat\bfm = \frac{\bff_2\times\bff_3}{|\bff_2\times\bff_3|}$ and that the area of the face is $|\bff_2\times\bff_3|$ in the first term, and similarly in the other terms.

To simplify further, we write $\bfE_0$ in terms of components in the basis $\{\bff_2\times\bff_3,\bff_3\times\bff_1,\bff_1\times\bff_2\}$ as:
\begin{equation}
\label{eqn:average2}
    \bfE_0 = E_0^1 (\bff_2\times\bff_3) + E_0^2 (\bff_3\times\bff_1) + E_0^3 (\bff_1 \times \bff_2)
\end{equation}
Using \eqref{eqn:average2} in \eqref{eqn:average1}, we find that:
\begin{equation}
\begin{split}
    - \int_\Omega \nabla\phi \dm\Omega
    & =
    E_0^1 (\bff_2\times\bff_3) (\bff_1\cdot\bff_2\times\bff_3)
    + E_0^2 (\bff_3\times\bff_1) (\bff_2\cdot\bff_3\times\bff_1)
    \\
    & \qquad + E_0^3 (\bff_1\times\bff_2) (\bff_3\cdot\bff_1\times\bff_2)
    \\
    & =
    \mathrm{vol}(\Omega) \bfE_0
\end{split}
\end{equation}

\section{Net Chain Dipole as a Derivative of the Free Energy} \label{app:polarization}

We show that a consequence of assuming \eqref{eq:density-eap} as the form of the monomer density function and enforcing the constraints given in \eqref{eq:cn-eap}, we arrive at the relationship: $\chainpolar = -\takepartialflat{\A}{\ezero}$.

	Taking derivatives of both sides of \eqref{eq:cn-eap} with respect to $\ezero$, we obtain:
	\begin{equation} \label{eq:dconstraints}
	\begin{split}
    	\takepartial{}{\ezero} \intoverSns{\density\left(\hat\bfv\right)} = \intoverSns{\takepartial{\density}{\ezero}} &= \takepartial{\N}{\ezero} = \bf{0}
    	\\
    	\takepartial{}{\ezero} \intoverSns{\density\left(\hat\bfv\right) \hat\bfv} = \intoverSns{\takepartial{\density}{\ezero} \hat\bfv} &= \takepartial{\rvec / \mlen}{\ezero} = \bf{0}
	\end{split}
	\end{equation}
	We are able to interchange the operations of derivation and integration because of the smoothness of the integrands; and in the last equalities we use the fact that neither the number of the monomers in the chain nor the end-to-end vector constraint depend on $\ezero$.
	
	Now, we obtain the desired result by taking derivatives of both sides of \eqref{eq:A-approx-eap}:
	\begin{align*}
    	-\takepartial{\A}{\ezero} &= -\takepartial{}{\ezero} \intoverSns{\left(\density \um + \kB \T \density \log \density\right)} \\
    	&= -\intoverSns{\left[\takepartial{\density}{\ezero} \um + \density \takepartial{\um}{\ezero} + \kB \T \takepartial{\density}{\ezero} \log \density + \density \left(\frac{1}{\density}\takepartial{\density}{\ezero} \right)\right]} \\
    	&= -\intoverSns{\left[\takepartial{\density}{\ezero} \um + \density \takepartial{\um}{\ezero} + \kB \T \takepartial{\density}{\ezero} \left(\log \C -\um / \kB \T + \mults \cdot \nvec\right) + \takepartial{\density}{\ezero}\right]} \\
    	&= -\intoverSns{\left[\density \takepartial{\um}{\ezero} + \kB \T \takepartial{\density}{\ezero} \left(\mults \cdot \nvec\right) + \left(\kB \T \log\C + 1\right)\takepartial{\density}{\ezero}\right]} \\
    	&= -\intoverSns{\density \takepartial{\um}{\ezero}} - \kB \T \mults \cdot \left(\intoverSns{\takepartial{\density}{\ezero} \nvec}\right) - \left(\kB \T \log\C + 1\right)\intoverSns{\takepartial{\density}{\ezero}}
    \end{align*}
    By \eqref{eq:dconstraints}, the last two terms vanish.
    Recalling \eqref{eq:dipole-response} and \eqref{eq:monomer-energy}:
    \begin{align*}
    	-\takepartial{\A}{\ezero} &= -\intoverSns{\density \takepartial{\um}{\ezero}} \\
    	&= \intoverSns{\density \takepartial{}{\ezero} \left(\frac{1}{2} \ezero \cdot \sustens \ezero\right)} \\
    	% &= \intoverSns{\density \frac{1}{2} \left(\sustens \ezero + \sustens^T \ezero \right)} \\
    	&= \intoverSns{\density \dipole}\\
    	&= \chainpolar
	\end{align*}
%\end{proof}
This result is not generally true but follows from our model that \emph{the energy to separate bound charges in the monomer is linear} such that $\tilde u = \frac{1}{2} \dipole \cdot \sustens^{-1} \dipole$, where $\sustens^{-1}$ is the generalized inverse of $\sustens$.
In the general setting, $\bfE_0$ and $\bfp$ are thermodynamic conjugates, and we use the appropriate free energies that are Legendre transform pairs.

We mention that the relation
\begin{equation*}
    \takepartial{\A}{\rvec} = \frac{\kB \T}{\mlen} \mults
\end{equation*}
follows from a similar series of arguments as above.
Physically, this means that $\mults$ is a nondimensional measure of the tension in the chain.

%%%%%%%%%%%%%%%%%%%%%%%%%%%%%%%%%%%%%%%%%%%%%%%%%%%%%%
%%%%%%%%%%%%%%%%%%%%%%%%%%%%%%%%%%%%%%%%%%%%%%%%%%%%%%
%%%%%%%%%%%%%%%%%%%%%%%%%%%%%%%%%%%%%%%%%%%%%%%%%%%%%%
%%%%%%%%%%%%%%%%%%%%%%%%%%%%%%%%%%%%%%%%%%%%%%%%%%%%%%
%%%END OF MAIN TEXT%%%


\balance


%%%REFERENCES%%%
\providecommand*{\mcitethebibliography}{\thebibliography}
\csname @ifundefined\endcsname{endmcitethebibliography}
{\let\endmcitethebibliography\endthebibliography}{}

\end{document}